\documentclass[aps,prd,twocolumn,nofootinbib,superscriptaddress,10pt]{revtex4-2} 

\usepackage{graphicx}  
\usepackage{dcolumn}   
\usepackage{amsmath,amssymb,amsxtra,bm,epsfig}   
\usepackage{bm} 
\usepackage{float}
\usepackage[table, dvipsnames]{xcolor}
\usepackage{xspace}
\usepackage{bbding}
\usepackage{hyperref}
\usepackage{cancel, soul, ulem}
\usepackage{array}
\usepackage{multirow}
\usepackage{tikzsymbols}
\usepackage{tikz}
\usetikzlibrary{arrows.meta}
\usepackage{orcidlink}

\usepackage{tikz-feynman}
\usepackage{stmaryrd}
\tikzfeynmanset{compat=1.1.0}
\usetikzlibrary{arrows}
\tikzset{	
	vertex/.style={circle,draw, minimum size=1.5em},	
	edge/.style={->,> = latex'}	
}
\hypersetup{
	colorlinks=true,
	linkcolor=red,
	filecolor=magenta,
	urlcolor=RoyalBlue,
	citecolor=ForestGreen}

\newcommand{\be}{\begin{eqnarray*}}
	\newcommand{\ee}{\end{eqnarray*}}

\newcommand{\bee}{\begin{eqnarray}}
	\newcommand{\eee}{\end{eqnarray}}
\newcommand{\beeq}{\begin{equation}}
	\newcommand{\eeq}{\end{equation}}

\usepackage{xspace}

\newcommand{\ba}{\begin{array}}
	\newcommand{\ea}{\end{array}}
\newcommand{\bd}{\begin{displaymath}}
	\newcommand{\ed}{\end{displaymath}}
\newcommand{\besub}{\begin{subequations}}
	\newcommand{\eesub}{\end{subequations}}
\newcommand{\bea}{\begin{eqnarray}}
	\newcommand{\eea}{\end{eqnarray}}

\def\q2 {q^2}

\def\r {\rightarrow}

\usepackage{tikz-feynman}
\tikzfeynmanset{compat=1.1.0}
\usetikzlibrary{arrows,automata,positioning}
\usetikzlibrary{snakes}
\tikzset{
  vertex/.style={circle,draw, minimum size=1.5em},
    edge/.style={->,> = latex'}
}
\tikzfeynmanset{
  fermion1/.style={
    /tikz/postaction={
      /tikz/decoration={
        markings,
        mark=at position 0.5 with {
          \node[
            transform shape,
            xshift=-0.5mm,
            fill,
            dart tail angle=100,
            inner sep=1.3pt,
            draw=none,
            dart
          ] { };
        },
      },
      /tikz/decorate=true,
    },
  },
  fermion2/.style={
    /tikz/postaction={
      /tikz/decoration={
        markings,
        mark=at position 0.5 with {
          \arrow{>[length=5pt, width=4pt]};
        },
      },
      /tikz/decorate=true,
    },
  },
}

\begin{document}

\title{\boldmath Exploring Leptogenesis, WIMP Dark Matter, and Gravitational Waves in an extended Scalar Framework}
  
\author{Subhaditya Bhattacharya\,\orcidlink{0000-0002-8841-603X}\,}
\email{subhab@iitg.ac.in}
\author{ Niloy Mondal\,\orcidlink{0009-0006-5837-9772}}
\email{niloy18@iitg.ac.in}
\author{and Arunansu Sil\,\orcidlink{0000-0001-9732-8958}\,}
\email{asil@iitg.ac.in}
\affiliation{Department of Physics, Indian Institute of Technology Guwahati,\\ North Guwahati, Assam-781039, India}

\begin{abstract}
We explore an extension of the Type-I seesaw framework in which a scalar mediator $\Phi$ connects the Standard Model (SM) to a dark sector consisting of right-handed neutrinos or RHNs ($N_i$) and a complex scalar $S$. The interactions are governed by a phenomenologically viable $\mathcal{Z}_4 \times CP$ symmetry, allowing a dimension-5 effective operator involving the SM lepton doublet, Higgs, RHNs, and $\Phi$. Once $\Phi$ acquires a vacuum expectation value, its couplings to the SM and RHNs affect the standard leptogenesis through $N_i$ decay, while electroweak symmetry breaking (EWSB) subsequently induces neutrino masses through the same structure.
Furthermore, a $\Phi$-portal interaction induces a vacuum expectation value for real (imaginary) component of $S$ opening additional annihilation channels for its imaginary (real) component, identified as dark matter (DM) in the GeV-TeV mass range (excluded in the simplest scalar singlet DM scenario). The spontaneous breaking of the associated discrete symmetry can lead to domain wall formation, whose required early annihilation may generate observable gravitational wave signals, providing a way to ideate the symmetry-breaking scale and its connection to neutrino mass, baryogenesis, and dark matter.

\end{abstract} 


\makeatletter
\gdef\@fpheader{}
\makeatother

\maketitle
\flushbottom
\section{Introduction}

A wide range of theoretical and experimental observations point compellingly toward the existence of physics beyond the Standard Model (BSM), with prominent motivations arising from the origin of tiny neutrino masses, the observed baryon asymmetry of the universe (BAU), and the nature of dark matter, which constitutes about $26\%$ of the universe’s energy density and roughly $85\%$ of its matter content. Despite extensive theoretical developments and experimental efforts, the lack of conclusive evidence for new physics (NP) suggests that such phenomena may lie at high-energy scales or be only weakly coupled to the SM, often necessitating effective field theory frameworks ~\cite{Buchmuller:1985jz,Grzadkowski:2010es,Masso:2014xra,Englert:2014cva,Calibbi:2017uvl,Aebischer:2021uvt,Azatov:2016sqh,saavedra2018interpretingtopquarklhcmeasurements,CMS:2020lrr,Degrande:2012wf,Coito:2022kif,Beneito:2023xbk} 
in which NP manifests through higher-dimensional operators rather than direct detection. These challenges have motivated complementary approaches to probing NP--most notably through gravitational-wave (GW) observations by ongoing and proposed experiments such as LISA~\cite{amaroseoane2017laserinterferometerspaceantenna}, LIGO~\cite{LIGOScientific:2007fwp}, and BBO~\cite{Crowder:2005nr}. In parallel, recent theoretical developments ~\cite{Bhattacharya:2018ljs, Asaka:2005an, Datta:2021elq, Bhattacharya:2021jli, Herrero-Garcia:2023lhv, Falkowski:2011xh, DuttaBanik:2020vfr, An:2009vq, Cosme:2005sb, Chun:2011cc, Barman:2021tgt, Ma:2006fn, Hambye:2006zn, Gu:2007ug, Gu:2008yj, Aoki:2008av, Aoki:2009vf, Josse-Michaux:2011sjn, Herrero-Garcia:2024tyh, Ghosh_2025, Brune:2025zwg, Datta:2024xhg, Kasai:2024diy, Borah:2023god, Bhattacharya:2023kws} have focused on minimal extensions of the SM that coherently address neutrino masses~\cite{KATRIN:2024cdt}, the DM relic abundance, and the BAU, as all three call for modifications to the electroweak sector; however, meeting these simultaneous requirements typically entails a delicate interplay between heavy BSM states and weakly coupled NP.

Type-I seesaw framework~\cite{Minkowski:1977sc, Gell-Mann:1979vob, Mohapatra:1979ia, Schechter:1980gr, Schechter:1981bd} with right handed neutrinos (RHN) acts as the minimal model that can address both neutrino mass generation and baryon asymmetry of the Universe via leptogenesis~\cite{Fukugita:1986hr} or coherent oscillations of nearly degenerate RHNs~\cite{Akhmedov:1998qx} popularly known as ARS mechanism, and DM in some special circumstances~\cite{Shi:1998km, Canetti:2012vf, Datta:2021elq, Gorbunov:2025nqs, akita2026maximalparameterspacesterile}. In this work we extend the type-I seesaw model by a scalar singlet $\Phi$ that acts like a mediator between the neutrino and the DM sector, primarily via a dimension-five opeartor, invariant under a suitably chosen dark symmetry $\mathcal{G}_{ D}$ while the conventional tree level Yukawa interaction between the SM lepton, Higgs, and RHN remains absent. This higher dimensional operator, after $\Phi$ gets a vacuum expectation value (VEV),  effectively generates the neutrino Yukawa interaction making it to be the primary source of neutrino mass generation. The same operator also enriches the standard picture of leptogenesis through right handed neutrino decay at a high enough scale with additional features. 

Building upon the framework developed in~\cite{Bhattacharya:2023kws} where the DM is a fermion, 
we address here a scalar DM instead, by introducing a complex scalar field $S$. Leptogenesis, DM genesis, and neutrino 
mass generation all activate only after the spontaneous breaking of $\mathcal{G}_{D} \equiv \mathcal{Z}_2\times\mathcal{Z}_2^{\prime}$, or equivalently $\mathcal{Z}_{4}$ in our case for reasons to be elaborated in the main text, through the VEV of $\Phi$. 
Interestingly, the VEV of $\Phi$ also induces a VEV to $S$ as a result of which, both of its components 
may not remain stable in general at the cosmological scale. It turns out that the situation can be improved by enhancing the imposed symmetry to $\mathcal{G}_{D}^\prime\equiv\mathcal{G}_{D} \times CP$ of the Lagrangian, which keeps one component of the complex scalar field $S$ stable rendering it as a DM candidate.
We treat this DM as a weakly interacting massive particle (WIMP)~\cite{Kolb:1988aj}, and adopt the standard thermal 
production mechanism for its relic abundance ($\Omega^{}_{\rm DM}h^2=0.12\pm0.0012$~\cite{ParticleDataGroup:2024cfk}). 
Our analysis reveals two distinct scenarios having respective allowed regions of the parameter spaces where the DM relic density exhibits different (opposite) dependence on the induced VEV of $S$ ($v^{}_{s}$). This turns out to be a salient feature of our work which has not been reported earlier in the literature.

\begin{table*}[htb]
\centering
\renewcommand{\arraystretch}{1.3}

\begin{tabular}{|c|c|c|c|c|c|c|c|}
\hline

\multirow{2}{*}{Symmetry} 
& \multicolumn{7}{c|}{Charges of BSM fields under the symmetry and allowed terms} \\
\cline{2-8}

& $\Phi$
&Scalar $\Phi$
& $N_i$ 
& $\rm S$ 
& $\frac{\bar{l}\tilde{H}\Phi N}{\Lambda}$ 
& $\overline{N^c_{}}N$ 
& Scalar DM \\
\hline

$\mathcal{Z}_2 \times \mathcal{Z}_2'$ ($\mathcal{G}_D$) 
& $(1,2)$ 
& Real/Complex
& $(1,2)$ 
& $(2,1)$ 
& \Checkmark 
& \Checkmark 
& Real/Complex \\
\hline

$\mathcal{Z}_3 \times \mathcal{Z}_2'$ ($\mathcal{G}_D$)
& $(1/2,2)$
& Complex
& $(2/1,2)$ 
& $(3,1)$ 
& \Checkmark 
& \XSolidBrush 
& Real/Complex \\
\hline

$\mathcal{Z}_4 \times \mathcal{Z}_2'$ ($\mathcal{G}_D$) 
& $(2,2)$ 
& Real/Complex
& $(2,2)$ 
& $(4,1)$ 
& \Checkmark 
& \Checkmark  
& Real/Complex \\
\hline

$\mathcal{Z}_2 \times \mathcal{Z}_2' \times CP$ ($\mathcal{G}_D^\prime$)
& $(1,2,\mathcal{I})$
& Real/Complex
& $(1,2,\mathcal{I})$ 
& $(2,1,\; \rm S \rightarrow S^\ast/-S^\ast)$ 
& \Checkmark  
& \Checkmark  
& $\mathrm{Im}[\rm S]/\mathrm{Re}[\rm S]$ \\
\hline

$\mathcal{Z}_4 \times CP$ ($\mathcal{G}_D^\prime$)
& $(2,\mathcal{I})$ 
& Real/Complex
& $(2,\mathcal{I})$ 
& $(3,\; \rm S \rightarrow S^\ast/-S^\ast)$ 
& \Checkmark  
& \Checkmark  
& $\mathrm{Im}[\rm S]/\mathrm{Re}[\rm S]$ \\
\hline

\end{tabular}

\caption{Charges of BSM fields under the symmetry groups $\mathcal{G}_{ D}$ (or $\mathcal{G}_{ D}^\prime$) and the corresponding allowed terms in the Lagrangian. We adopt the convention that, under the discrete symmetry $\mathcal{Z}_n$, the fields can carry charge $q \in \{1,\ldots,n\}$. For the $CP$ symmetry, a field is either invariant (denoted by $\mathcal{I}$) or transforms into its complex conjugate (with either $+$ or $-$ coefficient).}
\label{tab:k}

\end{table*}
Spontaneous breaking of a discrete symmetry in the early universe can create sheet-like two-dimensional topological defects known as domain walls (DW)~\cite{Vilenkin:2000jqa}. The energy density of the DWs scales with cosmic time as $\rho^{}_{_{DW}}\propto t^{-1}$, in contrast to the energy density of matter (radiation), which scales as 
$\rho^{}_{M(R)}\propto a^{-3}(a^{-4})\propto t^{-3/2}(t^{-2})$, $a$ denoting the scale factor of the Universe. This slower redshift causes DWs to eventually dominate the energy density of the universe (if they are created after inflation), leading to a cosmological evolution incompatible with standard observations~\cite{Zeldovich:1974uw}. This tension is commonly referred to as the domain wall problem. One can resolve this issue by introducing a small, explicitly symmetry-breaking term to the Lagrangian, which will lift the vacuum degeneracy and make the DW unstable. This will eventually help to annihilate the DW network before the overclosing of the universe. Crucially, the annihilation of the DW can produce a stochastic GW background, potentially detectable by the present and future GW detectors, as demonstrated for this model.

The paper is organized as follows. Section~\ref{sec:model} introduces the model. We discussed the scalar potential in Section~\ref{sec:Scalar sector}. Neutrino masses, leptogenesis and DM phenomenology are adressed consequently in Section~\ref{sec:Neutrino Masses}, Section~\ref{leptogenesis} and~Section\ref{sec:dm phenomenology}. Gravitational wave production from domain wall annihilations and prospects for testing the discrete symmetry scale are discussed in Section~\ref{sec:domain wall}. We conclude in Section~\ref{sec:summary}.

\section{The Model}\label{sec:model}

As mentioned in the introduction, our model is an extension of type I seesaw framework by a complex scalar field $S$ (one of the components of which acts as DM), and a scalar field $\Phi$ as mediator which connects the DM to neutrino mass generation and baryogenesis via leptogenesis. Hence, the scalar sector of our construction consists of a scalar potential $V(H,\Phi,S)$ with $\Phi$, $S$ and the usual SM Higgs doublet $H$.
Furthermore, we incorporate three Majorana RHNs ($N_i$) 
to generate neutrino masses via seesaw mechanism and to explain the observed baryon asymmetry of the universe through leptogenesis. 
The newly introduced fields are all charged under a generic discrete $\mathcal{Z}_n$ symmetry for example, 
while none of the SM fields have $\mathcal{Z}_n$ charge: requiring
the Lagrangian to remain invariant under it. A generic field $\chi$ with $\mathcal{Z}_n$ charge $q$ transforms as $\chi \rightarrow e^{i2\pi q/n} \chi$, 
a prescription that equally applies to $\Phi$, $N_i^{}$ and $S$. 
The RHNs charged under $\mathcal{Z}_n$ prohibit the standard  tree level Yukawa interaction $\overline{l}  \widetilde{H} N$. 
Further the $\mathcal{Z}_n$ symmetry breaks spontaneously by the VEV of $\Phi$, so that 
additional symmetries (such as $\mathcal{Z}_{2}^{\prime}$ or $\mathcal{Z}_2^{\prime}\times CP$) 
need to be imposed for the stability of the DM candidate $S$. 

The Lagrangian of the model, invariant under the $\mathrm{SM} \times \mathcal{G}_{D}$ symmetry encapsulate,
\begin{equation}\label{eq:lag}
 -\mathcal{L}\supset \frac{\mathcal{Y}_{\alpha i}}{\Lambda}\overline{l_{L_\alpha}}  \widetilde{H}\Phi N_i+\frac{{M}_{ij}}{2}\overline{N^c_i}N_j + V(H,\Phi,S)+ h.c.\;.
\end{equation}
Note the presence of an higher dimensional effective Yukawa interaction (a possible UV completion is discussed in Appendix~\ref{app:UVcomp}) term in Eq.~\ref{eq:lag}, involving $\Phi$ with suitable 
charges of $\mathcal{Z}_n$ symmetry assigned to $\Phi,S,N_i$. Here the SM lepton doublet is denoted by $l_L$, 
$\tilde{H} = i\sigma_2 H^\ast$, $\Lambda$ sets the cut-off scale of the theory, 
while $\{\alpha = e, \mu, \tau\}$ and $\{i,j=1,2,3\}$ correspond to flavor and generation indices respectively.  
This higher dimensional Yukawa interaction effectively generates the standard neutrino Yukawa term after $\Phi$ gets VEV, which holds the key to have the seesaw and leptogenesis.
The Yukawa coupling $\mathcal{Y}$ is a general complex $3 \times 3$ matrix, while the new physics scale $\Lambda$ 
must be heavier than the heaviest RHN. The possible realisations of the discrete symmetry groups $\mathcal{Z}_n$, under which the Lagrangian of Eq.~\ref{eq:lag} remains invariant, 
is listed in Table~\ref{tab:k}. However, the choice of dark symmetry $\mathcal{G}_{ D}\equiv\mathcal{Z}_{3}\times\mathcal{Z}_{2}^{\prime}$ (or $\mathcal{G}_{ D}\equiv\mathcal{Z}_{2p+1}\times\mathcal{Z}_{2}^{\prime}$, with $N_i$ transforming non-trivialy under $\mathcal{Z}_{2p+1}$) is not viable, 
as it forbids the Majorana mass term for the RHNs. Consequently, the Lagrangian of Eq.~\ref{eq:lag} is invariant only under symmetries of the form 
$\mathcal{G}_{ D}\equiv\mathcal{Z}_{2p}\times\mathcal{Z}_{2}^{\prime}$ (or $\mathcal{G}_{ D}^{\prime}\equiv\mathcal{Z}_{2p}\times\mathcal{Z}_{2}^{\prime}\times CP$). To accommodate the Majorana mass term, the RHNs must carry charge $p$ under the symmetry $\mathcal{Z}_{2p}$, with $p$ being a positive integer ($p\geq1$). The invariance of the term, $\overline{l}  \widetilde{H}\Phi N$, under this symmetry then requires the field $\Phi$ to carry the same charge $p$. It is important to note that the irreducible one-dimensional representations of $\mathcal{Z}_{2p}$ correspond to $\mathcal{Z}_{2p}\equiv\{g,g^2,\dots,g^{2p}\}$, with the generator $g=e^{i\frac{2\pi}{2p}}$, of which only $g^{2p}=e^{i2\pi 2p/2p}=1$, and $g^{p}=e^{i2\pi p/2p}=-1$ are real. Since a real field cannot transform under a complex one-dimensional representation, $\Phi$ can be considered as a real field only if it transforms under a real representation of $\mathcal{Z}_{2p}$ with characters $1$ or $-1$. In our case, $\Phi \rightarrow e^{i\pi}\Phi=-\Phi$, which is a real representation of $\mathcal{Z}_{2p}$. Hence, one can safely treat $\Phi$ as a real scalar field under this symmetry structure (see also Refs.~\cite{CentellesChulia:2016rms,Kim:2024cwp} for analogous arguments in $\mathcal{Z}_4$ frameworks). Promoting it to a complex field would introduce an additional degree of freedom without any essential benefit, and would merely constitute a non-minimal extension of the setup.

$V(H, \Phi, S)$ of Eq.~\ref{eq:lag} denotes the scalar potential involving the SM Higgs doublet $H$ and the dark sector scalars $\Phi$ and $S$. The distinction between different symmetries as mentioned in Table \ref{tab:k} actually arises in $V(H,\Phi,S)$, which we address shortly. Note that once the $\mathcal{Z}_{2p}$ symmetry is spontaneously broken through the VEV of $\Phi$, the active neutrino masses can be generated via the dimension-5 operator $(\mathcal{Y}/\Lambda)(\overline{l}\widetilde{H}\Phi N)$ in the Lagrangian, through the type-I Seesaw mechanism after the EWSB. Thus, the field $\Phi$ plays a crucial role in this framework, establishing a connection between the neutrino sector and the dark sector. The right-handed neutrino mass matrix $M$ is symmetric and typically complex. One can choose a basis where $M$ is diagonal ($M^{dia}=\text{diag}[M_1,M_2,M_3]$), but in this particular basis, the Yukawa coupling matrix $\mathcal{Y}$ is non-diagonal. It is important to note further that under the $\mathcal{Z}_{2p}\times\mathcal{Z}_{2}^{\prime}$ and/or $\mathcal{Z}_{2p}\times\mathcal{Z}_{2}^{\prime}\times CP$ symmetry transformation, with the charges assigned to $N_{i},~\Phi$ and $S$ as in Table~\ref{tab:k}, certain renormalizable terms such as
 $\overline{N^c} N \Phi$, and $\overline{N^c}N(S+S^{\ast})$ are forbidden. The exact choice of the symmetry will be elaborated below.
 
\section{Scalar sector, induced VEV, Dark Matter stability, and constraints}
\label{sec:Scalar sector}

We discuss here the rationale behind the symmetry chosen for $\mathcal{G}_{ D}$ and/or $\mathcal{G}_{ D}^\prime$ from the perspective of induced VEV and DM phenomenology. The detailed analysis 
of most of the possibilities mentioned in Table~\ref{tab:k} are done in Appendix \ref{sec:Scalar potential}. We provide a summary here, indicate the choice for phenomenological analysis, and the constraints applicable to the scenario.

\subsection{Choice of symmetry and Dark Matter stability}
\label{sec:Choice of symmetry}

To systematically address the optimal choice for the symmetry,
let us first examine the case where the DM particle is taken to be a real scalar $S$. The scalar potential, invariant under $\mathcal{Z}_{2p}\times\mathcal{Z}_2^{\prime}$, 
with real scalar $\Phi$ and $S$ is given in Eq.~\ref{eq:SM+2RSZ2}. After the EWSB, the DM mass takes the form
\begin{equation}\label{eq:mDM2RZ2}
\mathbf{M}_{S}^{2}=\mu_{S}^{2}+\frac{\lambda_{SH}}{2}v^2+\frac{\lambda_{\phi S}}{2}v_{\phi}^2\,.
\end{equation}
In presence of the interaction term $\overline{l}  \widetilde{H}\Phi N$, successful leptogenesis can only occur after the 
spontaneous symmetry breaking of $\mathcal{Z}_{2p}$ by requiring 
$v_{\phi}\geq10^{9}~\text{GeV}$~\cite{Bhattacharya:2023kws}. Unbroken $\mathcal{Z}_2^{\prime}$ ensures the stability of DM. Consequently, 
achieving DM mass in the GeV-TeV range necessitates a highly suppressed coupling, $\lambda_{\phi S}\lesssim10^{-12}$. Within this framework, 
if we aim to study thermal production of WIMP DM, the scenario effectively resembles to that of a scalar singlet Higgs portal DM model. However, as established in 
earlier studies~\cite{Beniwal:2020hjc, GAMBIT:2018eea, Cline:2013gha, Burgess:2000yq, EscuderoAbenza:2025cfj}, DM direct detection strongly constrains 
scalar singlet WIMP dark matter, pushing the DM mass into the multi-TeV regime and/or leaving a narrow window of Higgs resonance, where 
kinetic decoupling may occur before thermal freeze out~\cite{Binder:2017rgn, DiMauro:2023tho}. 

We now turn to the case where $S$ is taken to be a complex scalar. The scalar potential with one real scalar ($\Phi$) and one complex scalar 
($S\equiv(\mathbf{s}_1^{}+iS^{}_2)/\sqrt{2}$), invariant under $\mathcal{Z}_{2p}\times\mathcal{Z}_2^{\prime}$, is given by Eq~\ref{eq:scalarpot}. 
In this scenario after EWSB, the two components of $S$ acquire masses as
\begin{equation}\label{eq:DMmassZ2r}
\mathbf{M}_{S_{1,2}^{}}^{2}=\mu_{S}^{2}+\frac{\lambda_{SH}}{2}v^2\mp\frac{(\lambda_{\phi S}-|\lambda_{\phi S}^{\prime}|)}{2}v_{\phi}^2\,.
\end{equation}
The component of $S$ with smaller mass serves as the DM particle. 
In the limit $\lambda_{\phi S}\gg|\lambda_{\phi S}^{\prime}|$ with DM mass in the GeV-TeV range, both the components of $S$ become nearly degenerate, 
and the DM phenomenology essentially reduces to that of two real scalar singlet DMs. However, there exists a region of parameter 
space where one of the components of $S$ (either real or imaginary) develops an induced VEV. This occurs due to the spontaneous 
breaking of $\mathcal{Z}_{2p}$ symmetry in presence of the interaction terms $\lambda^{}_{\phi S}\Phi^2 (S^{\ast}S)$ and 
$\lambda^{\prime}_{\phi S}\Phi^2S^2$ (see Appendix~\ref{subsec: 1R1Csinglet scalar} for details). 
Such acquisition of induced VEV by one component of $S$ inevitably breaks the $\mathcal{Z}_{2}^{\prime}$ symmetry, 
leaving no symmetry to protect the stability of the remaining component of $S$ as DM. 

An additional symmetry is required to stabilise one of the components of complex scalar $S$ to became a viable DM candidate. A natural choice is the CP symmetry under which 
$S$ transforms as $S\rightarrow S^\ast$ (or equivalently $S\rightarrow-S^\ast$). In presence of the extended symmetry $\mathcal{G}_{D}^\prime\equiv\mathcal{Z}_{2p}\times\mathcal{Z}_{2}^{\prime}\times CP$, we can have a DM candidate even in the parameter space where an induced VEV is acquired by one component of $S$ (see Appendix~\ref{subsec: 1R1Csinglet scalar} for details). 
Importantly, this induced VEV satisfied region allows for a viable DM candidate with mass in the \text{GeV}-\text{TeV} range, which will be discussed in section~\ref{sec:dm phenomenology}.
The minimal symmetry structure for $\mathcal{G}_{D}^\prime$ must therefore be $\mathcal{Z}_{2}\times\mathcal{Z}_{2}^{\prime}\times CP$ to accommodate a stable scalar DM; 
or $\mathcal{G}_{D}^\prime\equiv\mathcal{Z}_{4}\times CP$, which are fully equivalent (for details, check Appendix~\ref{subsec: equivalence}), and we continue the rest of the discussion with it.

The most general scalar potential involving the $SU(2)_L$ isodoublet $H$ and $\mathcal{Z}_{4}$ charged real scalar $\Phi$ and complex scalar $S$ 
(charge assignments shown in Table~\ref{tab:k}), invariant under $\mathcal{Z}_{4}\times CP\times SM$ symmetry is given by:
\begin{equation}\label{eq:scalarpotZ4cp}
    \begin{split}
        V(H,\Phi,S)  = & -\mu_H^2H^{\dagger}H -\frac{1}{2}
\mu_\phi^2\Phi^2+\mu_S^2 S^{\ast}S+\lambda^{}_H(H^{\dagger}H)^2 \\
        & + \frac{1}{4!}\lambda^{}_\phi\Phi^4 + \lambda^{}_S (S^{\ast}S)^{2}+\lambda^{}_{S'}\big(S^4+(S^{\ast})^4\big)\\
        & +\lambda^{}_{S''}\big(S^2+(S^{\ast})^2\big)^2+\frac{1}{2}\lambda^{}_{\phi H}\Phi^2H^{\dagger}H\\
        &+\lambda^{}_{SH}(S^{\ast}S)H^{\dagger}H+\frac{1}{2}\lambda^{}_{\phi S}\Phi^2 (S^{\ast}S)\\
        &+\frac{\mu^{}_{\phi S}}{2}\Phi (S^2+S^{\ast 2})\,.\\
    \end{split}
\end{equation}
Here we choose $\lambda^{}_H, \lambda^{}_\phi > 0$ and $\mu_H^2, \mu_\phi^2 > 0$, which ensure that the scalar potential $ V(H, \Phi, S)$ acquires two stable minima along the $\Phi$ and $H$ directions. For the time being, we restrict ourselves to the case $\mu_S^2 > 0$. In the unitary gauge, the scalar fields take the form:\\
$H=\frac{1}{\sqrt{2}}\begin{pmatrix}
    0\\
    \mathcal{H}+v\\
\end{pmatrix},~~\Phi=(\mathit{\Phi}+v_{\phi}),~~S=(\mathbf{s}_1^{}+iS^{}_2)/\sqrt{2}$.  \\
Here, $v=246~\mathrm{GeV}$ denotes the VEV of the SM Higgs. 

The presence of the $\mu^{}_{\phi S}\Phi (S^2+S^{\ast 2})$ term allows one component of $S$ to acquire induced VEV ($v^{}_{s}$) as a consequence of $\mathcal{Z}_4$ symmetry breaking. The other ($\mathbf{s}^{}_1$ or $S^{}_2$) remains stable due to the imposed $CP$ symmetry under which $\mathbf{s}_{1}\rightarrow-\mathbf{s}_{1}$ (or $S_{2}\rightarrow-S_{2}$) and becomes a DM. The component which acquires induced VEV, introduces new additional annihilation channels for the DM, thereby enlarging the DM parameter space in the GeV-TeV range. The condition for $\mathbf{s}^{}_1$ to get induced VEV is given by:
\begin{equation}
     -\mu^{}_{\phi S}>\frac{\mu_S^2}{v_\phi^{}}+\frac{\lambda^{}_{\phi S}v_{\phi}^{}}{2}.
     \label{mu-1}
\end{equation}
 It is evident from the above expression that, if all the couplings are real and positive, $\mu^{}_{\phi S}$ must be negative for $\mathbf{s}^{}_1$ to develop an induced VEV. 
 Similarly, the condition for $S_2$ to acquire an induced VEV is given by:
\begin{equation}
     \mu^{}_{\phi S}>\frac{\mu_S^2}{v_\phi^{}}+\frac{\lambda^{}_{\phi S}v_{\phi}^{}}{2}.
     \label{mu-2}
\end{equation}
With $\mu_S^2<0$ and all real and positive couplings, it can be ensured that an induced VEV for $S^{}_2$ results naturally while $\mathbf{s}^{}_1$ does not acquire any VEV (a simultaneous satisfaction of Eqs. \ref{mu-1} and \ref{mu-2} is not possible) which is a manifestation of the imposed CP symmetry on $S$. This further ensures the stability of $\mathbf{s}_1$ making it a viable DM. 

The $S_{2}^{}$ can be expanded around the VEV as $S_2=(\tilde{S}_{2}+v_{s})$. 

The extremization conditions of the potential lead to the following relations,
\begin{align}
& \mu_{H}^{2}=\bigg(\lambda^{}_{H}v^{2}+\frac{\lambda^{}_{\phi H }}{2}v_{\phi}^{2}+\frac{\lambda^{}_{S H }}{2}v_{s}^{2}\bigg)\, ,\\
& \mu_{\phi}^2=\bigg(\frac{\lambda^{}_{\phi}}{6}v_{\phi}^{2}+\frac{\lambda^{}_{\phi H}}{2}v^{2}+\frac{\lambda^{}_{S \phi }}{2}v_s^{2}-\frac{\mu^{}_{S \phi}}{2v_\phi}v_s^2\bigg)\, ,\\
& \mu_{S}^2=\bigg(\mu^{}_{S \phi}v_{\phi}-\frac{\lambda^{}_{S H }}{2}v^{2}-\frac{\lambda^{}_{S \phi }}{2}v_\phi^{2}-\lambda^{}_{S^{}_{eff}}v_s^2\bigg)\,,
 \end{align}
 Here, the quantity $\lambda^{}_{S^{}_{eff}}=(\lambda^{}_{S}+2\lambda^{}_{S'}+4\lambda^{}_{S''})$ is the effective quartic coupling for $S$ field. Utilising the expression for $\mu_S^2$, the mass parameter $\mu^{}_{S\phi}$ can be expressed in terms of DM mass as
\begin{equation}\label{eq:mdm}
   \mu^{}_{S \phi}=\frac{M_{\mathbf{s}_1}^2+8(\lambda^{}_{S'}+\lambda^{}_{S''})v_s^2}{2v_\phi}.
\end{equation}
The spontaneous symmetry breaking induces mixing between the scalar fields with non-zero VEVs, resulting in the following mass matrix of the scalar sector, expressed in the $(\mathcal{H}~\mathit{\Phi}~\tilde{S}_{2})^T$ basis 
{\small{\bea
\textit{M}_{}^2 = \begin{pmatrix}
    2\lambda^{}_H v^2 & ~~\lambda^{}_{\phi H} v v_{\phi} & \lambda^{}_{SH}vv_s \\
    \lambda^{}_{\phi H} v v_{\phi} &  ~~2\bigg(\frac{\lambda^{}_\phi}{6}+\mu^{}_{S \phi}\frac{v_s^2}{4 v_\phi^3}\bigg)v_\phi^2 & \big(\lambda^{}_{S \phi}v_s v_\phi-\mu^{}_{S \phi}v_s\big)\\
    \lambda^{}_{SH}vv_s & \big(\lambda^{}_{S \phi}v_s v_\phi-\mu^{}_{S \phi}v_s\big) & 2\lambda^{}_{S^{}_{eff}}v_s^2
  \end{pmatrix}\,.\nonumber
\eea}}
The following transformation diagonalises the mass matrix and gives the eigenstates of the physical mass as $(h~\phi~\mathbf{s}^{}_2)^T$.
{\scriptsize\bea\label{eq:scalarmix}
\begin{pmatrix}
    h \\
    \phi\\
    \mathbf{s}^{}_2
  \end{pmatrix} = \begin{pmatrix}
    c^{}_{12}c^{}_{13}  &  s^{}_{12}c^{}_{13}  &  s^{}_{13}      \\
    -s^{}_{12}c^{}_{23}-c^{}_{12}s^{}_{13}s^{}_{23}  &  c^{}_{12}c^{}_{23}-s^{}_{12}s^{}_{13}s^{}_{23}   &  c^{}_{13}s^{}_{23}\\
    s^{}_{12}s^{}_{23}-c^{}_{12}s^{}_{13}c^{}_{23}   &    -c^{}_{12}s^{}_{23}-s^{}_{12}s^{}_{13}c^{}_{23}    &    c^{}_{13}c^{}_{23}
\end{pmatrix}^\dagger
  \begin{pmatrix}
    \mathcal{H} \\
    \mathit{\Phi}  \\
    \tilde{S}_2
  \end{pmatrix}.\nonumber
\eea}
The quantum field $h$ corresponds to the Standard Model Higgs boson with a mass of $M^{}_h=125.35$~GeV \cite{ATLAS:2012yve, CMS:2012qbp}. The mixing angles between the three scalars are denoted by 
$\theta_{12}, \theta_{23}$ and $ \theta_{13}$. For convenience, we use the shorthand $c_{ij}(s_{ij})=\text{cos}~\theta_{ij}(\text{sin}~\theta_{ij})$. The complete list of external and internal parameters is provided in Table~\ref{tab:p}. Note that the quartic and cubic couplings in the scalar potential can be expressed in terms of the scalar masses, their vacuum expectation values, and the mixing angles among them. 
\begin{table*}
\centering
\setlength{\tabcolsep}{3.5pt} 
\renewcommand{\arraystretch}{1.2} 
\resizebox{15cm}{!}{%
\begin{tabular}{|c|c|}
\hline
 External parameters &   Internal parameters \\
\hline
$M_\phi$, $M_{\mathbf{s}^{}_2}$, $M_{\mathbf{s}^{}_1}$, $M_i$, $\Lambda$, $\lambda^{}_{S'}=\lambda^{}_{S''}$ , $v_{\phi}$, $v_{s}$, $\theta^{}_{12}\equiv\theta$, $\theta^{}_{13}\equiv\delta$, $\theta^{}_{23}\equiv\omega$ &  $\lambda^{}_{H}$, $\lambda^{}_{S}$, $\lambda^{}_{\phi}$, $\lambda^{}_{\phi H}$, $\lambda^{}_{SH}$, $\mu^{}_{\phi S}$, $\lambda^{}_{\phi S}$\\
\hline
\end{tabular}
}\
\caption{\label{tab:p} {\small External and Internal parameters of the model.}}
\end{table*}
Let us remind finally the physical scalars that appear after $\mathcal{Z}_4$ symmetry breaking in this model, stemming from one iso-doublet, one real scalar singlet, and one complex scalar singlet,
\begin{eqnarray}
\text{Before EWSB}: && H~(\rm {Higgs~doublet}),\mathit{\Phi}, \tilde{S}_2,~\text{and}~\mathbf{s}_1\,,\nonumber\\
\text{Aftert EWSB}: && h~(\rm {SM~Higgs}),\phi, \mathbf{s}_2,~\text{and}~\mathbf{s}_1~(\text{DM})\,.\nonumber
\end{eqnarray}

We will now offer a brief glimpse into the timeline of the relevant phenomena, which will be examined in detail in the forthcoming sections. In the early universe, at a temperature much higher than the RHN mass ($T_{\mathcal{Z}^{}_4} \gg M^{}_1$), the discrete symmetry undergoes spontaneous breaking. As a result, both $\Phi$ and $S_{2}^{}$ acquire vacuum expectation values. The phenomenon of discrete symmetry breaking leads to the formation of domain walls, which can subsequently annihilate in the presence of a bias term, generating a stochastic gravitational wave signal. As the temperature of the universe approaches the mass scale of $M_{1}$, lepton asymmetry is generated through the decay and scattering processes involving the lightest right-handed neutrino $N_{1}^{}$. This lepton asymmetry is converted into the baryon asymmetry via the non-perturbative sphaleron transitions. At a later stage, when the temperature falls below the electroweak scale $T_{EW}^{}$ and the Standard Model Higgs acquires its mass, the dark matter candidate $\mathbf{s}^{}_{1}$ begins to decouple from the thermal bath, freezing out at a temperature approximately given by $T^{}_{fo}\sim M_{\mathbf{s}^{}_{1}}/25$. The dynamics of leptogenesis in this framework primarily depends on the parameters $\{M^{}_{i}, v^{}_{\phi}\}$, while dark-matter phenomenology is governed by the set of parameters $\{M_{\mathbf{s}^{}_{2}}, M_{\mathbf{s}^{}_{1}}, v^{}_{s}, c^{}_{ij}\}$. For the rest of the analysis, we assume that all scalar mixing angles are equal and remain small, typically of the order $\mathcal{O}(10^{-2})$ or less~($c_{ij}^{}\leq\mathcal{O}(10^{-2})$). Under this assumption, it is justified to take $M^{}_{\mathit{\Phi}}=M^{}_{\phi}$ and $M^{}_{\tilde{S}_{2}}=M_{\mathbf{s}^{}_{2}}$. It is also important to remind the mass hierarchy that we adhere to in this set up so that we obtain a thermal leptogenesis, a thermal DM after EWSB, and a consistent $\mathcal{Z}_4$ breaking,
\begin{equation}
\Lambda\gg M_3\gg M_2>M_1> M_\phi>M_{\mathbf{s}_2}\sim M_{\mathbf{s}_1}\,.
\end{equation}

\section{Neutrino Masses}\label{sec:Neutrino Masses}

We can get the active neutrino mass ($m_\nu$) from Eq. \ref{eq:lag} using the type-I seesaw mechanism, after 
$\Phi$ gets the VEV, as given by
\begin{equation}\label{eq:numass}
m_\nu= v^2 y_\nu M^{-1} y_\nu^T.
\end{equation}
Here the effective coupling $y_\nu$ can be written in terms of $\Phi$ VEV as ${y_\nu}_{\alpha i} \equiv (\mathcal{Y}_{\alpha i}v_\phi)/\Lambda$ while the RHN mass matrix $M$ can be considered as diagonal $M^{\rm diag}$ without loss of generality. This light neutrino mass matrix ${m_\nu}$  can then be diagonalized using  $U_{\rm PMNS}$ matrix (which embeds the neutrino oscillation data in terms of three mixing angles) via $m_\nu=U_{\rm PMNS}^\ast ~D_\nu~ U_{\rm PMNS}$, where $D_\nu$ corresponds to the diagonalized light neutrino mass matrix. 
Using the Cassas-Ibarra parametrization, we can express the coupling $y_\nu$ in terms of low-energy neutrino oscillation data (though $U_{\rm PMNS}$ and mass-squared differences for hierarchical neutrinos) as,
\begin{equation}\label{eq:ci}
    y_\nu =\frac{\sqrt{2}}{v}\,U_{\rm PMNS}^\ast\,\sqrt{D_\nu}\, R^T\,\sqrt{M^{\rm diag}}\,.
\end{equation}
Here $R$ is the complex $3\times 3$ orthogonal matrix parameterised by complex angle $z^{}_R=a+ib$. It is worth mentioning that a dimension-5 operator of the form $\Phi^2 \overline{N^c} N/\Lambda$, consistent with the model’s symmetries, could in principle be present. Once $\Phi$ acquires a VEV, this term would contribute to the effective light neutrino mass. However, such a contribution can be absorbed into a redefinition of $M$.

\section{Leptogenesis from decay and scattering of heavy RHN}\label{leptogenesis}

\begin{figure}[!htb]
    \centering
    \begin{tikzpicture}
    \begin{feynman}
    \vertex(i1) {$N_k$};
    \vertex[right=1.25 cm of i1] (i2);
    \vertex[right=1.25 cm of i2] (e);
    \vertex[above=0.5 cm of e] (a) {$l_j$};
    \vertex[below=0.5 cm of e](b) {$H$};
    \diagram* {
    (i1)-- (i2)--[fermion2] (a),
    (i2)-- [scalar](b),
    };
    \end{feynman}
    \end{tikzpicture}
    \hspace{2em}
    \begin{tikzpicture}
    \begin{feynman}
    \vertex(i1) {$N_k$};
    \vertex[right=1.25 cm of i1] (i2);
    \vertex[right=1.25 cm of i2] (e);
    \vertex[above=0.7 cm of e] (a);
    \vertex[below=0.7 cm of e](b);
    \vertex[right=1.0 cm of a](c) {$l_j$};
    \vertex[right=1.0 cm of b](d) {$H$};
    \diagram* {
    (i1)-- (i2)--[scalar] (a),
    (b)-- [fermion2, edge label=\(l_n\)](i2),
    (a)--[edge label =\(N_m\)] (b),
    (a)-- [fermion2](c),
    (b)-- [scalar](d)
    };
    \end{feynman}
    \end{tikzpicture}
    \hspace{2em}
    \begin{tikzpicture}
    \begin{feynman}
    \vertex(i1) {$N_k$};
    \vertex[right=1.0 cm of i1] (i2);
    \vertex[right=1.0 cm of i2] (f); 
    \vertex[right=0.75 cm of f] (g);
    \vertex[right=1.0 cm of g] (e);
    \vertex[above=0.5 cm of e] (a) {$l_j$};
    \vertex[below=0.5 cm of e](b) {$H$};
    \diagram* {
    (i1)-- (i2),(f)--[fermion2,half right,edge label'=\(l_n\)] (i2),(f) --[scalar, half left] (i2),
    (f)--[edge label' =\(N_m\)] (g),
    (a)-- [fermion2] (g)-- [scalar] (b),
    };
    \end{feynman}
    \end{tikzpicture}
\begin{tikzpicture}
    \begin{feynman}
    \vertex(i1) {$N_k$};
    \vertex[right=2.5cm of i1](i2) {$l_j$};
    \vertex[below=1.5 cm of i1](a) {$\mathit{\Phi}$};
    \vertex[below=1.5 cm of i2](b) {$H$};
    \vertex[below=0.75cm of i1](c);
    \vertex[right=1.25cm of c] (e);    
    \diagram* {
    (i1)-- (e)--[fermion2] (i2),
    (a)-- [scalar](e) --[scalar] (b),
    };
    \end{feynman}
    \end{tikzpicture}
    \hspace{2em}
    \begin{tikzpicture}
    \begin{feynman}
    \vertex(i1);
    \vertex[above=0.5cm of i1] (g) {$N_k$};
    \vertex[below=0.5cm of i1] (h) {$\mathit{\Phi}$};
    \vertex[right=1.0 cm of i1] (i2);
    \vertex[right=1.0 cm of i2] (e);
    \vertex[above=0.7 cm of e] (a);
    \vertex[below=0.7 cm of e](b);
    \vertex[right=1.0 cm of a](c) {$l_j$};
    \vertex[right=1.0 cm of b](d) {$H$};
    \diagram* {
    (g)-- (i2)--[scalar] (a),
    (h)--[scalar] (i2),(b)-- [fermion2, edge label=\(l_n\)](i2),
    (a)--[edge label =\(N_m\)] (b),
    (a)-- [fermion2](c),
    (b)-- [scalar](d)
    };
    \end{feynman}
    \end{tikzpicture} 
    \hspace{2em}
    \begin{tikzpicture}
    \begin{feynman}
    \vertex(i1);
    \vertex[above=0.5cm of i1] (k) {$N_k$};
    \vertex[below=0.5cm of i1] (h) {$\mathit{\Phi}$};
    \vertex[right=1.0 cm of i1] (i2);
    \vertex[right=1.0 cm of i2] (f); 
    \vertex[right=0.75 cm of f] (g);
    \vertex[right=1.0 cm of g] (e);
    \vertex[above=0.5 cm of e] (a) {$l_j$};
    \vertex[below=0.5 cm of e](b) {$H$};
    \diagram* {
   (f)--[fermion2,half right,edge label'=\(l_n\)](i2) ,(f) --[scalar, half left] (i2), (h)--[scalar](i2), (k) --(i2),
    (f)--[edge label' =\(N_m\)] (g),
    (a)-- [fermion2] (g)-- [scalar] (b),
    };
    \end{feynman}
    \end{tikzpicture} 
    \caption{Various tree and one-loop diagrams that contribute to the CP asymmetry parameters $\varepsilon_D$ and $\varepsilon_S$.}
    \label{fig:cpdecay}
\end{figure}
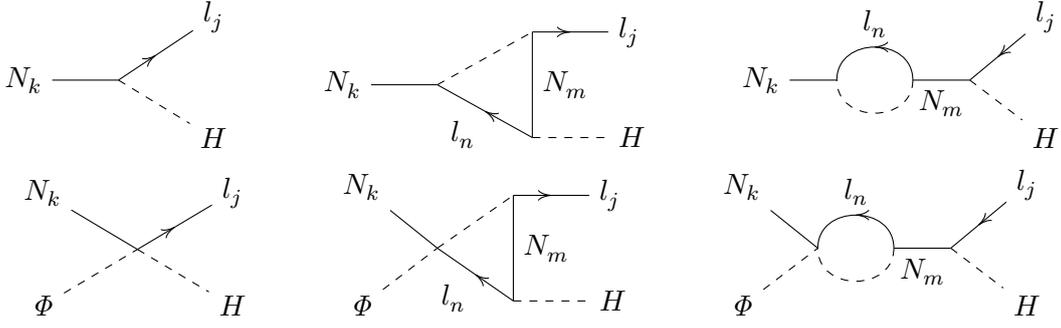
The lepton asymmetry in the present scenario is generated through both the decay 
($N\leftrightarrow lH$, $N\leftrightarrow \bar{l}\bar{H}$) and scattering ($N\mathit{\Phi}\leftrightarrow lH$, $N\mathit{\Phi}\leftrightarrow \bar{l}\bar{H}$) 
processes\footnote{ Note that there are other $2 \rightarrow 3$ and $2 \rightarrow 4$ process in the model which produces negligible CP asymmetry compared to the decay and scattering processes mentioned here, which we neglect for simplicity.}, similar to~\cite{Bhattacharya:2023kws}. It is important to note that the signifacnt amount of non-zero CP asymmetry from these processes arises only after the $\Phi$ field acquires VEV\footnote{One can have non-zero CP asymmetry generated from the scattering and three body decay ($N\leftrightarrow lH\mathit{\Phi}$, $lH\mathit{\Phi}\leftrightarrow N$) processes before the $\mathcal{Z}_4$ breaking, but the asymmetry is highly supressed due to the presence of two-loop diagrams.}. This requirement stems from the necessity of interference between tree-level and one loop-level diagrams, with the relevant $NlH$ vertex (appearing in both the loop contributions and in the tree-level scattering $N\mathit{\Phi} \leftrightarrow lH$, see Fig.~\ref{fig:cpdecay}) being generated only after the spontaneous breaking of the $\mathcal{Z}_{4}$ symmetry. Furthermore, assuming that the symmetry-breaking temperature satisfies $T_{\mathcal{Z}_{4}} \gg M_{1}$, the right-handed neutrino ($N_{1}$) remains in thermal equilibrium through inverse decay ($lH \rightarrow N$) and scattering interactions. We can quantify the decay and scattering-driven CP asymmetry as
\begin{equation}
\begin{split}
     &{\varepsilon_D}_i \equiv \frac{\Gamma(N_i\rightarrow l  H)-\Gamma(N_i \rightarrow \bar{l} \bar{H})}{\Gamma(N_i\rightarrow l  H)+\Gamma(N_i \rightarrow \bar{l} \bar{H})}\,,\quad\text{and}\\
     &{\varepsilon_S}_i \equiv \frac{\gamma(N_i
    \mathit{\Phi}\rightarrow l  \bar{H})-\gamma(N_i\mathit{\Phi} \rightarrow \bar{l} H)}{\gamma(N_i\mathit{\Phi}\rightarrow l  \bar{H})+\gamma(N_i\mathit{\Phi} \rightarrow \bar{l} H)}\,.
    \end{split}
\end{equation}
Here $\Gamma(N_i\rightarrow l  H)=(y_\nu^\dagger~y_\nu)M_1/16\pi$ is the decay width of $N_1$ and $\gamma(ab\rightarrow cd)$ is the reaction density of the $2\leftrightarrow 2$ process defined as,
{\small\begin{align}\label{eq:reactscat}
      \gamma(ab\rightarrow cd) 
      & =\frac{T}{512 \pi^6}\int \text{d}\tilde{E}\frac{|\textbf{P}||\textbf{Q}|}{\sqrt{\tilde{E}}}K_1\left(\frac{\sqrt{\tilde{E}}}{T}\right)\int d\Omega |\mathcal{M}|^2_{ab\rightarrow cd}\,. 
\end{align}}
In this expression, T denotes the bath temperature, $K_1$ is the order one modified Bassel's function of the second kind and $|\mathcal{M}|^2$ is the squared matrix element of the scattering amplitude (summed over initial states and averaged over final states). The symbols $\mathbf{P}$ and $\mathbf{Q}$ refer to the initial and final state momenta in the center-of-mass frame, respectively, while $\Omega$ represents the solid angle. The integration variable $\tilde{E}$ ranges from the lower limit $\tilde{E}_{min} = \text{max}[(m_a + m_b)^2, (m_c + m_d)^2]$ to the upper limit $\tilde{E}_{max} = \Lambda$. 
\begin{figure*}[htbp!]
    \centering
    \includegraphics[width=0.495\linewidth]{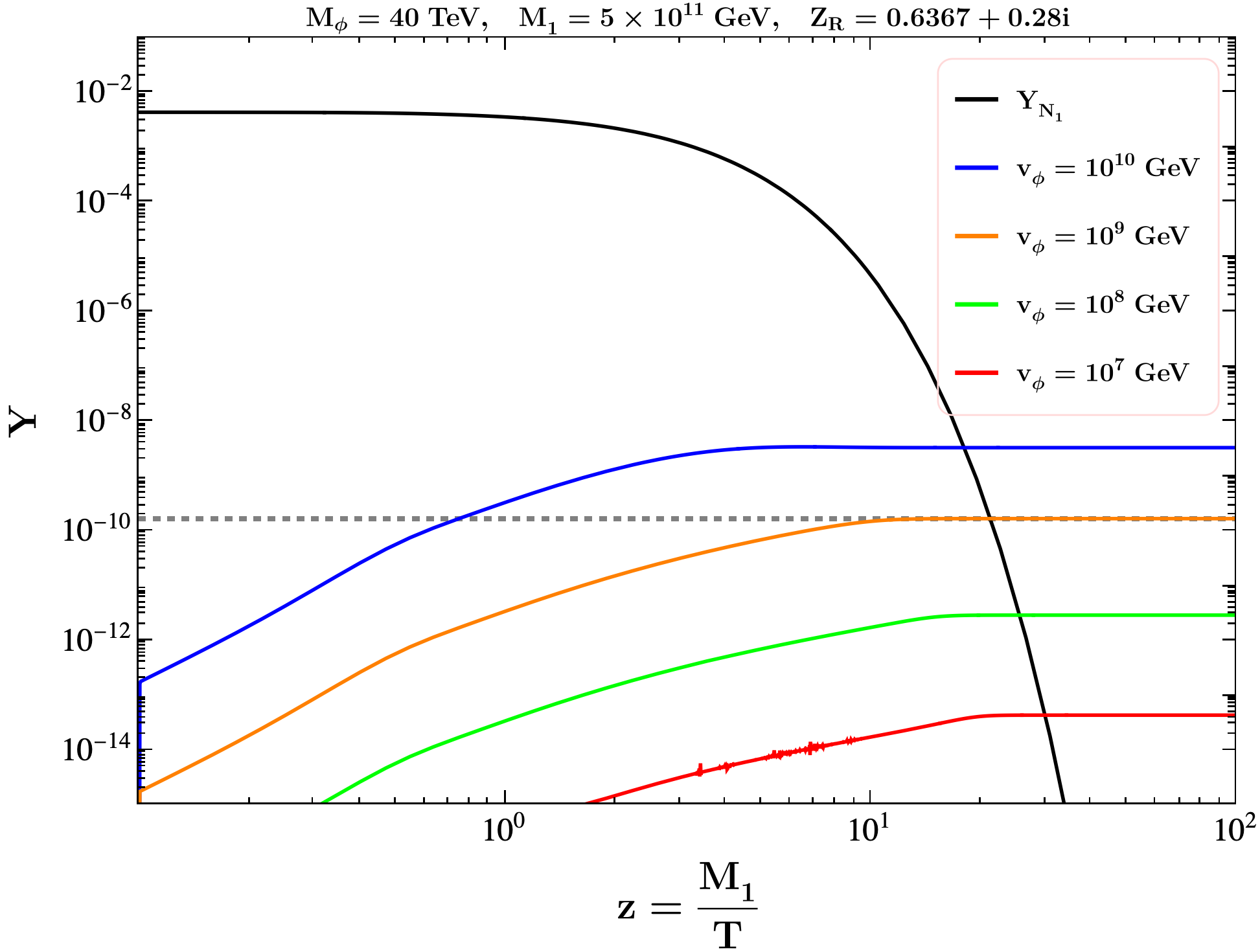}
     \includegraphics[width=0.495\linewidth]{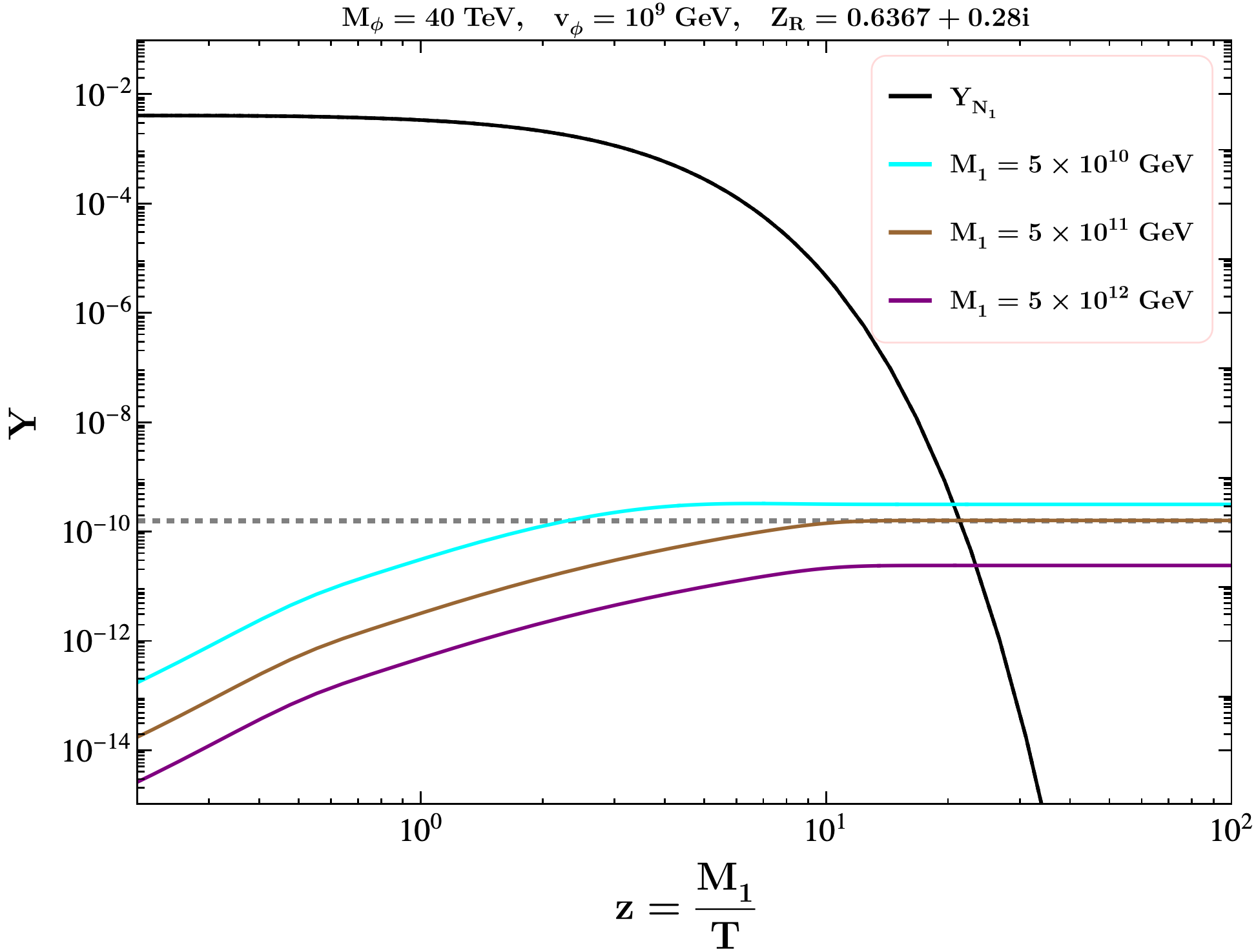}
      \caption{\emph{Left:} Plot for total lepton asymmetry $|Y_{\Delta L}|$ for different $v_{\phi}$ values. The dashed gray line indicates the correct lepton asymmetry necessary to create the observed baryon asymmetry. \emph{Right:} Plot for total lepton asymmetry $|Y_{\Delta L}|$ for different $M_1$ values.}
    \label{fig:BEQsol12}
\end{figure*} 
Finally\footnote{Here we do not include flavor effects in leptogenesis for simplicity; for details see \cite{Abada:2006fw, Nardi:2006fx, Blanchet:2006be, Dev:2017trv, Datta:2021gyi}. Moreover, the reheating temperature in this scenario is expected to be sufficiently high to ensure that the heavy seesaw states decay during the radiation-dominated epoch. If the decays were instead to occur during a prolonged reheating phase, the standard flavor regimes relevant for leptogenesis would be altered \cite{Datta:2022jic, Datta:2023pav, roshan2025delayedchargedleptonyukawa}.}, we can get simplified analytical forms for these CP asymmetry parameters as~\cite{Covi:1996wh},
\begin{equation}
\begin{split}
    {\varepsilon_D}_1 &= \frac{1}{8\pi} \sum_{k \neq 1} \frac{\text{Im}[(y_\nu^\dagger y_\nu)^2_{1k}]}{(y_\nu^\dagger y_\nu)_{11}}\bigg\{f_{v+s} \bigg(\frac{M_k^2}{M_1^2}\bigg)\bigg\}\\\nonumber
    &\approx \frac{3}{16\pi}\, \sum_{k \neq 1}\frac{1}{\sqrt{x}}\, \frac{\text{Im}[(y_{\nu}^\dagger y_{\nu})^2_{1k}]}{(y_{\nu}^\dagger y_{\nu})_{11}}\,~({\rm for}~x\gg 1)\,, 
\end{split}
\end{equation}
and~\cite{Bhattacharya:2023kws}
\begin{equation}
    {\varepsilon_S}_1 \approx \frac{2}{16\pi}\sum_{k \neq 1}\frac{1}{\sqrt{x}} \frac{\text{Im}[(y_{\nu}^\dagger y_{\nu})^2_{1k}]}{(y_{\nu}^\dagger y_{\nu})_{11}}({\rm for~} \tilde{E}\gg M_{i,k}^2,~x\gg1).
\end{equation}
Here $f_{v+s}=\sqrt{x}\bigg[1+\frac{1}{1-x}+(1+x)\ln{\frac{x}{1+x}}\bigg]$ is the loop function coming from both vertex and self-energy diagrams with $x=M_k^2/M_1^2$. Below are the relevant coupled Boltzmann equations to study the number density of $\mathit{\Phi}$, $N_1$ and $\Delta L=l-\bar{l}$
\begin{align}
    \frac{dY_{\mathit{\Phi}}}{dz}&=-\frac{1}{sHz}\left[\gamma_S \left(\frac{Y_{N}Y_{\mathit{\Phi}}}{Y_{N}^{eq}Y_{\mathit{\Phi}}^{eq}}-1\right)+\gamma_{\mathit{\Phi}}\left(\frac{Y_{\mathit{\Phi}}}{Y_{\mathit{\Phi}}^{eq}}-1\right)\right]\nonumber\\
    &\quad+\frac{s \langle\sigma v\rangle_{\mathit{\Phi} \mathit{\Phi}}}{Hz} \left(Y_{\mathit{\Phi}}^2-{Y_{\mathit{\Phi}}^{eq}}^2\right)\,,\label{eq:beqphi}\\
     \frac{dY_{N}}{dz}&=-\frac{1}{sHz}\left[\gamma_D\left(\frac{Y_{N}}{Y_{N}^{eq}}-1\right)+\gamma_S\left(\frac{Y_{N}Y_{\mathit{\Phi}}}{Y_{N}^{eq}Y_{\mathit{\Phi}}^{eq}}-1\right)\right],\label{eq:beqn}\\
    \frac{dY_{\Delta L}}{dz}&= \frac{1}{sHz}\Bigg[ \gamma_D \varepsilon_D\Bigg(\frac{Y_{N}}{Y_{N}^{eq}}-1\Bigg)+\gamma_S \varepsilon_S\Bigg(\frac{Y_{N}Y_{\mathit{\Phi}}}{Y_{N}^{eq}Y_{\mathit{\Phi}}^{eq}}-1\Bigg)\Bigg]\nonumber\\
    &\quad-\frac{1}{sHz}\frac{Y_{\Delta L}}{2Y_L^{eq}}\big(\gamma_D+\gamma_S\big)\,.\label{eq:beql}
\end{align}
Note that the yield is defined by $Y^{(eq)}=\frac{n^{(eq)}}{s}$, where $n^{(eq)}$ denotes the equilibrium number density and $s=0.44 g_{s} T^3$ is the total entropy density. The Hubble expansion of the universe is given by $H=1.66\sqrt{g_\rho} T^2/M_{Pl}$ with $g_{s}$ and $g_{\rho}$ representing the relative effective degrees of freedom for entropy and radiation energy density, respectively. The dimensionless parameter $z$ is defined as $z= {M_1} /T$ and and, $\gamma_{\mathit{\Phi}}$ denotes the total reaction density for $\mathit{\Phi}$ decaying into Higgs iso-doublet and $ \tilde{S}_2$ particles (at this era, before the EWSB, $H$ and $ \tilde{S}_2$ are the relevant fields), i.e., $\gamma_{\mathit{\Phi}} = \gamma_{\mathit{\Phi} HH} + \gamma_{\mathit{\Phi}  \tilde{S}_2  \tilde{S}_2}$. $\langle\sigma v\rangle_{\mathit{\Phi} \mathit{\Phi}}$ represents the sum of all the thermal average cross-sections of $\mathit{\Phi}$ ($\mathit{\Phi}\mathit{\Phi}\rightarrow HH, \tilde{S}_2 \tilde{S}_2,\mathbf{s}_{1}\mathbf{s}_{1}$). All other notations are the same as \cite{Bhattacharya:2023kws}.
The reaction density $\gamma$ is given by :
\bea
\gamma(a \rightarrow bc) = n^{eq}\frac{K_1(z)}{K_2(z)}\Gamma(a \rightarrow bc),
\eea
where $K_{1,2}$ are the modified Bessel's functions of 2nd kind. The expressions for the total reaction density of decay and scattering (for $T\gg M_{i}$) processes responsible for creating lepton asymmetry are (see Appendix~\ref{app:reation}),
\begin{align}
   & 2\gamma(N_i \mathit{\Phi} \r l H)=\gamma_{S}\approx \frac{(y_\nu^\dagger~y_\nu)}{32\pi^2}\frac{M_1^4}{z}f_c(M_1,v_\phi,z)\,,\nonumber\\
   & 2\gamma(N_i\rightarrow lH)=\gamma_{D}=\frac{(y_\nu^\dagger~y_\nu)}{32\pi^2}\frac{M_1^4}{z}K_1(z){\label{eq:DSre}}\,,
\end{align}
here $f_c(M_1,v_\phi,z)={M_1^2}/({v_\phi^2}{\pi^3 z^5})$ is the dimensionless quantity \footnote{ Note that while solving BEQ, we use the full numerical estimate of $\gamma_S$, while the approximate analytical forms highlight the parameter dependence.}. During the freeze-in of the lepton asymmetry, around $z\sim(10-20)$, if the scattering rate $\gamma_{s}$ dominates over the decay rate $\gamma_{D}$ ($\frac{\gamma_{S}}{\gamma_{D}}\approx\frac{f_c(M_1,v_\phi,z)}{K_1(z)}\gtrsim1$), a portion of the produced asymmetry is partially washed out by scattering processes (see Eq:~\ref{eq:beql}). We have shown $|Y_{\Delta L}|$ as a function of $z=\frac{M_1}{T}$ in Fig:~\ref{fig:BEQsol12} as the solution of the coupled BEQ \ref{eq:beql}. 
In the left panel of this figure, we fix the mass of the RHN $N_1$ at $5 \times 10^{11}$ GeV and vary the value of $v_{\phi}$, indicated by different colors: blue ($v_{\phi} = 10^{10}$ GeV), orange ($v_{\phi} = 10^{9}$ GeV), green ($v_{\phi} = 10^{8}$ GeV), and red ($v_{\phi} = 10^{7}$ GeV). As the value of $v_{\phi}$ decreases, the generated lepton asymmetry also decreases due to the enhancement of the scattering rate $\gamma_S$, which scales as $\gamma_S \propto M_1^2 / v_{\phi}^2$ from Eq:~\ref{eq:DSre}.  In the right panel of Fig:~\ref{fig:BEQsol12}, we fix $v_{\phi} = 10^{9}$ GeV and vary $M_1$ over three values: $10^{10}$ GeV (cyan), $10^{11}$ GeV (brown), and $10^{12}$ GeV (violet). As $M_1$ increases, $\gamma_S$ also becomes larger (see Eq:~\ref{eq:DSre}), resulting in a decreasing trend of the lepton asymmetry. Additionally, it is worth mentioning that for $M_1=5\times 10^{11}$ GeV, we require a minimum $v_{\phi}\sim10^9$ GeV, as discussed in \cite{Bhattacharya:2023kws}, to successfully reproduce the observed baryon asymmetry of the universe. The key physical parameters governing leptogenesis in this scenario are ${M_i, v_{\phi}}$. Since $M_{\phi} \ll M_i$ and $v_{s} \ll v_{\phi}$, their contributions to leptogenesis are negligible. Although variations in $v_{s}$, $M_{\mathbf{s}^{}_{1,2}}$, and $M^{}_{\phi}$ affect the scalar couplings of the model, these changes have no significant impact on leptogenesis as long as all scalars remain in thermal equilibrium. It is also worth noting that while Fig.~\ref{fig:BEQsol12} assumes an initial thermal abundance of both $N_i$ and $\mathit{\Phi}$, we have verified that the final asymmetry remains unaffected even if their initial abundances are set to zero. This is because inverse decay processes ($lH \rightarrow N_i$, $HH \rightarrow \mathit{\Phi}$) and annihilation channels ($HH \rightarrow \mathit{\Phi}\mathit{\Phi}$, $lH \rightarrow \mathit{\Phi} N$) can efficiently populate $N_i$ and $\mathit{\Phi}$ during the evolution, allowing them to quickly reach thermal equilibrium before the asymmetry freezes in, thanks to their interactions with the thermal bath via scalar couplings.

\section{WIMP Dark Matter and the role of induced VEV} 
\label{sec:dm phenomenology}

Owing to the sizeable coupling between the scalar DM ($\mathbf{s}^{}_1$) and other scalars ($\phi$, H and $\mathbf{s}_{2}^{}$), 
$\mathbf{s}^{}_1$ is likely to stay in equilibrium with the thermal bath particles, with its interaction exceeding the Hubble rate at $T=M_{\mathbf{s}_1^{}}^{}$, 
\begin{align}
    n^{eq}_{\mathbf{s}_1^{}} \langle\sigma v\rangle_{eff}^{}\geq H(T=M_{\mathbf{s}_1^{}}^{})\,,\label{eq:ineqequ}
\end{align}
and freezes out, yielding a WIMP like DM. The effective annihilation cross-section of DM is obtained by summing over all the final state particles that it can annihilate to 
(see the diagrams in Appendix \ref{sec:vertexf}, Fig:~\ref{fig:ann_diag})
\begin{align}
  \nonumber \langle\sigma v\rangle_{eff}^{} = &\langle\sigma v\rangle_{\mathbf{s}_{1}^{}\mathbf{s}_{1}^{}\rightarrow SMSM}^{}+\langle\sigma v\rangle_{\mathbf{s}_{1}^{}\mathbf{s}_{1}^{}\rightarrow\phi\phi}^{}+\langle\sigma v\rangle_{\mathbf{s}_{1}^{}\mathbf{s}_{1}^{}\rightarrow\phi h}^{}\\
  &+\langle\sigma v\rangle_{\mathbf{s}_{1}^{}\mathbf{s}_{1}^{}\rightarrow hh}^{}+\langle\sigma v\rangle_{\mathbf{s}_{1}^{}\mathbf{s}_{1}^{}\rightarrow \phi \mathbf{s}_{2}^{}}^{}+\langle\sigma v\rangle_{\mathbf{s}_{1}^{}\mathbf{s}_{1}^{}\rightarrow h\mathbf{s}_{2}^{}}^{}\nonumber\\
  &+\langle\sigma v\rangle_{\mathbf{s}_{1}^{}\mathbf{s}_{1}^{}\rightarrow \mathbf{s}_{2}^{}\mathbf{s}_{2}^{}}^{}\,,
\end{align}
where $\langle\sigma v\rangle_{ab\rightarrow cd}$ denotes the thermal average cross-section of the process 
$ab\rightarrow cd$. To obtain the freeze out of DM ($\mathbf{s}^{}_1$), one needs to solve the following BEQ \cite{Kolb:1990vq} 
(Eq:~\ref{eq:beqdm}), 
 {\footnotesize\begin{align}
   \nonumber \rm \frac{dn_{\mathbf{s}_{1}^{}}^{}}{dt}+3Hn_{\mathbf{s}_{1}^{}}^{}=&-\langle\sigma v\rangle_{eff}^{}(n_{\mathbf{s}_{1}^{}}^{2}-n_{\mathbf{s}_{1}^{}}^{eq^{{}_2}})-\langle\Gamma\rangle_{\phi\rightarrow \mathbf{s}_{1}^{}\mathbf{s}_{1}^{}}(n_{\mathbf{s}_{1}^{}}^{}-n_{\mathbf{s}_{1}^{}}^{eq^{{}}})\\
   &-\langle\Gamma\rangle_{h\rightarrow \mathbf{s}_{1}^{}\mathbf{s}_{1}^{}}(n_{\mathbf{s}_{1}^{}}^{}-n_{\mathbf{s}_{1}^{}}^{eq^{{}}})-\langle\Gamma\rangle_{\mathbf{s}_{2}^{}\rightarrow \mathbf{s}_{1}^{}\mathbf{s}_{1}^{}}(n_{\mathbf{s}_{1}^{}}^{}-n_{\mathbf{s}_{1}^{}}^{eq^{{}}})
   \,,\label{eq:beqdm}
\end{align}}
where $\langle\Gamma\rangle_{a\rightarrow bc}$ is the thermal average decay width of the process $a\rightarrow bc$. The relic density of the DM is mostly independent of the production processes ($\langle\Gamma\rangle_{\ast\rightarrow \mathbf{s}_{1}^{}\mathbf{s}_{1}^{}}$) as they help DM to equilibrate, and is inversely proportional to the effective thermal average of annihilation cross-section as $\Omega_{\mathbf{s}_{1}^{}}^{}h^{2}\propto\frac{1}{\langle\sigma v\rangle_{\rm eff}^{}}$.
\begin{figure*}[htb!]
    \centering
    \includegraphics[width=0.49\linewidth]{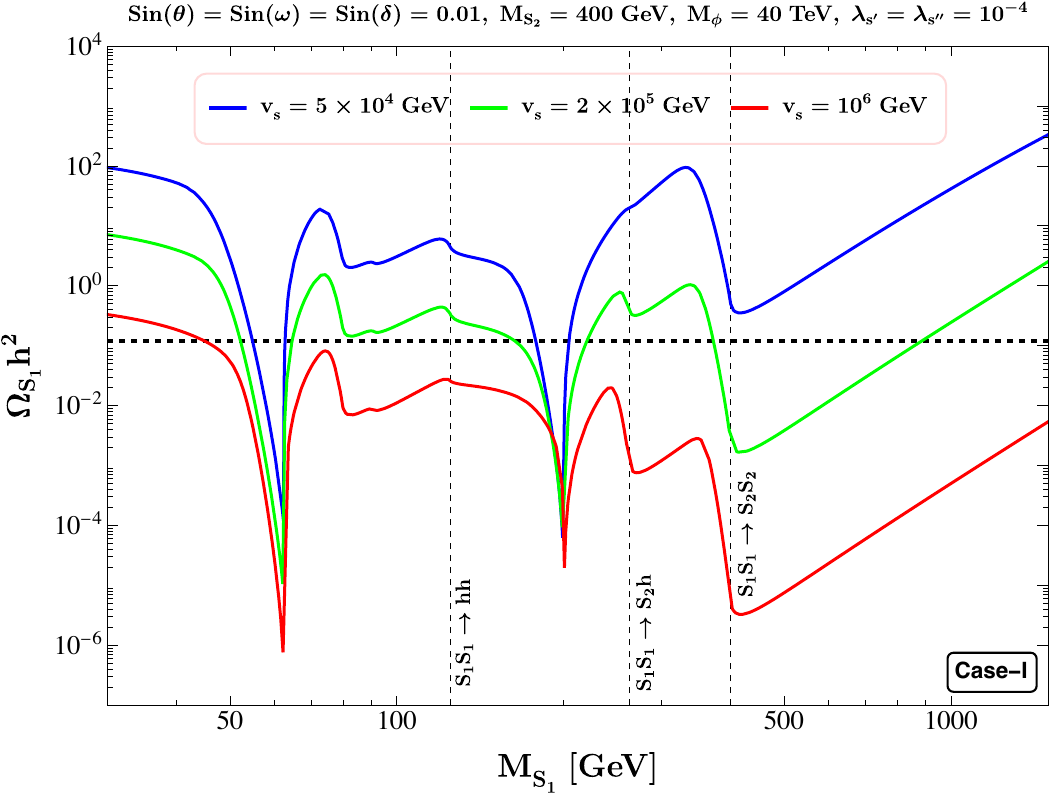}
     \includegraphics[width=0.49\linewidth]{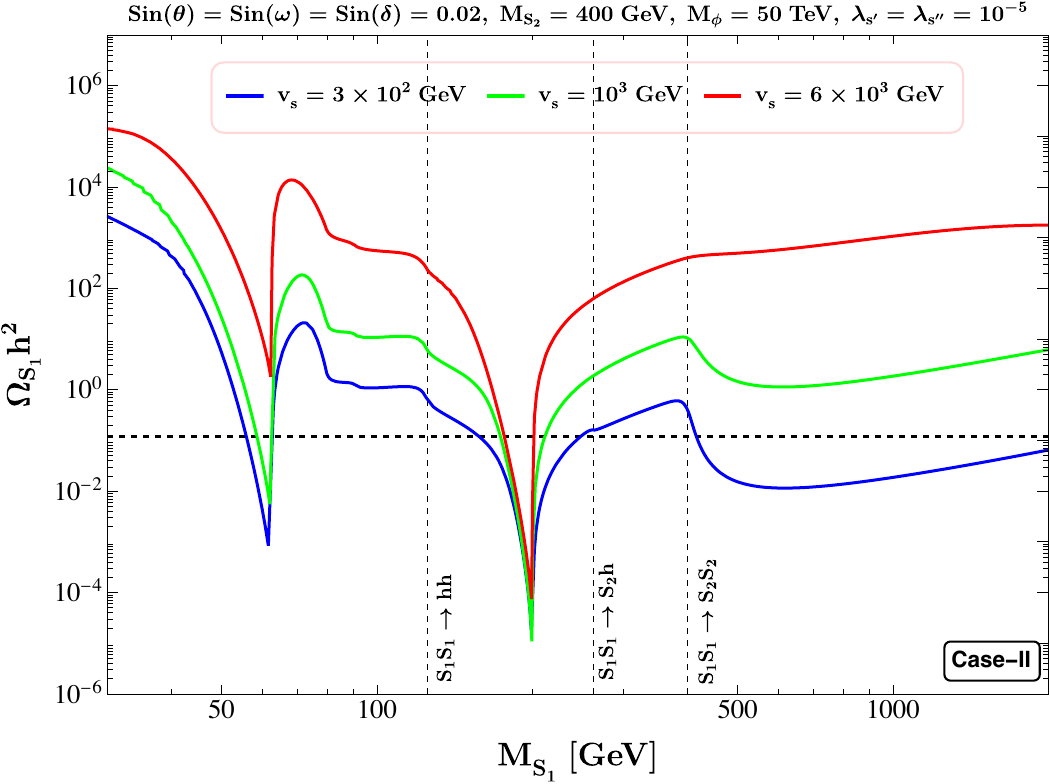}
    \caption{Variation of the DM relic density with DM mass ($M_{s_1}$) for two cases of our interest, {\bf case-I}: large $v_s$ on let left, and {\bf case-II}: small $v_s$ on the right. 
    We also varied the parameter $v_{s}$ for both cases, shown by red, green and blue representing increasing values of $v_{s}$. Parameters kept fixed are mentioned in the 
    heading and inset.}
    \label{fig:vartexrelic}
\end{figure*}
\subsection{Effect of the induced VEV on the Dark Matter relic abundance}
We focus on the DM phenomenology in the MeV-TeV mass range. $\lambda_{\mathbf{s}_{1}^{}\mathbf{s}_{1}^{}\mathbf{\phi}}$, $\lambda_{\mathbf{s}_{1}^{}\mathbf{s}_{1}^{}h}$, $\lambda_{\mathbf{s}_{1}^{}\mathbf{s}_{1}^{}\mathbf{s}_{2}^{}}$, $\lambda_{\mathbf{s}_{2}^{}\mathbf{s}_{2}^{}\mathbf{s}_{2}^{}}$ and $\lambda_{\mathbf{s}_{2}^{}\mathbf{s}_{2}^{}h}$ are the relevant vertex factors associated with the DM annihilation cross-sections used in the computation of the DM relic density. Approximated analytical form of these vertex factors are provided in Appendix~\ref{sec:vertexf}. 
\begin{table}[htb!]
\centering
\begin{tabular}{c c c c c c}
\hline \hline 
Scenarios & $M_\phi$  (GeV) &  $v_{s}$  (GeV) & $\lambda_{S'}$=$\lambda_{S''}$ &$s_{ij}$\\
 \hline 
 {\bf Case-I} & $4\times10^{4}$ & $\{10^4-10^6\}$ &  $10^{-4}$ &  $0.01$\\
 {\bf Case-II} &$5\times10^{4}$  & $\{10^2-10^4\}$ &   $10^{-5}$  & $0.02$\\
 \hline \hline
\end{tabular}
\caption{Values of $M_\phi$, $v_{s}$ and ($s_{ij}$) along with the corresponding value of $\lambda_{S'}$, $\lambda_{S'}$ for the 
explanation of DM relic for two specific scenarios,  {\bf case-I} and  {\bf case-II}. We fixed the value of $v_{\phi}$, i.e. $\mathcal{Z}_4$ breaking scale 
at $10^9$ GeV.}
\label{tab:case}
\end{table}
We investigate the DM relic abundance in two different parameter regimes of the model, showcasing contrasting dependencies on the 
induced VEV ($v_{s}$);  {\bf case-I}: where $v_{s}$ is large, and  {\bf case-II}: where $v_{s}$ is small, as indicated in Table~\ref{tab:case}. 
For {\bf case-I}, the triple coupling $\lambda_{\mathbf{s}_{1}^{}\mathbf{s}_{1}^{}h}$ (see Eq:~\ref{eq:vartex}) increases linearly with $v_{s}$ 
(due to the first term of the 2nd relation of Eq:~\ref{eq:vartex}), whereas for {\bf case-II}, the same coupling varies inversely with $v_{s}$ 
(due to the second term of the same relation). Since the DM relic depends on $\lambda_{\mathbf{s}_{1}^{}\mathbf{s}_{1}^{}h}$ as 
$\Omega_{\mathbf{s}_{1}^{}}^{}h^{2}\propto\frac{1}{(\lambda_{\mathbf{s}_{1}^{}\mathbf{s}_{1}^{}h})^{2}}$,  the behaviour of the DM relic density differs 
between {\bf case-I} and {\bf case-II}. In {\bf case-I} DM relic density decreases with increasing $v_{s}$, whereas in {\bf case-II}, it increases with increasing $v_{s}$. 
This phenomenon is illustrated in Fig.~\ref{fig:vartexrelic}. The left and right panels of Fig:~\ref{fig:vartexrelic} correspond to {\bf case-I} and {\bf case-II}, 
respectively, with the mass of $\mathbf{s}_{2}$ fixed at $400$ \text{GeV}. For both scenarios, we investigate the variation of the DM relic density across three 
distinct mass regimes of the DM candidate $\mathbf{s}_{1}$.\\ 
$\bullet~{\bf{Region1~(M_{\mathbf{s}_{1}^{}}<M_{h}):}}$ When DM mass is below the SM Higgs mass ($M_{h}=125.1$ Gev), all DM annihilation cross-sections to SM processes are mediated dominantly by SM Higgs. A sharp dip occurs in the relic line at $M_{\mathbf{s}_{1}^{}}=\frac{M_{h}}{2}$ due to the Higgs resonance, which significantly enhances the DM annihilation cross-section, thereby reducing the relic density. A wider dip near $M_{\mathbf{s}_{1}^{}}\sim 80$ GeV and a shallower dip near $M_{\mathbf{s}_{1}^{}}\sim 91$ GeV appear in both plots of the figure:~\ref{fig:vartexrelic}. This dips appear due to the presence of $\mathbf{s}_{1}\mathbf{s}_{1}\rightarrow W^{+}W^{-}$ and $\mathbf{s}_{1}\mathbf{s}_{1}\rightarrow Z^{}Z^{}$ processes which becomes kinematically accessible from this mass range.\\

$\bullet~{\bf{Region2~(M_{h}^{}\leq M_{\mathbf{s}_{1}^{}}< M_{\mathbf{s}_{2}^{}}):}}$ The process $\mathbf{s}_{1}\mathbf{s}_{1}\rightarrow hh$ becomes kinematically favorable when $M_{\mathbf{s}_{1}^{}}\geq M_{h}^{}$ resulting in a drop near $M_{\mathbf{s}_{1}^{}}\sim M_{h}^{}$ in both scenarios. For both the plots of figure~\ref{fig:vartexrelic}, this dip becomes shallower as $v_{s}$ increases. This occurs because the coupling $\lambda_{SH}\approx\frac{M_{\mathbf{s_{2}^{}}}^{2}}{v_{s}^{}v}$ weakens with higher $v_{s}^{}$. Since $\langle\sigma v\rangle\propto \lambda_{SH}^{2}$ the cross-section decreases with increasing $v_{s}^{}$, causing the dip to gradually diminish. We also observe a second sharp dip near $M_{\mathbf{s}_{1}^{}}\sim \frac{M_{\mathbf{s}_{2}^{}}}{2}$ due to the resonance enhancement of the process $\mathbf{s}_{1}\mathbf{s}_{1}\rightarrow hh$ mediated by $\mathbf{s}_{2}^{}$ in s channel interaction. A dip near $M_{\mathbf{s}_{1}^{}}\sim\frac{M_{\mathbf{s}_{2}^{}}+M_{h}^{}}{2}$ is observed due to the kinematically favorable process $\mathbf{s}_{1}\mathbf{s}_{1}\rightarrow\mathbf{s}_{2}h$. This dip is more pronounced in {\bf case-I} for higher $v_{s}^{}$ values, while in {\bf case-II}, this effect is negligible and only mildly visible for $v_{s}^{}=300$ GeV. Unlike the $\mathbf{s}_{1}\mathbf{s}_{1}\rightarrow hh$ process, the effect of this process becomes prominent at higher $v_{s}^{}$ in {\bf case-I}, whereas in {\bf case-II}, this only appears for lower $v_{s}^{}$. 
\begin{figure*}[htbp!]
    \centering
    \includegraphics[width=0.49\linewidth]{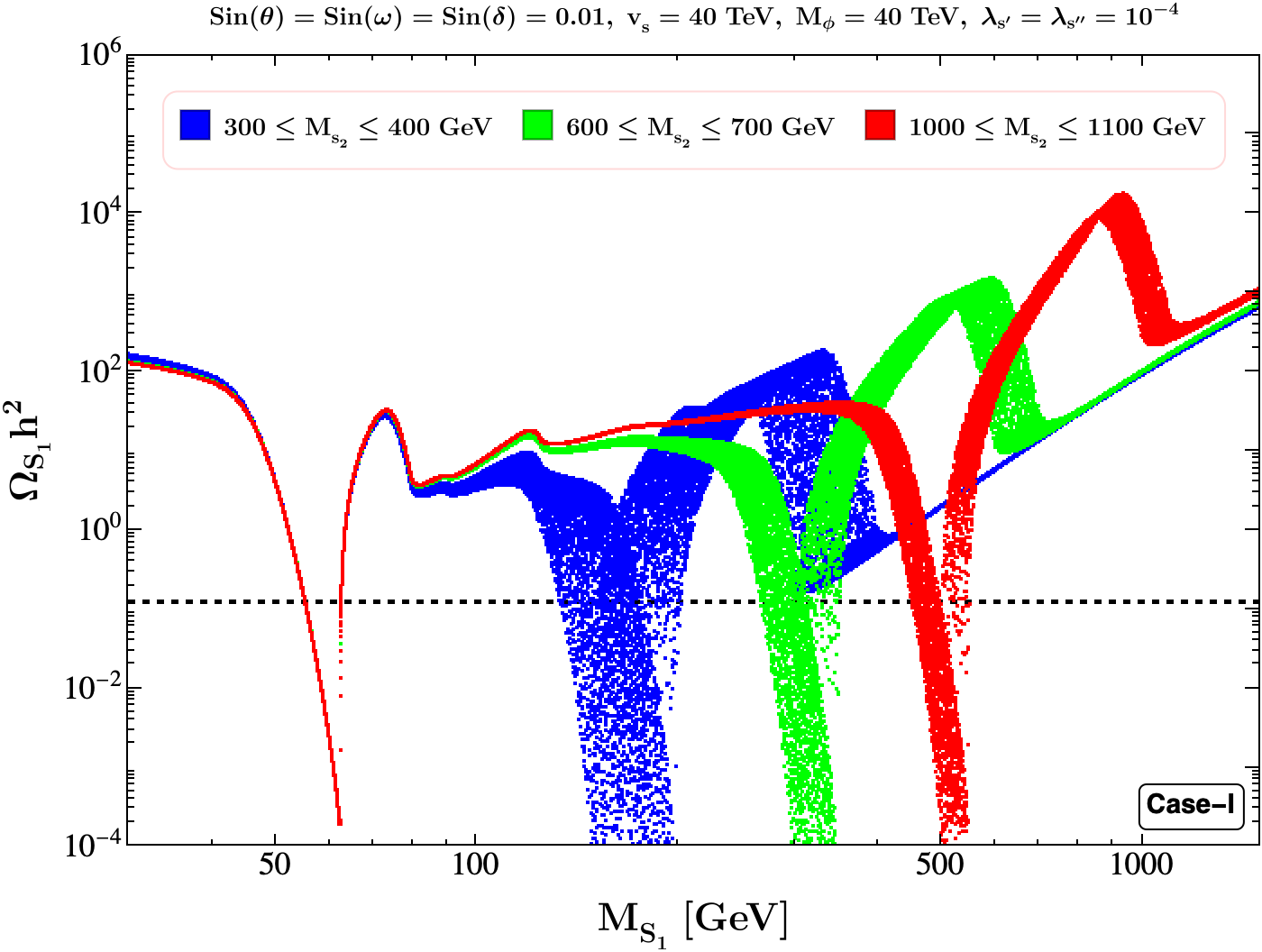}
     \includegraphics[width=0.49\linewidth]{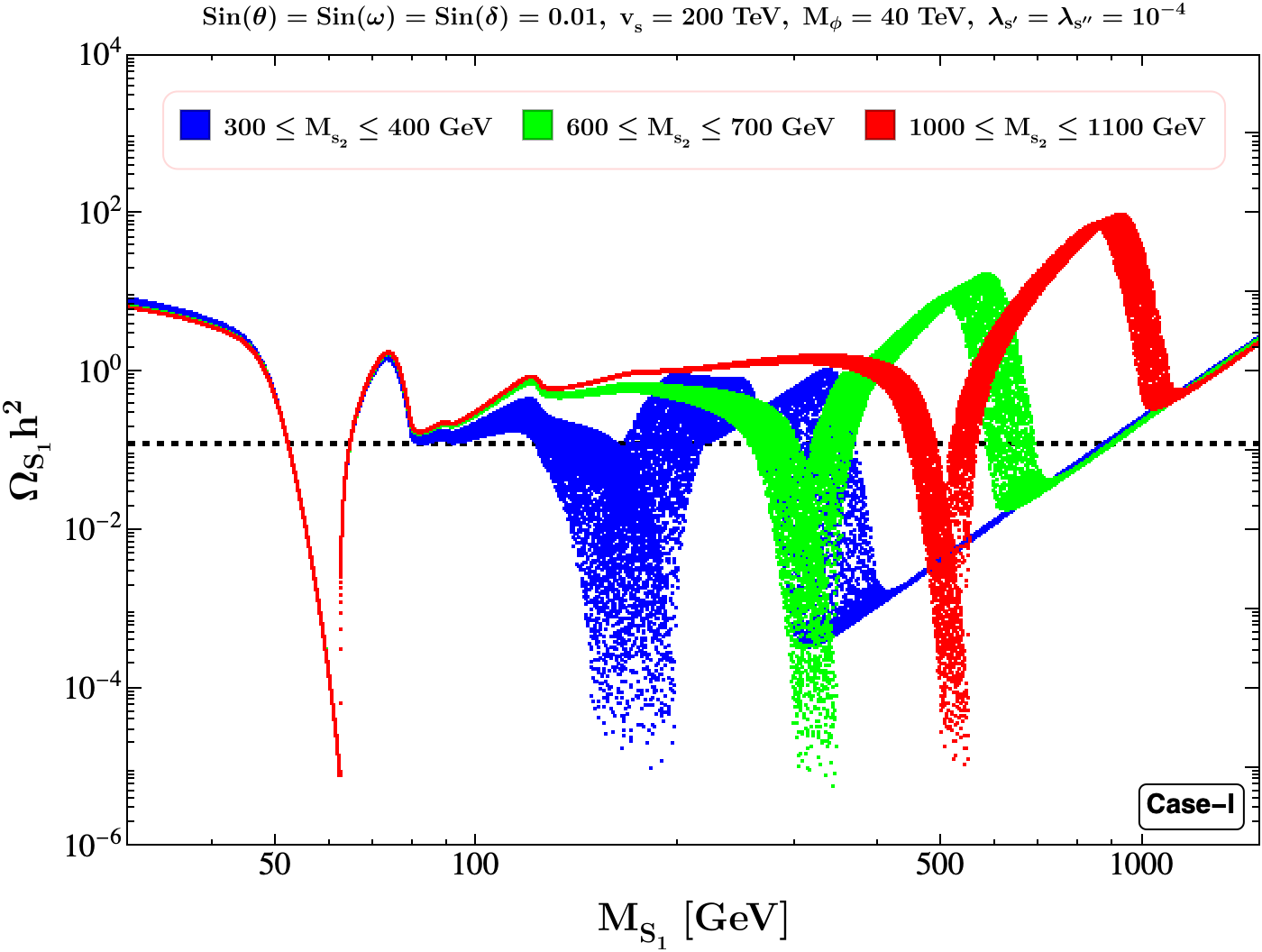}
     \includegraphics[width=0.49\linewidth]{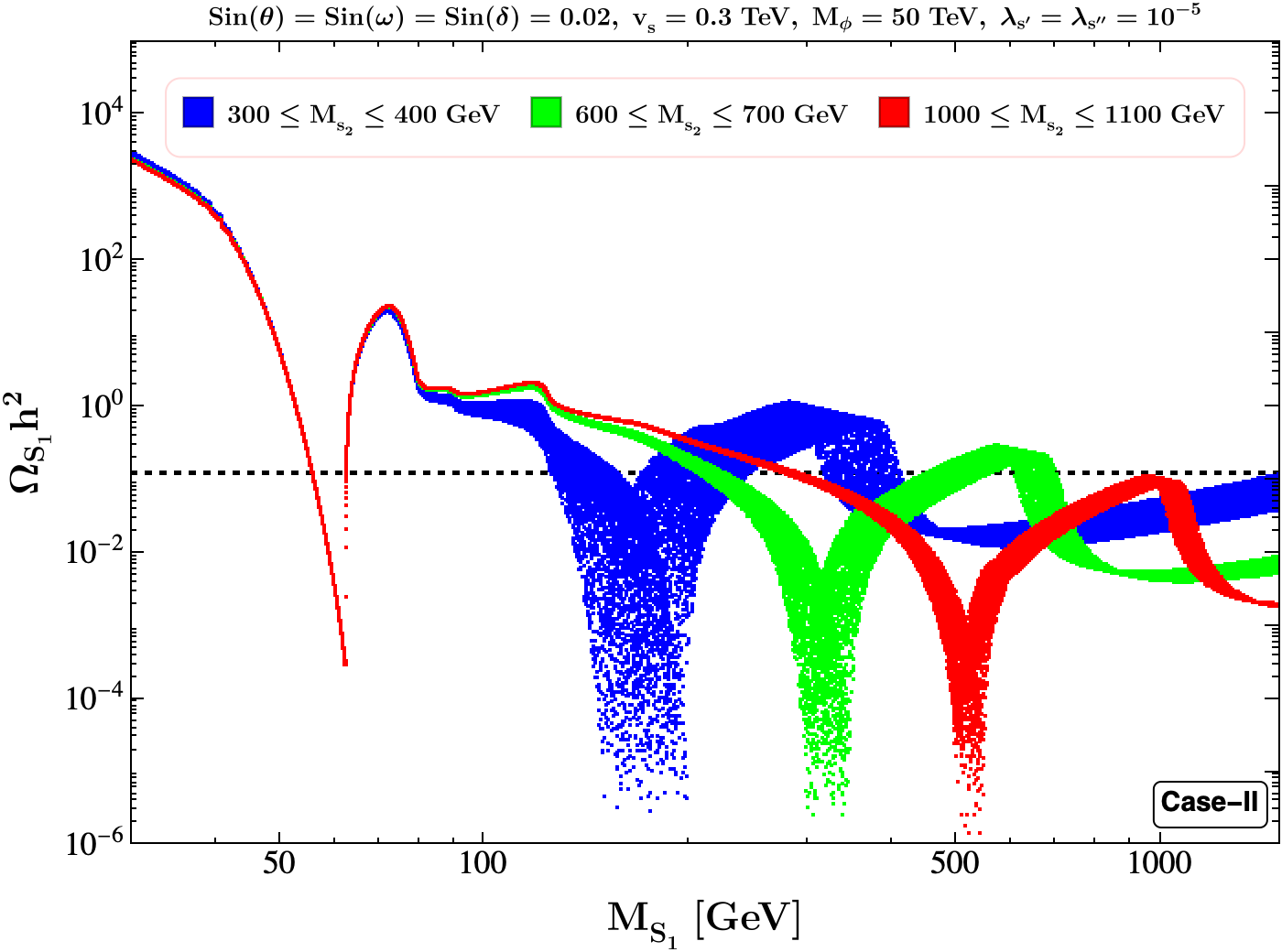}
     \includegraphics[width=0.49\linewidth]{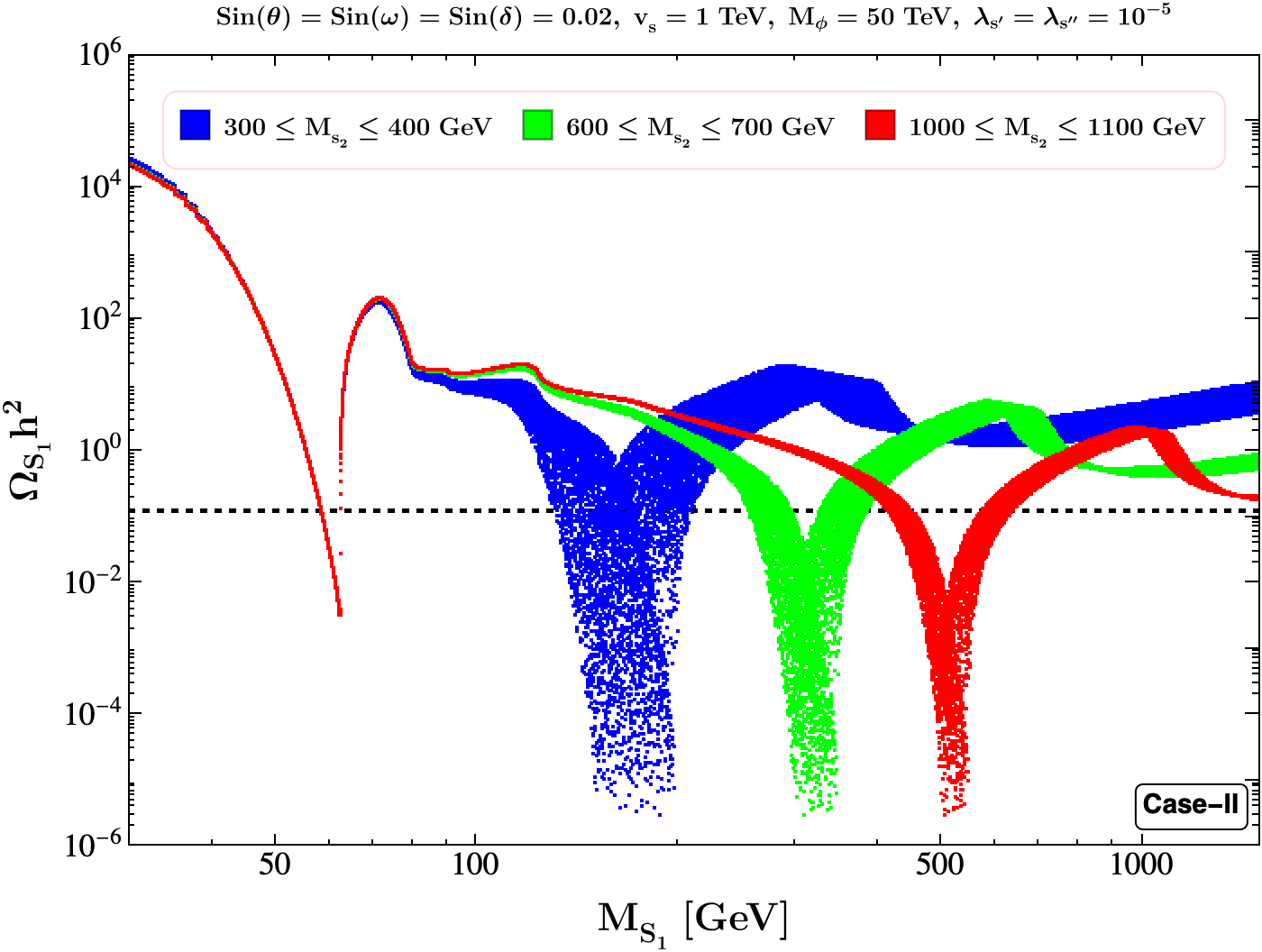}
    \caption{Plots for relic density as a function of DM mass are shown for different ranges of $M_{\mathbf{s}_{2}^{}}$. The upper row illustrates {\bf case-I}, with the left (right) panel corresponding to $v_{s}^{}= 40 (200)$ TeV. The lower row represents {\bf case-II}, with the left (right) panel corresponding to $v_{s}^{}= 0.3 (1)$ TeV.}
    \label{fig:relicMV}
\end{figure*}
The $\mathbf{s}_{2}$ pole plays a more significant role in this process, making the $\mathbf{s}_{2}$ mediated s channel diagram the dominant contributor to the cross-section, rather than the 4-point diagram ($\mathcal{M}_{\mathbf{s}_{2}^{}h}^{4p}\simeq\lambda_{SH}\approx\frac{M_{\mathbf{s_{2}^{}}}^{2}}{v_{s}^{}v}$). The matrix amplitude for the process $\mathbf{s}_{1}\mathbf{s}_{1}\rightarrow\mathbf{s}_{2}h$ via $\mathbf{s}_{2}$ mediated s channel diagram is provided in Eq:~\ref{eq:matrixamp1}. From Eq:~\ref{eq:matrixamp1} it is evident that for {\bf case-I} with $M_{\mathbf{s}_{2}^{}}=400$ GeV and $v_{s}^{}\geq20$ TeV the matrix amplitude $\mathcal{M}$ scales as $\mathcal{M}_{\mathbf{s}_{2}^{}h}^{s}\propto v_{s}^{}$. In contrast, for {\bf case-II}, $\mathcal{M}_{\mathbf{s}_{2}^{}h}^{s}\propto \frac{1}{v_{s}^{}}$. This scaling explains the dip near $M_{\mathbf{s}_{1}^{}}\sim\frac{M_{\mathbf{s}_{2}^{}}+M_{h}^{}}{2}$ in DM relic density. In {\bf case-I}, as $v_{s}$ increases, the matrix amplitude $\mathcal{M}$ grows, enhancing the cross-section of the process $\mathbf{s}_{1}\mathbf{s}_{1}\rightarrow\mathbf{s}_{2}h$. This leads to a more prominent dip at higher $v_{s}^{}$. Conversely, in {\bf case-II}, $\mathcal{M}_{\mathbf{s}_{2}^{}h}^{s}$ decreases with increasing $v_{s}$, suppressing the cross-section and making the dip vanishing only except for lower $v_{s}$.

$\bullet~{\bf{Region3~(M_{\mathbf{s}_{1}^{}}\geq M_{\mathbf{s}_{2}^{}}):}}$ Here we should mention that we vary the DM mass up to $2$ TeV and neglect the effect of heavy scalar $M_{\phi}$ in DM phenomenology for simplicity. We observe a dip near $M_{\mathbf{S}_{1}^{}}\sim M_{\mathbf{S}_{2}^{}}$ due to the new contribution from the DM annihilation process of $\mathbf{s}_{1}\mathbf{s}_{1}\rightarrow\mathbf{s}_{2}\mathbf{s}_{2}$. Similar to the $\mathbf{s}_{1}\mathbf{s}_{1}\rightarrow\mathbf{s}_{2}h$ process in this process also the more dominant contribution comes from the $\mathbf{s}_{2}$ mediated s channel diagram. The matrix amplitude for the  $\mathbf{s}_{2}$ mediated s channel diagram of the process $\mathbf{s}_{1}\mathbf{s}_{1}\rightarrow\mathbf{s}_{2}\mathbf{s}_{2}$ is $\mathcal{M}_{\mathbf{s}_{2}^{}\mathbf{s}_{2}^{}}^{s}$.

\begin{figure*}[htbp!]
    \centering
    \includegraphics[width=0.49\linewidth]{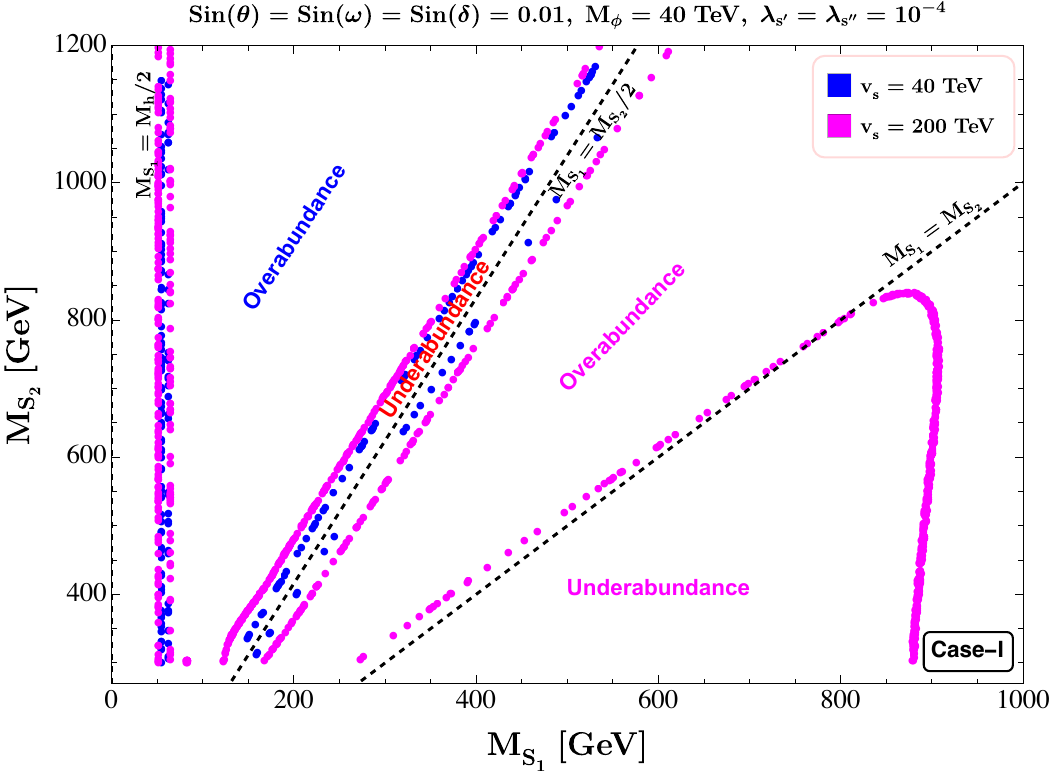}
    \includegraphics[width=0.49\linewidth]{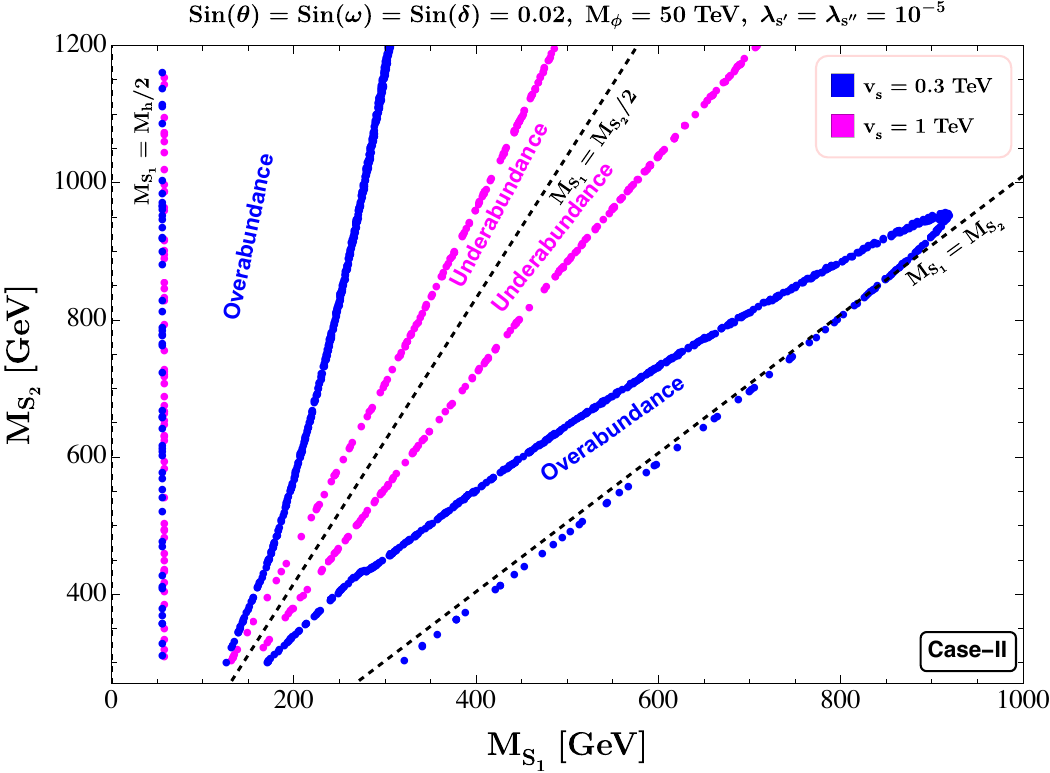}
   \caption{Plots for parameter space in the $M_{\mathbf{s}_{1}^{}}-M_{\mathbf{s}_{2}^{}}$ plane satisfying the observed DM relic ($\rm\Omega_{DM}h^{2}=0.1200\pm 0.0012$) for {\bf case-I} and {\bf case-II}. The left panel plot corresponds to {\bf case-I} with $v_{s}^{}=40$ TeV (represented by blue points) and $v_{s}^{}=200$ TeV (represented by magenta points). The right panel plot corresponds to {\bf case-II} with $v_{s}^{}=0.3$ TeV (represented by blue points) and $v_{s}^{}=1$ TeV (represented by magenta points).}
    \label{fig:rlicMM}
\end{figure*} 

In {\bf case-I}, $\mathcal{M}_{\mathbf{s}_{2}^{}\mathbf{s}_{2}^{}}^{s}\propto v_{s}^{}$ (see Eq:~\ref{eq:matrixamp2}), while in {\bf case-II}, $\mathcal{M}_{\mathbf{s}_{2}^{}\mathbf{s}_{2}^{}}^{s}\propto\frac{1}{v_{s}^{}}$ (see Eq:~\ref{eq:matrixamp2}). In {\bf case-I}, the dip becomes more pronounced at higher $v_{s}^{}$ due to the enhanced cross-section, while in {\bf case-II}, it diminishes with increasing $v_{s}^{}$. The influence of $\mathbf{s}_{2}^{}$ mass on the DM relic is illustrated in Fig.~\ref{fig:relicMV}, where $M_{\mathbf{s}_{2}^{}}$ is varied in three specific ranges for both {\bf case I} and {\bf case II}. For {\bf case-I}, $v_{s}^{}$ values of $40$ TeV (left panel) and $200$ TeV (right panel) are shown in the upper row of Fig.~\ref{fig:relicMV}. For {\bf case-II}, $v_{s}^{}$ values of $0.3$ TeV (left panel) and $1$ TeV (right panel) are displayed in the lower row of Fig.~\ref{fig:relicMV}. For both cases, when the DM mass is below \text{W} boson mass ($M_{W}^{}$), most DM annihilation processes occur predominantly through SM Higgs-mediated s-channel processes. As a result, the influence of $M_{\mathbf{s}_{2}^{}}^{}$ on the DM relic is not significant when $M_{\mathbf{s}_{1}^{}}\lesssim M_{W}^{}$. As discussed above, for a higher DM mass ($M_{\mathbf{s}_{1}^{}}\geq\frac{M_{\mathbf{s}_{2}^{}}+M_{h}^{}}{2}$) the processes $\mathbf{s}_{1}\mathbf{s}_{1}\rightarrow\mathbf{s}_{2}h$ and $\mathbf{s}_{1}\mathbf{s}_{1}\rightarrow\mathbf{s}_{2}\mathbf{s}_{2}$ are the dominant DM production channels. In {\bf case-I} (where $\mathcal{M}_{\mathbf{s}_{2}^{}h}^{s}(\mathcal{M}_{\mathbf{s}_{2}^{}\mathbf{s}_{2}^{}}^{s})\propto v_{s}$), as $M_{\mathbf{s}_{2}^{}}$ increases, the matrix amplitude $\mathcal{M}_{\mathbf{s}_{2}^{}\mathbf{s}_{2}^{}}^{s}$ and $\mathcal{M}_{\mathbf{s}_{2}^{}h}^{s}$ decrease (see Eq.~\ref{eq:matrixamp1} and Eq.~\ref{eq:matrixamp2}) leading to a reduction in the cross-section. Consequently, the relic density increases for a higher mass range of $\mathbf{s}_{2}^{}$. A similar pattern is visible in the right panel of the upper row of Fig~\ref{fig:relicMV}. In case of the top right plot, $v_{s}$ is increased, which leads to a reduction of the relic abundance, primarily driven by the enhancement of $\lambda_{SH}^{}$, which in turn boosts the annihilation cross section. In {\bf case-II} (with $v_{s}^{}\leq10^{4}$ GeV and $M_{\mathbf{s}_{2}^{}}\geq200$ GeV), the second terms of the couplings $\lambda_{\mathbf{s}_{1}^{}\mathbf{s}_{1}^{}\mathbf{s}_{2}^{}}$ and $\lambda_{\mathbf{s}_{1}^{}\mathbf{s}_{1}^{}h}$ (see Eq.~\ref{eq:vartex}) dominate over the first terms. The matrix amplitude $\mathcal{M}_{\mathbf{s}_{2}^{}\mathbf{s}_{2}^{}}^{s}\propto\frac{M_{\mathbf{s}_{2}^{}}^{2}}{v_{s}^{}}$ and $\mathcal{M}_{\mathbf{s}_{2}^{}h}^{s}\propto\frac{M_{\mathbf{s}_{2}^{}}^{2}}{v_{s}^{}}$ increase with $M_{\mathbf{s}_{2}^{}}$, enhancing the cross-section for $\mathbf{s}_{1}\mathbf{s}_{1}\rightarrow\mathbf{s}_{2}h$ and $\mathbf{s}_{1}\mathbf{s}_{1}\rightarrow\mathbf{s}_{2}\mathbf{s}_{2}$. This leads to a decrease in relic density (for $M_{\mathbf{s}_{2}^{}}\geq M_{\mathbf{s}_{1}^{}}$), as clearly shown in the lower row of Fig.~\ref{fig:relicMV}. Additionally, a rise in $v_{s}^{}$ (in the lower right panel plot for {\bf case-II}) diminishes the coupling $\lambda_{\mathbf{s}_{1}^{}\mathbf{s}_{1}^{}h}$ (see Eq.~\ref{eq:vartex}), reducing the DM annihilation cross-section to SM particles. Consequently, for a fixed $M_{\mathbf{s}_{1}}$, the DM relic density is higher in the right panel plot of {\bf case-II} compared to the left one.

In Fig~\ref{fig:rlicMM} we present the correct relic-allowed points in the $M_{\mathbf{s}_{1}^{}}-M_{\mathbf{s}_{2}^{}}$ plane for both {\bf case-I} (left panel) and {\bf case-II} (right panel). All other parameters are fixed as previously mentioned to explain Fig~\ref{fig:relicMV}. Here, $M_{\mathbf{s}_{1}^{}}$ is varied across the range $\{200-1200\}$ GeV. In the left panel of Fig.~\ref{fig:rlicMM}, where for $v_{s}^{}=40$ TeV (shown in blue colour), the correct DM relic density is achieved near the vicinity of the SM Higggs pole (depicted as the $M_{\mathbf{s}_{1}^{}}=M_{h}^{}/2$) and $\mathbf{s}_{1}^{}$ pole (represented by $M_{\mathbf{s}_{1}^{}}=M_{\mathbf{s}_{2}^{}}/2$ line), complying with Fig.~\ref{fig:relicMV}. Across the narrow region on either side of the $M_{\mathbf{s}_{1}^{}}=M_{\mathbf{s}_{2}^{}}/2$ line lies the underabundance zone, while the rest of $M_{\mathbf{s}_{1}^{}}-M_{\mathbf{s}_{2}^{}}$ parameter space represents the overabundance region for $v_{s}=40$ TeV. For $v_{s}=200$ TeV (represented by magenta line), apart from the two poles, relic-satisfied points also emerge near $M_{\mathbf{s}_{1}^{}}=M_{\mathbf{s}_{2}^{}}$ line and $M_{\mathbf{s}_{1}^{}}\sim900$ GeV for the parameters used in this analysis (see the upper right panel of Fig.~\ref{fig:relicMV}). However, since the cross-section of  $\mathbf{s}_{1}\mathbf{s}_{1}\rightarrow\mathbf{s}_{2}\mathbf{s}_{2}$ increases with increasing $M_{\mathbf{s}_{1}^{}}$ this effect is only observable upto $M_{\mathbf{s}_{2}^{}}\sim880$ GeV. Also, for $v_{s}=400$ TeV, an under-abundance region appears between $M_{\mathbf{s}_{1}^{}}=M_{\mathbf{s}_{2}^{}}$ and $M_{\mathbf{s}_{1}^{}}\sim900$ GeV (for $M_{\mathbf{s}_{1}^{}}\leq900$ GeV), along with the narrow zones on both sides of the $M_{\mathbf{s}_{1}^{}}=M_{\mathbf{s}_{2}^{}}/2$ line. The right panel of Fig.~\ref{fig:rlicMM} demonstrates the {\bf case II} scenario. For $v_{s}=0.3$ TeV (shown in blue), a wider \textbf{V} shape (divided by $M_{\mathbf{s}_{1}^{}}=M_{\mathbf{s}_{2}^{}}/2$ line) region is observed compared to $v_{s}=1$ TeV (shown in magenta). This occurs because in {\bf case II}, as $v_{s}$ increases, the cross section of the process $\mathbf{s}_{1}\mathbf{s}_{1}\rightarrow\mathbf{s}_{2}\mathbf{s}_{2}$ decreases (see Eq.~\ref{eq:matrixamp2}) leading to an increase in relic density. Inside this \textbf{V} shaped zone lies the underabundance region similar to {\bf case-I}, while the overabundance region lies outside. For $v_{s}=0.3$ TeV, the overabundance area forms a disc-like region near the $M_{\mathbf{s}_{1}^{}}=M_{\mathbf{s}_{2}^{}}$ line. As $M_{\mathbf{s}_{2}}$ increases the cross-section of the process $\mathbf{s}_{1}\mathbf{s}_{1}\rightarrow\mathbf{s}_{2}\mathbf{s}_{2}$ also rises, reducing the DM relic density. As a result for $M_{\mathbf{s}_{2}}\geq916$ GeV, all regions satisfying $M_{\mathbf{s}_{1}}\geq M_{\mathbf{s}_{2}}/2$ fall into the under abundance category for $v_{s}=0.3$ TeV. 
\subsection{Direct detection prospects}
\begin{figure}[ht]
   \begin{center}
    \begin{tikzpicture}[line width=0.6 pt, scale=2.1]
\draw[dashed] (-1.8,1.0)--(-0.8,0.5);
\draw[solid] (-1.8,-1.0)--(-0.8,-0.5);
\draw[dashed] (-0.8,0.5)--(-0.8,-0.5);
\draw[dashed] (-0.8,0.5)--(0.2,1.0);
\draw[solid] (-0.8,-0.5)--(0.2,-1.0);
\node at (-2.1,1.1) {$\mathbf{s}_{1}$};
\node at (-2.1,-1.1) {$n$};
\node at (-0.5,0.07) {$h,\phi,\mathbf{s}^{}_{2}$};
\node at (0.5,1.2) {$\mathbf{s}_{1}$};
\node at (0.5,-1.2) {$n$};
     \end{tikzpicture}
 \end{center}
\caption{Relevant processes for spin-independent DM-nucleon scattering.} 
\label{fig:DD}
 \end{figure}
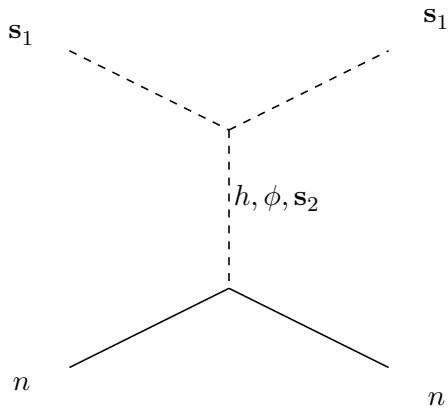

Direct detection (DD) experiments provide a powerful avenue to probe thermal DM through its elastic scattering with nucleons or electrons in the underground detectors. Although no positive signal has been observed so far, stringent upper bounds on the DM-nucleon scattering cross-section have been set by experiments such as \text{XENON1T}~\cite{XENON:2018voc}, \text{XENONnT}~\cite{XENON:2023cxc}, \text{LUX-ZEPLIN}~\cite{LZ:2022lsv}, and the projected sensitivities of \text{DARWIN/XLZD}~\cite{Baudis:2024jnk} and \text{PandaX-xT}~\cite{PANDA-X:2024dlo}. In our model, DM interactions with nucleons relevant to direct detection primarily proceed via spin-independent (SI) scattering, mediated by t-channel exchange of three scalar particles ($h$, $\phi$ and $\mathbf{s}_{2}^{}$), as depicted in Fig:~\ref{fig:DD}. 

The spin-independent DM-nucleon scattering cross-section is~\cite{Jungman:1995df, Bertone:2004pz, Ghosh:2017fmr}
 \begin{widetext}
     \begin{eqnarray}
    \sigma_{\rm DD}^{\rm SI} &=& \frac{f_{\mathbf{s}_{1}^{}}}{4\pi} \bigg(  \frac{f_n \mu_n m_{n}}{M_{\mathbf{s}_{1}^{}}v}    \bigg)^2  ~\bigg[\frac{\lambda_{\mathbf{s}_{1}^{}\mathbf{s}_{1}^{}h} U_{11} }{t-M_{h}^{2}}+\frac{\lambda_{\mathbf{s}_{1}^{}\mathbf{s}_{1}^{}\mathbf{s}_{2}^{}}U_{13}}{t-M_{\mathbf{s}_{2}^{}}^{2}}+\frac{\lambda_{\mathbf{s}_{1}^{}\mathbf{s}_{1}^{}\phi}U_{12}}{t-M_{\phi}^{2}}\bigg]^{2} \nonumber \\
    &\overset{t \to 0}{=}&  \frac{f_{\mathbf{s}_{1}^{}}}{4\pi} \bigg(  \frac{f_n \mu_n m_{n}}{M_{\mathbf{s}_{1}^{}}v}    \bigg)^2   ~\bigg[\frac{\lambda_{\mathbf{s}_{1}^{}\mathbf{s}_{1}^{}h} U_{11}}{M_{h}^{2}}+\frac{\lambda_{\mathbf{s}_{1}^{}\mathbf{s}_{1}^{}\mathbf{s}_{2}^{}}U_{13}}{M_{\mathbf{s}_{2}^{}}^{2}}+\frac{\lambda_{\mathbf{s}_{1}^{}\mathbf{s}_{1}^{}\phi}U_{12}}{M_{\phi}^{2}}\bigg]^{2}\nonumber \\
    &\approx&\frac{f_{\mathbf{s}_{1}^{}}}{4\pi} \bigg(  \frac{f_n \mu_n m_{n}}{M_{\mathbf{s}_{1}^{}}v}    \bigg)^2   ~\left[16~s_{13}(\lambda_{s'}+\lambda_{s''})v_{s}\bigg(\frac{1}{M_{h}^{2}}+\frac{1}{M_{\mathbf{s}_2^{}}^{2}}\bigg)\right]^{2},
    \label{eq:ddX}
\end{eqnarray}
 \end{widetext}
where $f_{\mathbf{s}_{1}^{}}=\frac{\Omega_{\mathbf{s}_{1}^{}}}{\Omega_{DM}^{}}$ is the fractional DM density in our case $f_{\mathbf{s}_{1}^{}}=1$, expressions for all the vertex factor is already given in Eq:~\ref{eq:vartex}, $m_{n}^{}=0.946$ MeV is the mass of neutron, $\mu_{n}^{}=\frac{m_{n}^{}M_{\mathbf{s}_{1}^{}}}{m_{n}^{}+M_{\mathbf{s}_{1}^{}}}$ is the reduced mass of the DM-neutron system and $f_{n}=0.308\pm0.018$~\cite{Hoferichter:2017olk} is the nucleon form factor. Also, U is the unitary matrix (see Eq:\ref{eq:scalarmix}) that diagonalises the mass matrix and gives the physical mass to the scalars $h,\phi,\mathbf{s}_{2}^{}$. While deriving Eq:~\ref{eq:ddX}, we assume the hierarchy of scalar masses $M_{\phi} \gg M_{h},\, M_{\mathbf{s}_2^{}}$ and the the hierarchy between the scalar VEVs $v_{\phi}\gg v_s$. 

\begin{figure*}[htbp!]
    \centering
    \includegraphics[width=0.49\linewidth]{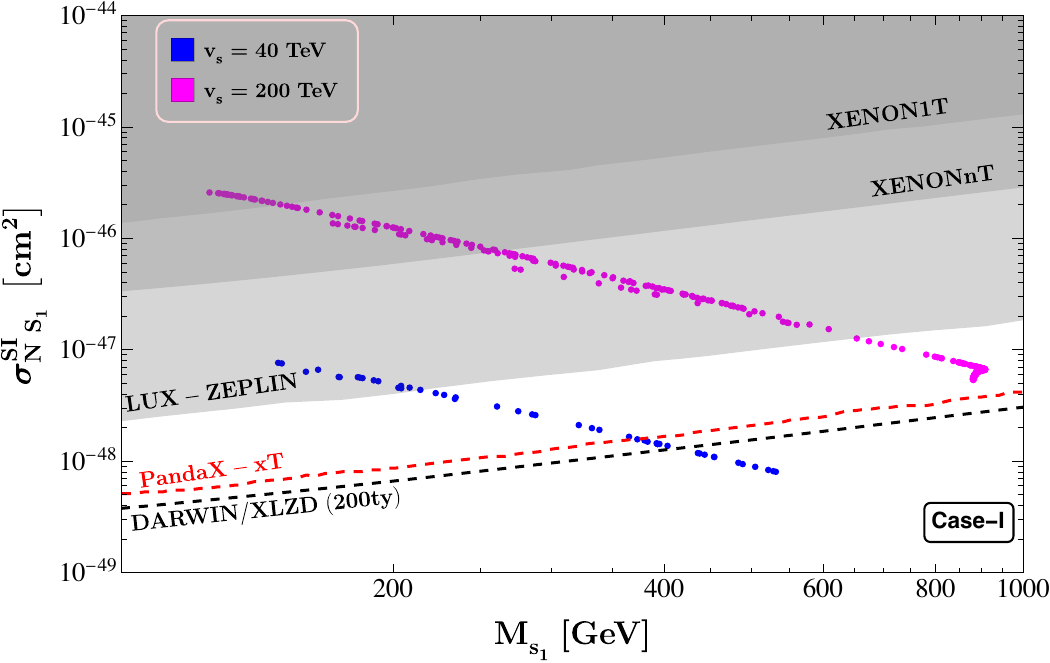}
    \includegraphics[width=0.49\linewidth]{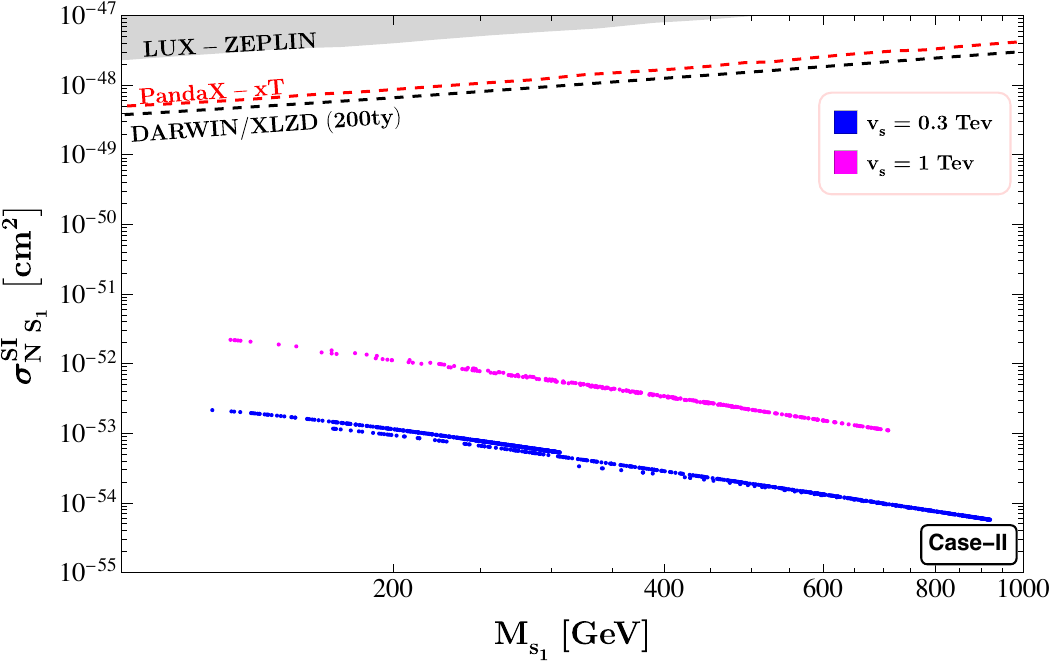}
   \caption{The relic allowed parameter space is represented in $M_{\mathbf{s}_{1}^{}}-\sigma^{\text{SI}}_{\text{N}~\mathbf{s}_{1}^{}}$ plane for both {\bf case-I} and {\bf case-II} scenarios.}
    \label{fig:rlicDD}
\end{figure*}

The dependence of the spin-independent direct detection cross-section, $\sigma_{\text{DD}}^{\text{SI}}$, on the parameters $M_{\mathbf{s}_2^{}}$ and $v_s$ is clearly illustrated in Fig.~\ref{fig:rlicDD}. For both {\bf case I} and {\bf case II}, $\sigma_{\text{DD}}^{\text{SI}}$ increases with increasing $v_s$, while it decreases with increasing $M_{\mathbf{s}_2^{}}$, as also evident in Eq:~\ref{eq:ddX}. The left panel of Fig:~\ref{fig:rlicDD} clearly demonstrates that in {\bf case-I}, with $v_{s}=200$ \text{TeV} and $\theta_{ij}\sim10^{-2}$, the majority of the parameter space is excluded by direct detection constraints, leaving only the higher mass region ($M_{\mathbf{s}_{2}^{}}$) as experimentally viable. For $v_{s}=40$ \text{TeV}, we have a larger parameter space allowed by DD experiments, but future experiments can also rule out most of that. A viable approach to evade the DD constraints in {\bf case-I} is to consider smaller mixing angles, such as $\theta_{ij} \sim 10^{-4}$ or lower (see Eq:~\ref{eq:ddX}). This region of parameter space will be particularly relevant when we investigate the gravitational wave signatures arising from domain wall annihilation. It is worth mentioning that the DM analysis has been carried out with $v_{\phi}=10^9~\text{GeV}$. We have also verified that increasing $v_{\phi}$ does not alter our conclusions, as the $\phi$ field, being heavier, decays entirely into other scalars of the model well before the DM freeze-out, rendering it irrelevant to the DM relic abundance.
\section{Domain wall annihilation and gravitational wave signals}\label{sec:domain wall}
\begin{figure*}[htbp!]
\centering
\includegraphics[width=0.48\linewidth]{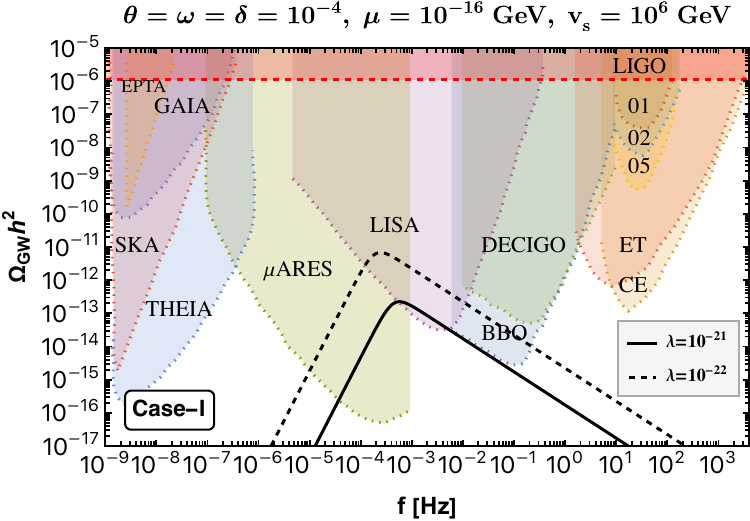}
\includegraphics[width=0.48\linewidth]{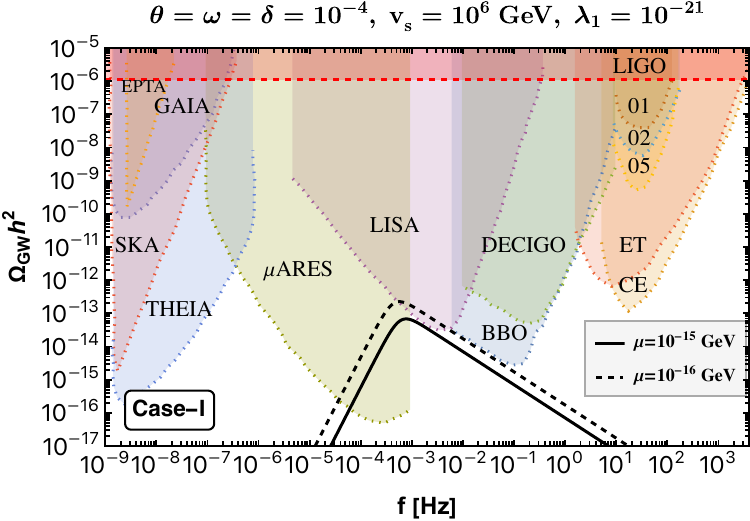}
\includegraphics[width=0.48\linewidth]{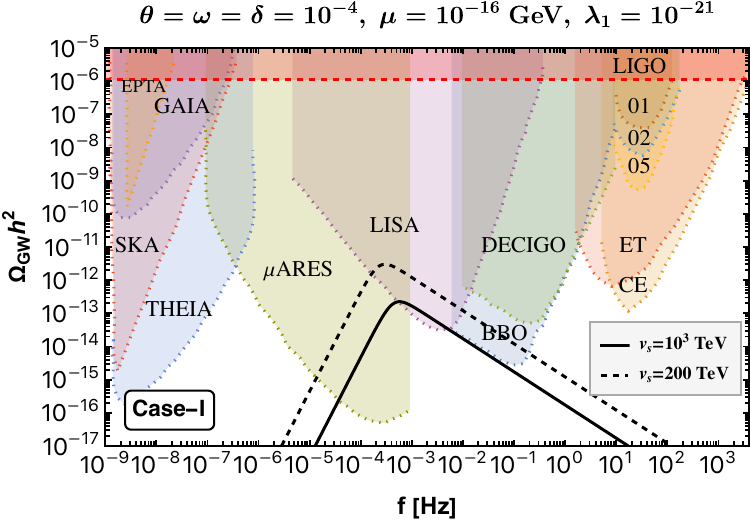}
\includegraphics[width=0.48\linewidth]{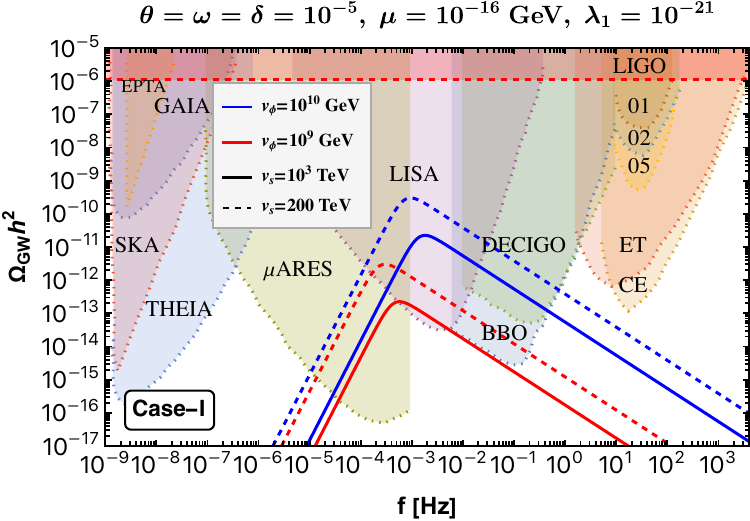}
\caption{We fixed $v_\phi=10^{9}$ GeV for the top left, top right, and bottom left panels. Bottom right panel: $\lambda_{1}=10^{-21}$, $\mu=10^{-16}$ GeV. For all of the above plots, we fixed $M_{\phi}=4\times10^{4}$ GeV.}
\label{GW_spectrum}
\end{figure*}

This section examines the phenomenological consequence of discrete $\mathcal{Z}_{4}$ symmetry breaking and its potential as a probe to test the model via gravitational wave signals. As mentioned earlier, initially our model respects a global $\mathcal{Z}_{4}$ symmetry, which is spontaneously broken when the field $\Phi$ acquires a non-zero VEV, $\langle\Phi\rangle=\pm v_{\phi}$. Spontaneous breaking of the discrete symmetry leads to the formation of a network of domain walls~\cite{Vilenkin:2000jqa, Zeldovich:1974uw, Kibble:1976sj}. These walls were formed at the boundaries that separate the regions where the field settles into different vacuum states, $+v_{\phi}$ and $-v_{\phi}$. The DW energy density, given by $\rho_{_{_{DW}}}=\frac{\sigma}{L}\simeq\frac{\mathcal{A}~\sigma}{t}$ (here, $\sigma$ is the surface energy density of the DW), decreases more slowly than the radiation energy density, $\rho_{_{R}}\simeq\frac{\pi^2}{30}~g_{\rho}~T^4\propto\frac{1}{t^{2}}$. As a result, DWs eventually dominate the universe's energy density when $\rho_{_{_{DW}}}\simeq3M_{P}^{2}H$ ($M_{P}$ is the reduced Planck mass), indicating the time scale of the order of $\mathcal{O}(\frac{M_{P}^{2}}{\sigma})$~\cite{Gouttenoire_2025} for the domination of DW to begin. This alters the universe's evolution, which is in contrast with CMB observation.\\
One possible approach to resolving the stable DW problem is to incorporate soft $\mathcal{Z}_{4}$ breaking terms, which will create an energy difference between these minima. The inclusion of the energy bias term makes the DW unstable, leading to its eventual collapse before overclosing the universe. With the three scalar fields in our hands, we can have various combinations for the energy bias term. One of the cases is the one below \footnote{Although we remain agnostic about the origin of the domain-wall bias term, one can get it from higher-dimensional operators (with cut-off scale, $\Lambda_{QG}\geq10^{17}$ GeV) induced by quantum-gravity effects that explicitly break discrete symmetries~\cite{Kallosh:1995hi,Abbott:1989jw}}
\begin{align}
    \Delta V=&\mu\Phi^3+\mu_{\phi H}\Phi H^\dagger H+\mu_{\phi ss}\Phi S^\ast S+(\lambda^{}_{1}~\Phi^3S+h.c.)\nonumber\\
    &+(\lambda^{}_{2}~S^\ast S~S \Phi+h.c.)+\frac{C_1}{\Lambda}\Phi^5+\frac{C_2}{\Lambda_{}}\Phi^3H^\dagger H\nonumber\\
    &+\frac{C_3}{\Lambda}\Phi (H^\dagger H)^2+\cdots,
    \label{eq:delV}
\end{align}
where $\mu,\mu_{\phi H}$ and $\mu_{\phi ss}$ are the couplings with mass dimension, while $\lambda_{1}$ and $\lambda_{2}$ can generally be complex quantities. In our analysis, we showed that we need $v_{\phi}\geq10^{9}$ GeV to obtain the correct order of the asymmetry, as predicted by PLANCK. The parameter $v_{s}$ plays a crucial role in WIMP-like DM phenomenology of our model. We can have the correct DM relic with $300~\text{GeV}\leq v_{s}\leq10^6~\text{GeV}$. This indicates that in our model, we have the condition $v_{s}<<v_{\phi}$. With the assumption of $C_{i}$ very small, we can safely drop dimension-5 non-renormalizable terms from Eq:~\ref{eq:delV}. Now, we can write the energy bias term as
\begin{eqnarray}
    \mathbf{V}_{bias}\equiv\mu v_{\phi}^{3}+2~\text{Re}[\lambda_{1}]v_{\phi}^{3}v_{s}^{}.
     \label{eq:Vbias}
\end{eqnarray}
As long as $\mu\approx2~\text{Re}[\lambda_{1}]v_{s}^{}$, we have to take into account both the term in Eq:~\ref{eq:Vbias} while performing gravitational wave signal analysis. However, for $\mu\gg~\text{Re}[\lambda_{1}]v_{s}^{}$ the peak amplitude and peak frequency of the GW will predominantly depend on two parameters $v_{\phi}$ and $\mu$. Below the elecro-weak scale, the presence of $(\lambda^{}_{1}~\Phi^3S+h.c.)$ and $(\lambda^{}_{2}~(S^{\ast}S)S\Phi+h.c.)$ terms will make DM unstable ($\mathcal{Z}_{2}$ breaking term). With the introduction of these terms, DM $\mathbf{s}_{1}^{}$ can decay to $\phi, \mathbf{s}_{2}$ and SM Higgs (h). Assuming the SM Higgs is the lightest scalar, the most dominant decay process will be $\mathbf{s}_{1}\rightarrow hh$. 
\begin{figure*}[htbp!]
    \centering
    \includegraphics[width=0.49\linewidth]{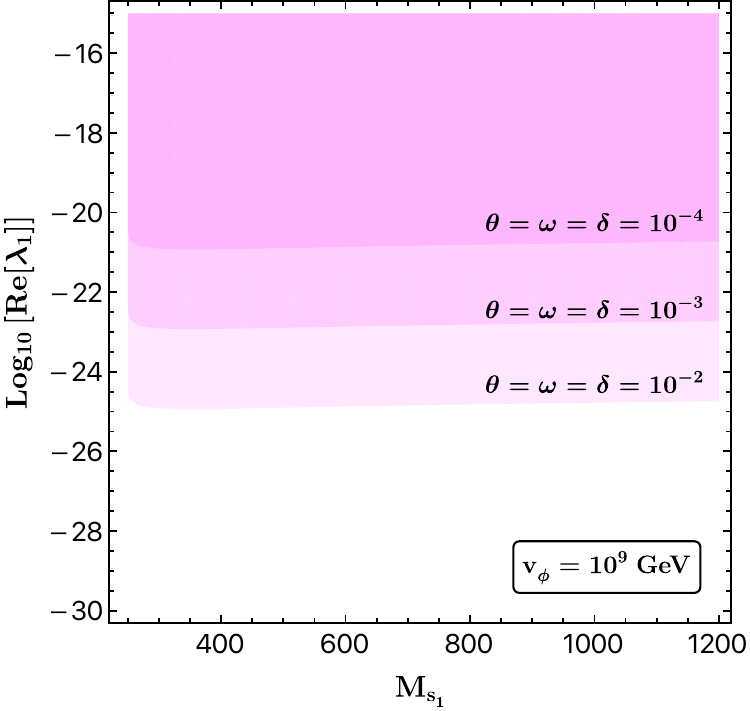}
    \includegraphics[width=0.49\linewidth]{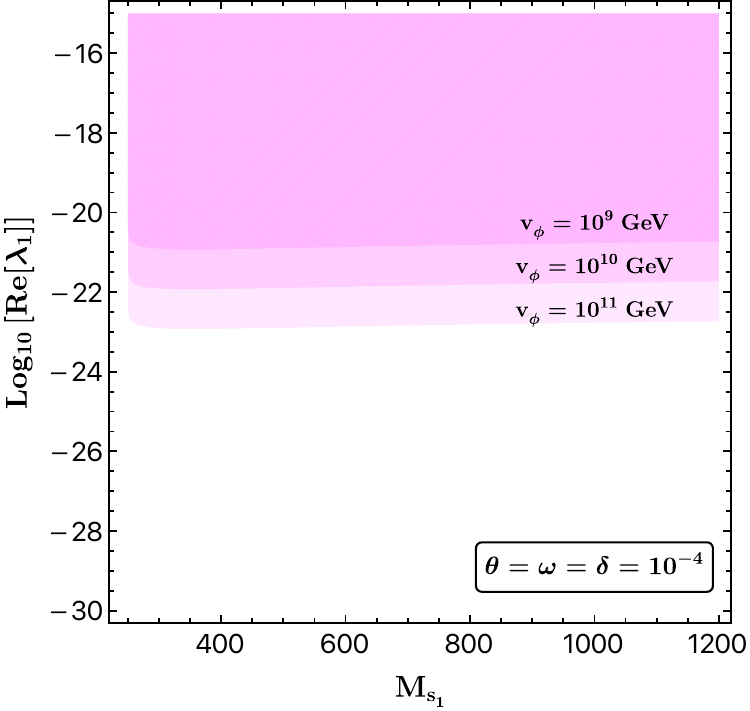}
   \caption{The above plots depict the upper bound on $\lambda_{}$ imposed by the dark matter stability condition, considering the universe's lifetime. In the left panel, different shaded regions represent different mixing angles with $v_{\phi}$ fixed at $10^9$ \text{GeV}, while in the right panel, they correspond to different values of $v_{\phi}$, with the scalar mixing angles ($\theta, \delta, \omega$) fixed at $10^{-4}$.}
    \label{fig:cconst}
\end{figure*}
To make the DM stable during the universe's time scale, we have to satisfy the condition 
\begin{equation}
    \Gamma_{\mathbf{s}_{1}\rightarrow hh}~\leq\frac{1}{\tau_{}}\times 6.58\times10^{-25}~\text{GeV},
    \label{eq:DMst}
\end{equation}
where $\tau$ is the lifetime of the universe ($\tau\approx4.3\times 10^{17}$ seconds). From Eq:~\ref{eq:DMst} we get an upper bound of the coupling $\lambda_{1}^{}$ which is depicted in Fig:~\ref{fig:cconst} (for simplicity we take $\lambda_{1}^{}=\lambda_{2}^{}=\lambda_{}^{}$). For the left panel (of Fig:~\ref{fig:cconst}) we consider the VEV of $\Phi$, $\langle\Phi\rangle=10^9~\text{GeV}$, increasing $v_{\phi}$ will increase the decay width $\Gamma_{\mathbf{s}_{1}\rightarrow hh}$ which will eventually decrease the upper bound on the coupling $\lambda_{1}$ as depicted in the right panel of Fig:~\ref{fig:cconst}. Introducing $V_{bias}$ will resolve the vacuum degeneracy and make the DW unstable. The energy difference between the two minima will create a volume pressure force, $p^{}_{V}\sim \mathbf{V}_{bias}$, which acts upon the walls, causing the negative VEV ($\langle\phi\rangle=-v^{}_{\phi}$) region to shrink. When the volume pressure force overcomes the tension force $p_{T}^{}$ of the wall, the domain walls begin to collapse and annihilate, producing a significant amount of GW, which may persist as a stochastic background in the current universe. The peak amplitude ($\Omega_{GW}h^{2}$) and the peak frequency ($f_{peak}$) of the GWs at the current time can be expressed as~\cite{Hiramatsu:2013qaa,Hiramatsu:2010yz, Borboruah:2024lli, Borboruah:2022eex}
 \begin{widetext}
 \begin{align}\label{fpeak}
     \Omega_{peak}h^2&\simeq 1.5\times10^{-10}\times\bigg(\frac{\tilde{\epsilon}^{}_{\text{GW}}}{0.7}\bigg)\bigg(\frac{\mathcal{A}}{1.41}\bigg)^4\bigg(\frac{10^7~\text{GeV}^4}{\mathbf{V}_{bias}}\bigg)^{2}\bigg(\frac{\mathit{\sigma}^{1/3}}{10^7~\text{GeV}}\bigg)^{12}\,,\\
     f_{peak}&\simeq 1.4\times10^{-5}~\text{Hz}\times\bigg(\frac{1.41}{\mathcal{A}}\bigg)^{1/2}\bigg(\frac{\mathbf{V}_{bias}}{10^7~\text{GeV}^4}\bigg)^{1/2}\bigg(\frac{10^7~\text{GeV}}{\mathit{\sigma}^{1/3}}\bigg)^{3/2}.
     \label{omegapeak}
 \end{align}
 \end{widetext}
 The area parameter is selected as $\mathcal{A}=1.41\pm0.3$ based on the axion model $N=4$~\cite{Kawasaki:2014sqa}. In the scaling regime, the efficiency factor $\tilde{\epsilon}=0.7$~\cite{Hiramatsu:2013qaa} is treated as constant and extracted from the numerical simulation. The surface energy density, denoted as $\mathit{\sigma}$ and also known as the tension of the domain wall, is expressed as $\mathit{\sigma}=\frac{\sqrt{8\lambda_{\phi}}v_{\phi}^{3}}{3}$. While calculating the above expressions (Eq:~\ref{fpeak}, Eq:~\ref{omegapeak}), we consider two crucial assumptions. DWs collapse during the radiation-dominated epoch of the universe, and the collapse happens instantaneously. The shape of the GW spectrum can be approximately represented by the broken power law, as demonstrated in the following equation~\cite{Caprini:2019egz, NANOGrav:2023hvm}
 \begin{eqnarray}
 \Omega_{\rm GW}h^2_{} = \Omega_{peak} h^2 \frac{(\alpha+\beta)^\gamma}{\left(\alpha x^{\beta / \gamma}+\beta x^{-\alpha / \gamma}\right)^\gamma} \ ,
\label{eq:spec-par}
\end{eqnarray}
where $x=\frac{f}{f_{peak}}$ and $\alpha, \beta$ are the two real positive numbers, represent the spectral slopes of the GWs at $f<<f_{peak}$ and $f>>f_{peak}$ respectively. The parameter $\gamma$ represents the width around the peak of the GW. Although $\alpha=3$ is fixed from causality, we need help from numerical simulations to determine $\beta$ and $\gamma$. Recent studies suggest $\beta \simeq \gamma \simeq 1$~\cite{Hiramatsu:2013qaa}. The GW signals corresponding to various choices of model parameters are shown in Fig:~\ref{GW_spectrum}. The experimental sensitivities of SKA~\cite{Weltman:2018zrl}, GAIA~\cite{Garcia-Bellido:2021zgu}, EPTA~\cite{Moore:2014eua}, THEIA~\cite{Garcia-Bellido:2021zgu}, $\mu$ARES~\cite{Sesana:2019vho}, LISA~\cite{amaroseoane2017laserinterferometerspaceantenna}, DECIGO~\cite{Seto:2001qf, Kawamura:2006up, Yagi:2011wg}, BBO~\cite{Crowder:2005nr, Corbin:2005ny, Harry:2006fi}, ET~\cite{Punturo:2010zz, Hild:2010id, Sathyaprakash:2012jk, Maggiore:2019uih}, CE~\cite{LIGOScientific:2016wof, Reitze:2019iox}, and LIGO~\cite{LIGOScientific:2016wof, LIGOScientific:2014qfs, LIGOScientific:2016jlg} are represented by shaded regions in different colours. The region shaded in red is excluded from $N_{\text{eff}}^{}$~\cite{Planck:2018vyg,Cyburt:2015mya,abazajian2019cmbs4sciencecasereference}
\footnote{The energy density of the GW ($\rho_{\tiny \text{GW}}^{}$) varies with the scale factor ($a$) of the universe as, $\rho_{\tiny \text{GW}}^{}\propto a^{-4}$~\cite{Caprini:2018mtu}. Hence, GW can act as an extra relativistic species (in addition to those of the SM) in the radiation-dominated era and $\rho_{\tiny\text{GW}}^{}$ should be less than $20\%$ of the total energy density of the universe at BBN ($T_{\tiny\text{BBN}}^{}\sim0.1~\text{MeV}$). This constraint can be written as,
\begin{equation}
h^{2}\Omega_{\tiny\text{GW}}^{0}\leq h^{2}\Omega_{\gamma}^{0}\frac{7}{8}\Big(\frac{4}{11}\Big)^{\frac{4}{3}}\Delta N_{eff}=5.6\times10^{-6}\Delta N_{eff},\nonumber
\end{equation}
where, the superscript $0$ indicates the value of the quantity at present time and $\Omega_{\gamma}$ is the photon relic density of the universe. Here, we use $\Delta N_{eff}\leq0.2$ from the PLANCK 2018~\cite{Planck:2018vyg} data.}.

\begin{figure*}[!htb]
\centering
\includegraphics[scale=0.5]{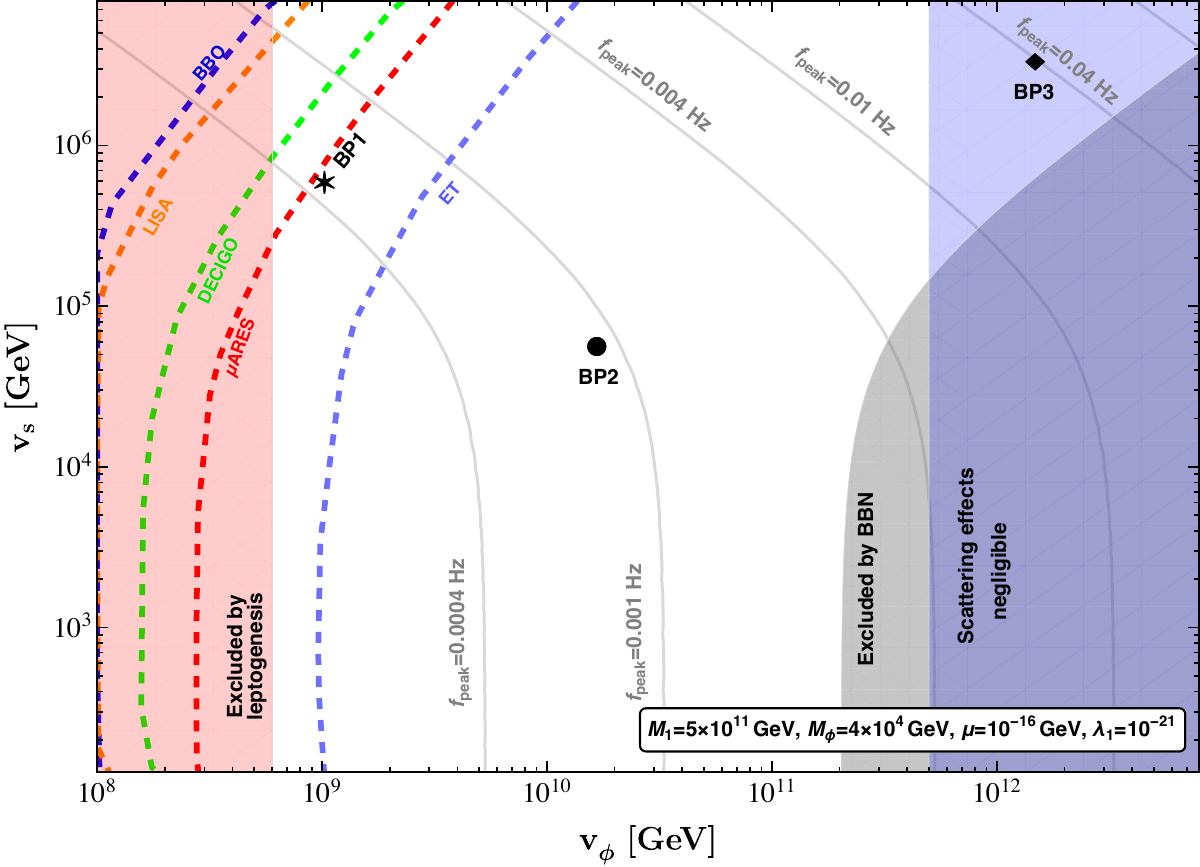}
\caption{Parameter space of the model to simultaneously explain neutrino masses, baryon asymmetry and dark matter abundance, for $M_1 = 5 \times 10^{11}$ GeV, $\mu = 10^{-8}$ and $\lambda_{1}=10^{-21}$. The solid gray lines represent $f_{peak}$ contours for the GW, whereas the dashed lines signify the sensitivity curves of various gravitational wave detection experiments (current and future). The black region is excluded due to overproduction of GWs, which is incompatible with the BBN prediction. In the blue region, $N\phi$ scatterings can be neglected, while the grey and red shaded regions are excluded by BBN and leptogenesis, respectively.}
\label{fig:Summary}
\end{figure*}

We have fixed the scalar mixing angles $\theta=\omega=\delta=10^{-4}$ and set $v_{\phi}=10^9~\text{GeV}$ for the two panels in the upper row and the left panel in the lower row of Fig:~\ref{GW_spectrum}. $M_{\mathbf{S_{2}^{}}}=200~\text{GeV}$ and $M_{\phi}=40~\text{TeV}$ are taken as constant for all the plots of Fig:~\ref{GW_spectrum}. In the top-left panel, we show the effect of varying the coupling $\lambda = \lambda_{1}$, while keeping the model parameters fixed at $\mu = 10^{-16}$ \text{GeV} and $v_{s} = 10^{6}$ \text{GeV}. Changing $\lambda_{1}$ affects the bias term ($\mathbf{V}_{\text{bias}}$), but does not alter the surface energy density ($\sigma$) of the domain wall (DW), since $v_{\phi}$ is fixed and $\lambda_{\phi} \propto \frac{M_{\phi}^{2}}{v_{\phi}^2}$ also remains constant. With increasing $\lambda_{1}$, $\mathbf{V}_{\text{bias}}$ will also increase, which in turn leads to lower $\Omega_{peak}$, as predicted by~Eq:\ref{omegapeak} and observed in this plot. In contrast, with increasing $\lambda_{1}$, $f_{\text{peak}}$ also increases (see~Eq:\ref{fpeak}), as is evident from the plot. Similar reasoning applies to the cases of varying $\mu$ and $v_{s}$, as illustrated in the top-right and bottom-left panels of Fig.~\ref{GW_spectrum}. In the bottom-right panel of the plot, the solid and dashed lines correspond to $v_{s} = 10^{3}$ \text{TeV} and $v_{s} = 200$ \text{TeV}, respectively. The blue and red curves represent $v_{\phi} = 10^{9}$ \text{GeV} and $v_{\phi} = 10^{10}$ \text{GeV}, respectively. In the Fig:~\ref{fig:Summary}, as $v_{\phi}$ increases, both $\mathbf{V}_{\text{bias}}$ and $\sigma$ increase. However, since $\Omega^{}_{peak}$ is primarily governed by $\sigma$, scaled as $\sigma^4$, a larger $v_{\phi}$ corresponds to a higher $\Omega^{}_{peak}$, as shown in the graph. Depending on the specific combinations of $\{\lambda_{1},\mu,v_{s},v_\phi\}$, the gravitational wave (GW) spectrum lies within the sensitivity range of current and upcoming GW experiments. 
\begin{table}[htbp!]
\centering
\begin{tabular}{c c c c c}
\hline \hline 
Benchmarks & $v_{s}$  (GeV)& $v_\phi$ (GeV) &$M_{\mathbf{S}_{1}{}}$ (GeV)&$M_{\mathbf{S}_{2}{}}$ (GeV)\\
 \hline 
 BP1 & $5.8\times10^{5}$ & $10^{9}$ & $335$ & $400$\\
 BP2 &$5.7\times10^{4}$ & $1.6\times10^{10}$  & $395$ & $328$\\
  BP3 & $3.4\times10^6$ & $1.5\times10^{12}$  &$180$ &$354$\\
 \hline \hline
\end{tabular}
\caption{Set of benchmark values of $v_s, v_{\phi}, M_{\mathbf{S}_{1}{}}$ and $M_{\mathbf{S}_{2}{}}$ that satisfy the observed baryon asymmetry as well as the corresponding DM mass that reproduces the observed DM relic abundance and also satisfied the direct detection bound.}
\label{tab:bpf}
\end{table}
Fig:~\ref{fig:Summary} depicts the parameter space allowed for our model that is accessible to current and upcoming GW detectors. The sensitivity curves for various GW experiments are shown as dashed lines in the $v_{\phi}-v_{s}$ plane with $\mu$ and $\lambda_{1}$ fixed at $10^{-18}$ \text{GeV} and $10^{-21}$, respectively (with $\theta=\omega=\delta=10^{-5}$). Gray colored contour lines indicate different GW peak frequencies across the parameter space. As discussed earlier, the plot further confirms that below a certain value of $v_{s}$ (where $\mu>>2\text{Re}[\lambda_{1}]v_{s}$) the peak frequencies of the GW become independent of $v_{s}$ and depend only on $v_{\phi}$ (see Eq:\ref{eq:Vbias}). The blue region on the right denotes the high $v_{\phi}$ regime where $N\Phi$ scattering becomes negligible, effectively restoring the canonical type-I leptogenesis. The red region on the left indicates parameter values incapable of generating the observed baryon asymmetry due to strong wash-out from $N\phi$ scatterings for low $v_{\phi}$ (see Eq:\ref{eq:DSre}). These exclusion regions correspond to the specific choice of $M_{1}=5\times10^{11}$ \text{GeV}. For larger (smaller) $M_{1}$, these regions shift to higher (lower) values of $v_{\phi}$, since for fixed $v_{\phi}$, the reaction density $\gamma_{S}$ grows with $M_{1}$, enhancing scattering-induced wash-out (see Eq:\ref{eq:DSre}). In Fig:~\ref{fig:Summary}, we showed three benchmark points (listed in Table:\ref{tab:bpf}) that simultaneously account for DM abundance, neutrino mass and baryon asymmetry. $\text{BP}1$ and $\text{BP}2$ inherits the effect of scattering in achieving correct lepton asymmetry, whereas $\text{BP}3$ stands for usual decay dominated leptogenesis scenario. We would also like to mention that different sensitivity curves for different GW detectors in Fig:~\ref{fig:Summary}, has been evaluated using the signal-to-noise ratio ($\varrho=20$) \cite{Maggiore:1999vm, Allen:1997ad}
\begin{equation}
    \varrho=\left[n_{\mathrm{det}} t_{\mathrm{obs}} \int_{f_{\min }}^{f_{\max }} d f\left(\frac{\Omega_{\text {signal }}(f)}{\Omega_{\text {noise }}(f)}\right)^2\right]^{1 / 2} \, .
\end{equation}
\section{Summary and conclusions}\label{sec:summary}
We propose a unified framework that simultaneously accounts for neutrino mass generation, baryogenesis via leptogenesis, and dark matter production by extending the type-I seesaw mechanism with a scalar mediator $\Phi$ that couples a complex dark scalar $S$ to right-handed neutrinos $N_i$. In particular, two features make this model special, one, an effective dimension five term $\overline{l}  \widetilde{H}\Phi N$, second the scalar potential ($V(H,\Phi,S)$). 
The model inherits a dark sector symmetry, which governs the form $V(H,\Phi,S)$ and dictates the phenomenology of the model. One of the most interesting choice of the symmetry can be $\mathcal{Z}_4\times CP$, which spontaneously gets broken by the VEV of $\Phi$ ($v_\phi$), which in turn instigates neutrino mass generation, and lepton number asymmetry. $v_\phi$ induces a VEV to the complex scalar ($v_s$), keeping one component $\mathbf{s}^{}_1$ stable, as the DM candidate due to the imposed CP symmetry, while allowing a rich phenomenology beyond the simplest Higgs portal framework. 

DM phenomenology is crucially dictated by $v_s$. When we assume the DM to be a thermal WIMP, the 
annihilation cross-sections are dictated by the triple scalar couplings, which depend on $v_s$. When $v_s$ is large, the contribution to the couplings are directly proportional to $v_s$, whereas when $v_s$ is small, the effect comes from inverse dependence, marking a sharp distinction between these cases. As a 
result, relic density plus direct search allowed parameter space favors small $v_s \lesssim 10^4$ GeV, while the larger $v_s$ region partially lies within exclusion limits. Leptogenesis, on the other hand, is governed by the standard RHN decay, as well as the scattering term involving $\Phi$ which contributes to wash out. Importantly, the saturation of lepton asymmetry is also crucially governed by the VEV of $\Phi$ ($v_\phi$). To address the observed baryon asymmetry, we achieve a lower limit on $v_\phi\gtrsim 10^9$ GeV. Neutrino mass generation relies on the $\mathcal{Z}_4$ symmetry breaking, 
although the active neutrino masses do not pertain to a specific $v_\phi$ having the Yukawa coupling on board. 

The discrete symmetry breaking creates cosmological domain walls that can over close the energy budget of the Universe, so that we need to incorporate soft $\mathcal{Z}_4$ breaking terms to annihilate domain walls. This in turn create gravitational wave signals which also therefore depend on $v_\phi$ and $v_s$. The detectability of such GW signals at future experimental sensitivities may provide crucial 
indication to such symmetry breaking mechanism to govern the simultaneous existence of baryon asymmetry and DM. The extended scalar sector, apart from the DM and SM Higgs, 
particularly $\mathbf{s}_2$ having mass in the GeV-TeV range, can in principle be probed at the future colliders, 
which in turn may provide a complementarity to the GW probe for the model. We will address some such 
features in our future endeavours.

\acknowledgments

NM thanks the Council of Scientific \& Industrial Research (CSIR), Govt. of India, for the senior research fellowship. NM also wants to thanks Rishav Roshan for the various fruitful discussion related to the GW from DWs. We would also like to thank Peter Maták for his insightful comments and suggestions regarding the CP asymmetry calculations.

\appendix

\onecolumngrid

\section{Scalar potential for real and complex scalar DM invariant under $\mathcal{G}_{D}$ or $\mathcal{G}_{D}^{\prime}$}
\label{sec:Scalar potential}

In this appendix section, we outline the various possible structure of the scalar potential that remain invariant under the $\mathcal{G}_{D}\equiv\mathcal{Z}_{2p}\times\mathcal{Z}_{2}^{\prime}$ (or $\mathcal{G}_{D}^{\prime}\equiv\mathcal{Z}_{2p}\times\mathcal{Z}_{2}^{\prime}\times CP$) symmetry group structure. The charge assignments of the newly introduced fields under this group structure is already specified in Table~\ref{tab:k}.  

\subsection{SM with two real singlet scalars}
\label{subsec: two real singlet scalar}
If two real singlet scalars ($\Phi$ and $S$) are added with the SM Higgs, the resulting scalar potential invariant under $\mathcal{Z}_{2p}\times\mathcal{Z}_{2}^{\prime}$, can be expressed as follows
\begin{equation}\label{eq:SM+2RSZ2}
\begin{split}
\widetilde{V}(H,\Phi,S)=&-\mu_H^2H^{\dagger}H+\lambda^{}_H(H^{\dagger}H)^2 -\frac{\mu_\phi^2}{2}
\Phi^2+ \frac{\lambda^{}_\phi}{4!}\Phi^4 
+\mu_S^2 S^{2}+\frac{\lambda_{S}}{4!}S^4_{}\\
&+\frac{\lambda^{}_{\phi H}}{2}\Phi^2H^{\dagger}H+\frac{\lambda^{}_{S H}}{2}S^2H^{\dagger}H+\frac{\lambda^{}_{\phi S}}{4}\Phi^2 S^{2}\,.
\end{split}
\end{equation}
Here and through the other scalar potentials discussed in the Appendix, we impose $\lambda^{}_H, \lambda^{}_\phi > 0$ and $\mu_H^2, \mu_\phi^2 > 0$, which ensure that the scalar potential $ \widetilde{V}(H, \Phi, S)$ acquires two stable minima along the $\Phi$ and $H$ directions. In addition we restrict ourselves to the case $\mu_S^2 > 0$.

Once $\Phi$ acquires the VEV through the spontaneous breaking of the $\mathcal{Z}_{2p}$ symmetry, the scalar $S$ may also develop an induced VEV. The condition for $S$ to get an induced VEV is given by
\begin{equation}\label{eq:iVEvc1R}
-2\mu_{S}^{2}-\lambda_{\phi S}v_{\phi}^2>0\,.
\end{equation}  
If the parameter space  satisfies above inequality, the $\mathcal{Z}_{2}^{\prime}$ symmetry will also broken via the induced VEV of $S$, as a result, $S$ no longer possesses any residual stabilizing symmetry to become a viable DM candidate.

\subsection{SM with one real singlet and one complex singlet scalar}
\label{subsec: 1R1Csinglet scalar}

If, instead of introducing two real singlet scalars, the scalar sector is extended by one real ($\Phi$) and one complex scalar singlet ($S\equiv(\mathbf{s}_1^{}+iS^{}_2)/\sqrt{2}$) in addition to the SM Higgs, then the corresponding scalar potential, preserving the $\mathcal{Z}_{2p}\times\mathcal{Z}_{2}^{\prime}$ can be written as
\begin{equation}\label{eq:scalarpot}
    \begin{split}
        \widetilde{V}(H,\Phi,S)  = & -\mu_H^2H^{\dagger}H+\lambda^{}_H(H^{\dagger}H)^2 -\frac{\mu_\phi^2}{2}
\Phi^2+ \frac{\lambda^{}_\phi}{4!}\Phi^4 
+\mu_S^2 S^{\ast}S+\Bigg[\frac{\mu_{SS}^2S^2}{2}+h.c. \Bigg] \\
        & +\lambda^{}_S (S^{\ast}S)^{2}+\Bigg[\frac{\lambda_{S^\prime}}{2} (S^{\ast}S)S^2+h.c.\Bigg]+\Bigg[\frac{\lambda_{S^{\prime\prime}}}{4!}S^4+h.c.\Bigg]+\frac{\lambda^{}_{\phi H}}{2}\Phi^2H^{\dagger}H\\
        &+\lambda^{}_{SH}(S^{\ast}S)H^{\dagger}H+\Bigg[\frac{\lambda^{\prime}_{SH}}{2}H^{\dagger}H S^2+h.c.\Bigg]+\frac{\lambda^{}_{\phi S}}{2}\Phi^2 (S^{\ast}S)+\Bigg[\frac{\lambda^{\prime}_{\phi S}}{4}\Phi^2S^2+h.c.\Bigg]\,,\\
    \end{split}
\end{equation}
where the mass coupling $\mu_{SS}^{}$ and the dimensionless couplings $\lambda_{S^{\prime}},\lambda_{S^{\prime\prime}},\lambda^{\prime}_{SH}$ and $\lambda^{\prime}_{\phi S}$ can, in general, take complex values. For the above scalar potential, the viable DM candidate is identified as the component of $S$ that attains the smallest mass after the electroweak (EW) symmetry breaking, which can be quantified as  
\begin{equation}\label{eq:DMmassZ2}
\mathbf{M}_{S}^{2}=\mu_{S}^{2}+\frac{\lambda_{SH}}{2}v^2+\frac{(\lambda_{\phi S}-|\lambda_{\phi S}^{\prime}|)}{2}v_{\phi}^2\,.
\end{equation}
In deriving the above expression for the DM mass, we have assumed that $|\mu_{SS}^{}\ll v_{\phi}^{}|$. This condition naturally follows from the leptogenesis requirement, which imposes the constraint $v_{\phi}\geq10^9$ \text{GeV} ~\cite{Bhattacharya:2023kws} in order to achieve successful leptogenesis.

Once $\Phi$ acquires a VEV, here also arises the possibility that one of the components of $S$ (either its real or imaginary part) develops an induced VEV as a consequence of the spontaneous breaking of $\mathcal{Z}_{2p}$, driven by the presence of the interaction terms $\lambda^{}_{\phi S}\Phi^2 (S^{\ast}S)$ and $\lambda^{\prime}_{\phi S}\Phi^2S^2$. In particular, the condition for real component of $S$ to acquire induced VEV after $\Phi$ obtains it VEV is given by
\begin{equation}\label{eq:inVEVZ2}
-2(\mu_{S}^2+\text{Re}[\mu_{SS}^{2}])-(\lambda_{\phi S}+\text{Re}[\lambda_{\phi S}^{\prime}])>0\,.
\end{equation}
At first glance, one might expect that the remaining component of $S$, which does not develop an induced VEV, could serve as a viable DM candidate. However, this possibility is ruled out by the presence of terms containing $\mathbf{s}_1^{}S^{}_2$ (arising from $S^2+\text{h.c.}$ terms). Once one of the component of $S$ gets induced VEV, this $\mathbf{s}_1^{}S^{}_2$ term introduce bilinear mixing between the remaining components of $S$ (which doesn't develop induced VEV) with the other scalars that acquire VEV. Consequently, the remaining component of $S$ can decay into the other scalars, thus disqualifying it as a candidate for DM. This outcome arises because the acquisition of a VEV by one component of $S$ breaks the $\mathcal{Z}_{2}^{\prime}$ symmetry, leaving no residual symmetry to stabilise either $\mathbf{s}_1^{}$ or $S^{}_2$ in the regions of parameter space where the induced VEV condition is satisfied.

From the above two examples, it becomes evident that if the induced VEV condition is satisfied, $S$ can not serve as viable DM candidate under the $\mathcal{Z}_{2p}\times\mathcal{Z}_{2}^{\prime}$ symmetry, regardless of whether it is real or complex. In the case of a complex scalar, an additional symmetry is required to stabilise one of its component. A natural choice is the CP symmetry under which $S$ transforms as $S\rightarrow S^\ast$ (or $S\rightarrow-S^\ast$). In presence of $\mathcal{Z}_{2p}\times\mathcal{Z}_{2}^{\prime}\times CP$, the scalar potential given in Eq:~\ref{eq:scalarpot} is modified and takes the following forms:
\begin{equation}\label{eq:scalarpotcp}
    \begin{split}
        \widetilde{V}(H,\Phi,S)  = & -\mu_H^2H^{\dagger}H+\lambda^{}_H(H^{\dagger}H)^2 -\frac{\mu_\phi^2}{2}
\Phi^2+ \frac{\lambda^{}_\phi}{4!}\Phi^4 
+\mu_S^2 S^{\ast}S+\frac{\mu_{SS}^2}{2}\Big(S^2+(S^\ast)^2\Big)\\
        & +\lambda^{}_S (S^{\ast}S)^{2}+\frac{\lambda_{S^{\prime}}}{2} (S^{\ast}S)\Big(S^2+(S^\ast)^2\Big)+\frac{\lambda_{S^{\prime\prime}}}{4!}\Big(S^4+(S^\ast)^4\Big)+\frac{\lambda^{}_{\phi H}}{2}\Phi^2H^{\dagger}H\\
        &+\lambda^{}_{SH}(S^{\ast}S)H^{\dagger}H+\frac{\lambda^{\prime}_{SH}}{2}H^{\dagger}H \Big(S^2+(S^\ast)^2\Big)+\frac{\lambda^{}_{\phi S}}{2}\Phi^2 (S^{\ast}S)+\frac{\lambda^{\prime}_{\phi S}}{4}\Phi^2\Big(S^2+(S^\ast)^2\Big)\,.\\
    \end{split}
\end{equation}
Here, the condition for real component of $S$ to acquire induced VEV is given by
\begin{equation}
-2(\mu_{S}^2+2\mu_{SS}^{2})-(\lambda_{\phi S}+\lambda_{\phi S}^{\prime})>0\,.
\end{equation}
Owing to the imposed CP symmetry, only terms of the form $(S^2+(S^\ast)^2)$ and $(S^4+(S^\ast)^4)$ are permitted in the scalar potential. This restriction forbids the appearance of $\mathbf{s}_1^{}S^{}_2$ term, thereby preventing any mixing between the real and imaginary components, $\mathbf{s}_1^{}$ and $S^{}_2$. Even if the $\mathcal{Z}_{2}^{\prime}$ symmetry is broken through the induced VEV, the remaining $CP$ symmetry ensures that one component of $S$ remains stable, allowing it to serve as a DM candidate. 
\subsection{Equivalence between $\mathcal{Z}_{2}\times\mathcal{Z}_{2}^{\prime}$ and $\mathcal{Z}_{4}$ for the case of one real and one complex scalar}
\label{subsec: equivalence}
In this subsection of the appendix, we demonstrate that the $\mathcal{Z}_4$ symmetry leads to a situation analogous to that obtain under $\mathcal{Z}_{2}\times\mathcal{Z}_{2}^{\prime}$ when considering one real ($\Phi$) and one complex scalar ($S$). The charge assignments for BSM fields under $\mathcal{Z}_4$ is provided in Table:~\ref{tab:k}. The corresponding scalar potential, invariant under this $\mathcal{Z}_{4}$, takes the form 
\begin{equation}\label{eq:scalarpotZ4ncp}
    \begin{split}
        \widetilde{V}(H,\Phi,S)  = & -\mu_H^2H^{\dagger}H+\lambda^{}_H(H^{\dagger}H)^2 -\frac{1}{2}
\mu_\phi^2\Phi^2+ \frac{1}{4!}\lambda^{}_\phi\Phi^4 
+\mu_S^2 S^{\ast}S+ 
\lambda^{}_S (S^{\ast}S)^{2}\\
        & +\left(\lambda_{S^\prime} (S^{\ast}S)S^2+h.c.\right)+\big(\lambda^{}_{S^{\prime\prime}}S^4+h.c.\big)+\frac{\lambda^{}_{\phi H}}{2}\Phi^2H^{\dagger}H+\lambda^{}_{SH}(S^{\ast}S)H^{\dagger}H\\
        & +\frac{\lambda^{}_{\phi S}}{2}\Phi^2 (S^{\ast}S)+\left(\frac{\mu^{}_{\phi S}}{2}\Phi S^2+h.c.\right)\,,\\
    \end{split}
\end{equation}
where the couplings $\lambda^{}_{S^\prime},\,\lambda^{}_{S^{\prime\prime}}$ and $\mu^{}_{\phi S}$ can, in general, take complex values. Once the $\mathcal{Z}_{4}$ symmetry is spontaneously broken through the VEV of $\Phi$, our Lagrangian retains a remnant $\mathcal{Z}_2$ symmetry under which $S$ is odd (transform as $S\rightarrow-S$). This residual symmetry ensures that the lighter component of $S$ serves as the DM candidate, in exact analogy with the $\mathcal{Z}_{2}\times\mathcal{Z}_{2}^{\prime}$ case. The corresponding DM mass is given by
\begin{equation}\label{eq:DMmassZ4}
\mathbf{M}_{S}^{2}=\mu_{S}^{2}+\frac{\lambda_{SH}}{2}v^2+\frac{\lambda_{\phi S}}{2}v_{\phi}^2-|\mu_{\phi S}^{}|v_{\phi}^{}\,,
\end{equation}
which is structurally identical to Eq:~\ref{eq:DMmassZ2}, with the only difference being that the term $\frac{|\lambda_{\phi S}^{\prime}|v_{\phi}^2}{2}$ is replaced by $|\mu_{\phi S}^{}|v_{\phi}^{}$.

As in the $\mathcal{Z}_{2}\times\mathcal{Z}_{2}^{\prime}$ case, there exist the possibility that one component of $S$ develops an induced VEV following the spontaneous breaking of $\mathcal{Z}_4$ via $v_{\phi}^{}$. The condition for the real component of $S$ to acquire such an induced VEV is
\begin{equation}\label{eq:inVEVZ4}
-2\mu_{S}^2-\big(2\text{Re}[\mu_{\phi S}]+\lambda_{\phi S}^{}v_{\phi}^{}\big)v_{\phi}^{}>0\,,
\end{equation}
which closely resembles the condition in Eq:~\ref{eq:inVEVZ2}. Once one of the component of $S$ develops induced VEV, the remnant $\mathcal{Z}_2$ symmetry inherited from parent $\mathcal{Z}_4$ symmetry is also broken. Therefore, as in the previous, an additional symmetry (in our framework, a $CP$ symmetry) is required to ensure the stability of the DM candidate (see Eq:~\ref{eq:scalarpotZ4cp}).
\section{Expressions for quartic couplings associated with $\mathcal{Z}_4\times CP$ symmetric scalar potential}
\label{sec: couplings}
Below, we present the expressions for the scalar couplings in terms of the external parameters listed in Table:~\ref{tab:p}
\begin{align}
&\lambda_{H}^{}=\frac{M_{H}^2+M_{\phi}^2+2M_{\mathbf{s}_{2}^{}}^2+2\Big(M_{H}^2-M_{\phi}^2\Big)c2_{12}^{}c_{13}^2+\Big(M_{H}^2+M_{\phi}^2-2M_{\mathbf{s}_{2}^{}}^2\Big)c2_{13}^{}}{8v^2}\,,\\
&\lambda_{S_1^{}}^{}=\frac{6M_{H}^2+6M_{\phi}^2+4M_{\mathbf{s}_{2}^{}}^2-64\Big(\lambda^{}_{S^\prime_{}}+2\lambda^{}_{S^{\prime\prime}_{}}\Big)v_{\phi}^2-2\Big(M_{H}^2+M_{\phi}^2-2M_{\mathbf{s}_{2}^{}}^2\Big)\big(c2_{13}^{}+2c_{13}^{}c2_{23}^{}\big)}{32v_{s}^2}\,,\\
&\lambda_{S_2^{}}^{}=\frac{-2\Big(M_{H}^2-M_{\phi}^2\Big)\big(1+c2_{13}^{}-3c2_{13}^{}+c2_{23}^{}c2_{13}^{}\big)c2_{12}^{}-8\Big(M_{H}^2-M_{\phi}^2\Big)s2_{12}^{}s2_{23}^{}s_{13}^{}}{32v_{s}^2}\,,\\
&\lambda_{\phi_{1}^{}}=\frac{6\Big(M_{H}^2-M_{\phi}^2\Big)v_{\phi}+4M_{\mathbf{s}_{2}^{}}^2v_{\phi}-8\mu_{\phi S}^{}v_{s}^{2}-2\Big(M_{H}^2+M_{\phi}^2-2M_{\mathbf{s}_{2}^{}}^2\Big)\big(c2_{13}^{}-2c_{13}^2c2_{23}^{}\big)v_{\phi}}{16v_{\phi}^3}\,,\\
&\lambda_{\phi_{2}^{}}=\frac{2\Big(M_{H}^2-M_{\phi}^2\Big)\big(c2_{13}^{}c2_{23}^{}-3c2_{23}^{}-2c_{13}^{2}\big)c2_{12}^{}v_{\phi}^{}+8\Big(M_{H}^2-M_{\phi}^2\Big)v_{\phi}^{}s2_{12}^{}s2_{23}^{}s_{13}^{}}{16v_{\phi}^3}\,,
\end{align}
\begin{align}
&\lambda_{\phi H}^{}=\frac{c_{13}^{}\Big[s2_{12}^{}c_{23}^{}\Big(M_{\phi}^2-M_H^2\Big)+\Bigl \{2M_{\mathbf{s}_{2}^{}}^2-M_{\phi}^2-M_{H}^2+\Big(M_{\phi}^2-M_H^2\Big)c2_{12}^{}\Bigl \}s_{13}^{}s_{23}^{}\Big]}{2vv_{\phi}^{}}\,,\\
&\lambda_{SH}^{}=\frac{c_{13}^{}\Big[c_{23}^{}s_{13}^{}\Bigl \{2M_{\mathbf{s}_{2}^{}}^2-M_{\phi}^2-M_{H}^2+\Big(M_{\phi}^2-M_H^2\Big)c2_{12}^{}\Bigl \}-\Big(M_{\phi}^2-M_H^2\Big)s2_{12}^{}s_{23}^{}\Big]}{2vv_{S}^{}}\,,\\
&\lambda_{\phi S1}^{}=\frac{s2_{23}^{}\Big[2c_{13}^2\Big(2M_{\mathbf{s}_{2}^{}}^2-M_{\phi}^2-M_{H}^2\Bigl)+c2_{12}\Big(M_{\phi}^2-M_H^2\Big)\big(c2_{13}^{}-3\big)\Big]}{8v_{S}^{}v_{\phi}^{}}\,,\\
&~~~~~~~~~~~~\lambda_{\phi S2}^{}=\frac{8\mu_{\phi S}^{}v_{\phi}^{}-4\Big(M_{\phi}^2-M_{H}^2\Big)c2_{23}^{}s2_{12}^{}s_{13}^{}}{8v_{s}^{}v_{\phi}^{}};\quad\quad \quad\lambda_{\phi S}^{}=\lambda_{\phi S1}^{}+\lambda_{\phi S2}^{};\\
&~~~~~~~~~~~~~~~\lambda_{S}^{}=\lambda_{S_1^{}}^{}+\lambda_{S_2^{}}^{};\quad\quad\quad~~~~~~~~~~~~~~~~~~~~~~~~~~~~~~~~~~~~\lambda_{\phi}=3(\lambda_{\phi_{1}^{}}+\lambda_{\phi_{2}^{}})\,,
\end{align}
where $\cos{\theta_{ij}}\equiv c_{ij}$, $\sin{\theta_{ij}}\equiv s_{ij}$, $\cos{2\theta_{ij}}\equiv c2_{ij}^{}=c^2_{ij}-s^2_{ij}$ and $\sin{2\theta_{ij}}\equiv s2_{ij}^{}=2s_{ij}^{}c_{ij}^{}$.
\section{Theoretical and experimental constraints of the model}\label{sec: theoexcon}

This section outlines the theoretical and experimental constraints on the model parameters that are pertinent to our analysis.
\begin{itemize}
    \item {\bf stability criteria of the scalar potential}\\
    To ensure that the scalar potential $V(H,\Phi, S)$ remains bounded from below, the quartic couplings must satisfy the following co-positivity conditions~\cite{Elias-Miro:2012eoi,Kannike:2016fmd,Chakrabortty:2013mha, Bhattacharya:2021rwh}:
    \begin{align}
    &\lambda^{}_{H}\ge0,\quad\quad\quad\lambda_{\phi}\geq0,\quad\quad\quad\lambda^{}_{S^{}_{eff}}\geq0,\quad\quad\quad\tilde{\lambda}^{}_{\phi H}\equiv\lambda^{}_{\phi H}+2\sqrt{\lambda^{}_{H}\lambda^{}_{\phi}}\geq0,\nonumber\\
    & \quad\quad\tilde{\lambda}^{}_{SH}\equiv\lambda^{}_{SH}+2\sqrt{\lambda^{}_{H}\lambda^{}_{S^{}_{eff}}}\geq0,\quad\quad\tilde{\lambda}^{}_{S\phi }\equiv\lambda^{}_{S\phi}+2\sqrt{\lambda^{}_{\phi}\lambda^{}_{S^{}_{eff}}}\geq0,\nonumber\\
    &\sqrt{\lambda^{}_{H}\lambda^{}_{\phi}\lambda^{}_{S^{}_{eff}}}+\lambda^{}_{\phi H}\sqrt{\lambda^{}_{S^{}_{eff}}}+\lambda^{}_{S\phi}\sqrt{\lambda^{}_{H}}+\lambda^{}_{SH}\sqrt{\lambda^{}_{\phi}}+\sqrt{2\tilde{\lambda}^{}_{\phi H}\tilde{\lambda}^{}_{SH}\tilde{\lambda}^{}_{S\phi }}\geq0
    \end{align}\label{eq:copos}
    \item {\bf Unitarity constraint}\\
    Perturbative unitarity constraint on the quartic couplings of the model can be derived by analysing the scattering amplitude of all possible $2\rightarrow2$ scalar processes. Ensuring that these amplitudes remain within the perturbative regime imposes an upper bound on the scalar couplings, thereby maintaining the consistency of the theory at high energies. We compute the perturbative bounds on the quartic couplings of the scalar potential, which are summarised below~\cite{Bhattacharya:2019fgs, Bhattacharyya:2015nca, Horejsi:2005da}
    \begin{equation}
       \lambda^{}_{H}<8\pi,~~\lambda^{}_{S^{}_{eff'}}<8\pi,~~\lambda^{}_{\phi H}<8\pi,~~\lambda_{SH}<8\pi,~~\lambda^{}_{S\phi}<8\pi,~~\text{and}~~x^{}_{1,2,3}<16\pi.
    \end{equation}\label{eq:untarity}
    Here, $\lambda^{}_{S^{}_{eff'}}=(\lambda^{}_{S}-6\lambda^{}_{S'}-4\lambda^{}_{S''})$ and $x^{}_{1,2,3}$ are the roots of the following equation
    \begin{align}
        &2x^3-x^2(12\lambda^{}_{H}-6\lambda^{}_{S^{}_{eff}}-2\lambda^{}_{S^{}_{eff'}}+\lambda^{}_{\phi})+x\big(36\lambda^{}_{H}\lambda^{}_{S^{}_{eff}}+3\lambda_{\phi}\lambda^{}_{S^{}_{eff}}+12\lambda_{H}\lambda^{}_{S^{}_{eff'}}\nonumber\\
        &+\lambda_{\phi}\lambda^{}_{S^{}_{eff'}}+6\lambda^{}_{H}\lambda^{}_{\phi}-2\lambda^{2}_{\phi H}-\lambda_{S\phi}^{2}-4\lambda_{SH}^{2}\big)+6\lambda^{}_{H}\lambda_{S\phi}^{}-4\lambda_{\phi H}^{}\lambda_{SH}^{}\lambda_{s\phi}^{}+2 \lambda_{\phi}^{}\lambda_{SH}^{2}\nonumber\\
        &~~~~~\quad\quad\quad\quad\quad\quad\quad-18\lambda_{H}^{}\lambda_{\phi}^{}\lambda^{}_{S^{}_{eff}}+6\lambda_{\phi}^{2}\lambda^{}_{S^{}_{eff}}-6\lambda_{H}^{}\lambda_{\phi}^{}\lambda^{}_{S^{}_{eff'}}+2\lambda_{\phi}^{2}\lambda^{}_{S^{}_{eff'}}.
    \end{align}
    \item {\bf Collider constraints on scalar mixing angles and Higgs invisible decay} \\
    Scalar extensions of the standard model can alter the SM Higgs couplings to the SM particles at tree level through scalar mixing, characterised by angles $\theta^{}_{ij}$. Combined analyses from the ATLAS~\cite{ATLAS:2022vkf} and CMS~\cite{CMS:2022dwd} collaborations, incorporating multiple decay channels ($\gamma\gamma$, $ZZ$, $WW$, $\gamma Z$, $b\bar{b}$, $\mu\mu$, $\tau\tau$), have placed a stringent upper bound on this mixing, yielding $|\sin{\theta}| \leq 0.29$~\cite{Lane:2024vur} under the assumption of a single scalar mixing with the SM Higgs. In our framework, three independent mixing angles ($\theta^{}_{ij}$ here $i,j\equiv\{1,2,3\}$) arise due to the mixing among three scalar fields. For consistency with current experimental constraints, we adopt a conservative approach by keeping all mixing angles small, typically $\sin{\theta_{ij}} \leq \mathcal{O}(10^{-2})$, ensuring that our scenario safely abides by the constraint.
    
\end{itemize}
\section{Approximate expressions for vertex factors and matrix amplitudes}
\label{sec:vertexf}
This section of the Appendix provides the approximate expressions of the vertex factors essential for evaluating various DM annihilation cross-sections, as well as the corresponding approximations for the matrix amplitudes involved in these processes. Using Eq.\eqref{eq:scalarmix} in the scalar potential (Eq.\eqref{eq:scalarpot}) part of the Lagrangian, we obtain the following list of interaction vertices involving three scalars ($h$, $\phi$, $\mathbf{s}_{2}^{}$) and the DM $\mathbf{s}_{1}^{}$
\begin{align}\label{eq:vartex}
\lambda_{\mathbf{s}_{1}^{}\mathbf{s}_{1}^{}\mathbf{\phi}}&\propto (c_{12}^{}s_{23}^{}+s_{12}^{}s_{13}^{}c_{23}^{})\Bigg(\frac{M_{\phi}^{2}}{v_{s}^{}}-16(\lambda_{s'}+\lambda_{s''})v_{s}\Bigg),\nonumber\\
\lambda_{\mathbf{s}_{1}^{}\mathbf{s}_{1}^{}h}&\propto(c_{12}^{}c_{23}^{}s_{13}^{}+s_{12}^{}s_{23}^{})\Bigg(16(\lambda_{s'}+\lambda_{s''})v_{s}-\frac{M_{h}^{2}}{v_{s}^{}}\Bigg),\nonumber\\
\lambda_{\mathbf{s}_{1}^{}\mathbf{s}_{1}^{}\mathbf{s}_{2}^{}}&\propto\Bigg(16(\lambda_{s'}+\lambda_{s''})v_{s}-\frac{M_{\mathbf{s}_{2}^{}}^{2}}{v_{s}^{}}\Bigg),\nonumber\\
\lambda_{\mathbf{s}_{2}^{}\mathbf{s}_{2}^{}\mathbf{s}_{2}^{}}&\propto -3M_{S_{2}^{}}^{2}\Bigg(\frac{c_{13}^{3}c_{23}^{3}}{v_{s}^{}}+\frac{s_{13}^{}}{v}\Bigg)\nonumber,\\ 
\lambda_{\mathbf{s}_{2}^{}\mathbf{s}_{2}^{}h}&\propto 2M_{\mathbf{s}_{2}^{}}^{2}\Bigg( \frac{c_{13}^{2}c_{23}^{2}s_{23}^{}}{v_{s}}+\frac{c_{23}^{3}c_{13}^{2}c_{12^{}}s_{13}^{}}{v_{s}^{}}+\frac{c_{13}^{3}-c_{13}^{}}{v}\Bigg)\nonumber.\\ 
\end{align}
\begin{figure}[htb!]
\begin{center}
    \begin{tikzpicture}[line width=0.5 pt, scale=1.5]
        \draw[dashed] (-5,1)--(-4,0);
	\draw[dashed] (-5,-1)--(-4,0);
	\draw[dashed] (-4,0)--(-3.2,0);
	\draw[dashed] (-3.2,0)--(-2.2,1);
	\draw[dashed] (-3.2,0)--(-2.2,-1);
	\node  at (-5.3,-1) {$\mathbf{s}^{}_1$};
	\node at (-5.3,1) {$\mathbf{s}^{}_1$};
	\node [above] at (-3.6,0) {$\mathbf{s}^{}_2,\phi,h$};
	\node at (-2.0,1.0) {$\phi$};
	\node at (-2.0,-1.0) {$\mathbf{s}^{}_2$};
        \draw[dashed] (-1.4,0)--(-0.4,0);
    \draw[dashed] (-0.4,0)--(0.4,1);
    \draw[dashed] (-0.4,0)--(0.4,-1);
    \node at (-1.7,0) {$\phi,\mathbf{s}^{}_2,h$}; 
    \node at (0.5,1) {$\mathbf{s}^{}_1$};
    \node at (0.5,-1) {$\mathbf{s}^{}_1$};
        \draw[dashed] (1.2,1)--(2.0,0);
	\draw[dashed] (1.2,-1)--(2.0,0);
	\draw[dashed] (2.0,0)--(3.0,0);
	\draw[dashed] (3.0,0)--(4.0,1);
	\draw[dashed] (3.0,0)--(4.0,-1);
	\node  at (1.2,1.2) {$\mathbf{s}^{}_1$};
	\node at (1.2,-1.2) {$\mathbf{s}^{}_1$};
	\node [above] at (2.5,0) {$h$};
	\node at (4.0,1.2) {$h$};
	\node at (4.0,-1.2) {$\phi,\mathbf{s}^{}_2$};
     \end{tikzpicture}
 \end{center}
 \begin{center}
    \begin{tikzpicture}[line width=0.5 pt, scale=1.1]
        \draw[dashed] (-5,1)--(-4,0);
	\draw[dashed] (-5,-1)--(-4,0);
	\draw[dashed] (-4,0)--(-3,1);
	\draw[dashed] (-4,0)--(-3,-1);
	\node at (-5.4,1) {$\mathbf{s}^{}_1$};
	\node at (-5.4,-1) {$\mathbf{s}^{}_1$};
	\node at (-2.6,1) {$h,\phi,\mathbf{s}^{}_2$};
	\node at (-2.6,-1) {$h,\phi,\mathbf{s}^{}_2$};
        \draw[dashed] (-1,1)--(0,0);
	\draw[dashed] (-1,-1)--(0,0);
	\draw[dashed] (0,0)--(1.5,0);
	\draw[dashed] (1.5,0)--(2.5,1);
	\draw[dashed] (1.5,0)--(2.5,-1);
	\node  at (-1.2,-1) {$\mathbf{s}^{}_1$};
	\node at (-1.2,1) {$\mathbf{s}^{}_1$};
	\node [above] at (0.75,0) {$\phi,\mathbf{s}^{}_2,h$};
	\node at (2.9,1) {$h,\phi,\mathbf{s}^{}_2$};
	\node at (2.9,-1) {$h,\phi,\mathbf{s}_2$};
	\draw[dashed] (5,1)--(6,0);
	\draw[dashed] (6,0)--(6,-0.6);
	\draw[dashed] (5,-1.4)--(6,-0.6);
	\draw[dashed] (6,0)--(7,1);
	\draw[dashed] (6,-0.6)--(7,-1.4);
	\node  at (4.6,1) {$\mathbf{s}^{}_1$};
	\node at (7.4,1) {$\phi,\mathbf{s}^{}_2,h$};
	\node [right] at (6,-0.3) {$\mathbf{s}^{}_1$};
	\node at (7.2,-1.6) {$\phi,\mathbf{s}^{}_2,h$};
	\node at (4.6,-1.6) {$\mathbf{s}^{}_1$};
     \end{tikzpicture}
 \end{center}
\caption{The above diagrams contribute to the DM relic after EWSB in $\mathcal{Z}_{4}\times CP$ invariant scenario. }
\label{fig:ann_diag}
 \end{figure}
The matrix amplitude for the process $\mathbf{s}_{1}\mathbf{s}_{1}\rightarrow\mathbf{s}_{2}h$ via $\mathbf{s}_{2}$ mediated s channel diagram is
\begin{align}\label{eq:matrixamp1}
    \mathcal{M}_{\mathbf{s}_{2}^{}h}^{s}=\frac{\lambda_{\mathbf{s}_{1}^{}\mathbf{s}_{1}^{}\mathbf{s}_{2}^{}}~~\lambda_{\mathbf{s}_{2}^{}\mathbf{s}_{2}^{}h}}{M_{\mathbf{s}_{2}^{}}^{2}}\approx\Bigg(16(\lambda_{s'}+\lambda_{s''})v_{s}-\frac{M_{\mathbf{s}_{2}^{}}^{2}}{v_{s}^{}}\Bigg)\Bigg( \frac{c_{13}^{2}c_{23}^{2}s_{23}^{}}{v_{s}}+\frac{c_{23}^{3}c_{13}^{2}c_{12^{}}s_{13}^{}}{v_{s}^{}}+\frac{c_{13}^{3}-c_{13}^{}}{v}\Bigg).
\end{align}
Simillarly, the matrix amplitude for the process $\mathbf{s}_{1}\mathbf{s}_{1}\rightarrow\mathbf{s}_{2}\mathbf{s}_{2}$ via $\mathbf{s}_{2}$ mediated s channel diagram is
\begin{align}\label{eq:matrixamp2}
    \mathcal{M}_{\mathbf{s}_{2}^{}\mathbf{s}_{2}^{}}^{s}=\frac{\lambda_{\mathbf{s}_{1}^{}\mathbf{s}_{1}^{}\mathbf{s}_{2}^{}}~~\lambda_{\mathbf{s}_{2}^{}\mathbf{s}_{2}^{}\mathbf{s}_{2}^{}}}{M_{\mathbf{s}_{2}^{}}^{2}}\approx\Bigg(16(\lambda_{s'}+\lambda_{s''})v_{s}-\frac{M_{\mathbf{s}_{2}^{}}^{2}}{v_{s}^{}}\Bigg)\Bigg(\frac{c_{13}^{3}c_{23}^{3}}{v_{s}^{}}+\frac{s_{13}^{}}{v}\Bigg).
\end{align}
\section{UV Completion of the EFT Operator}
\label{app:UVcomp}
On top of the BSM fields ($N_i,\,\Phi$ and $S$ with same $\mathcal{Z}_4$ charge) present in our framework we add three BSM vector like heavy neutral fermion $\mathcal{F}_{L_\beta,R_\beta}\sim (1,1,0)$ (where $\beta$ varies from 1 to 3) singlet under SM gauge group and also transform trivially under the $\mathcal{Z}_4$ symmetry. The Lagrangian responsible for generating the active neutrino mass in this model is given by
\begin{equation}\label{eq:lagUV}
 -\mathcal{L}_\nu\supset \mathbb{Y}_{\alpha\beta}\overline{l_{L_\alpha}}  \widetilde{H}\mathcal{F}_{R_\beta}+\mathbb{K}_{\beta i}\overline{\mathcal{F}_{L_\beta}}\Phi N_{i}+\mathbb{M}_{\beta\gamma}\overline{\mathcal{F}_{L_\beta}}\mathcal{F}_{R_\gamma}+\frac{{M}_{ij}}{2}\overline{N^c_i}N_j + h.c.\;.
\end{equation}

\begin{figure}[htb!]
\begin{center}

\begin{tikzpicture}[scale=1.0,line width=0.9pt]

\draw[->] (-3.5,0)--(-2.5,0);
\draw[->] (-2.5,0)--(-1.6,0);
\draw[->] (-1.6,0)--(-0.9,0);
\draw[-] (-0.9,0)--(0,0);

\draw[->] (2.5,0)--(1.7,0);
\draw[-] (1.4,0)--(1.7,0);
\draw[->] (3.5,0)--(2.5,0);
\draw[-] (0.0,0)--(1.0,0);
\draw[->] (1.4,0)--(1.0,0);

\draw[dashed] (-2.2,1)--(-2.2,0);
\draw[dashed] (2.2,1)--(2.2,0);

\draw[dashed] (-0.7,-1)--(-0.7,0);
\draw[dashed] (0.7,-1)--(0.7,0);

\draw[line width=1pt] (-1.5,0.13)--(-1.3,-0.13);
\draw[line width=1pt] (-1.5,-0.13)--(-1.3,0.13);

\draw[line width=1pt] (-0.12,0.12)--(0.12,-0.12);
\draw[line width=1pt] (-0.12,-0.12)--(0.12,0.12);

\draw[line width=1pt] (1.3,0.13)--(1.5,-0.13);
\draw[line width=1pt] (1.3,-0.13)--(1.5,0.13);

\node[left] at (-3.5,0) {$\nu_L^{}$};
\node[right] at (3.5,0) {$\nu_L^{}$};

\node[above] at (-2.2,1) {$\langle H\rangle$};
\node[above] at (2.2,1) {$\langle H\rangle$};

\node[below] at (-0.7,-1) {$\langle\Phi\rangle$};
\node[below] at (0.7,-1) {$\langle\Phi\rangle$};

\node[below] at (-1.85,-0.05) {$\mathcal F_R$};
\node[below] at (-1.0,-0.05) {$\mathcal F_L$};
\node[below] at (1.95,-0.05) {$\mathcal F_R$};
\node[below] at (1.05,-0.05) {$\mathcal F_L$};
\node[below] at (-0.4,-0.05) {$N$};
\node[below] at (0.4,-0.05) {$N$};

\end{tikzpicture}

\end{center}

\caption{
Tree level diagram of active neutrino mass generation for the UV complete model.
}

\label{fig:uvcompletion}

\end{figure}
After the EWSB, with the discrete symmetry $\mathcal{Z}_4$ already broken at a much earlier stage through the VEV of $\Phi$, i.e., $\langle\Phi\rangle=v_{\phi}$. The mass terms involving  $\nu_L^{}\,, \mathcal{F}_L,\,\mathcal{F}_R$ and $N$ can be expressed as 
\begin{equation}
-\mathcal{L}_{\nu}
=
\frac{1}{2}
\begin{pmatrix}
\overline{\nu_L} &
\overline{\mathcal{F}_L} &
\overline{N^C} &
\overline{\mathcal{F}_R^C}
\end{pmatrix}
\begin{pmatrix}
0 & 0 & 0 & m_D \\
0 & 0 & m_K &  \mathbb{M}\\
0 & m_K^T & M & 0 \\
m_D^T & \mathbb{M}^T & 0 & 0
\end{pmatrix}
\begin{pmatrix}
\nu_L^C \\
\mathcal{F}_L^C \\
N \\
\mathcal{F}_R
\end{pmatrix}\,,
\end{equation}
where $m_D=\mathbb{Y}\,v/\sqrt{2}$ and $m_K=\mathbb{K}\,v_{\phi}$ are the Dirac mass terms for $\nu$ and $N$, respectively. After the block diagonalization of this mass matrix one can find form of active neutrino mass matrix as
\begin{align}\label{eq:numassUV}
    m_{\nu}^{\rm UV}\simeq v\,v_\phi \mathbb{Y}\,\mathbb{M}^{-1}\,\mathbb{K}\,M^{-1}\mathbb{K}^{T}\left(\mathbb{M}^{-1}\right)^{T}\mathbb{Y}^T
\end{align}
It is important to note that in this model we assume the following mass hierarchy: $\mathbb{M}\gg M>m_K\gg m_D$. After integrating out the heavy degrees of fields $\mathcal{F}$ one can get the desired EFT operator ($\overline{l}  \widetilde{H} \Phi N$) with the corresponding matching condition
\begin{align}
    \frac{\mathcal{Y}}{\Lambda}\equiv\frac{\mathbb{Y}\,\mathbb{K}}{\mathbb{M}}\,.
\end{align}
\section{Approximated form of the reaction density $\gamma_S$ and comparison between various $\Delta L=1$ processes}
\label{app:reation}
\begin{figure}[!htb]
\centering
\includegraphics[scale=0.35]{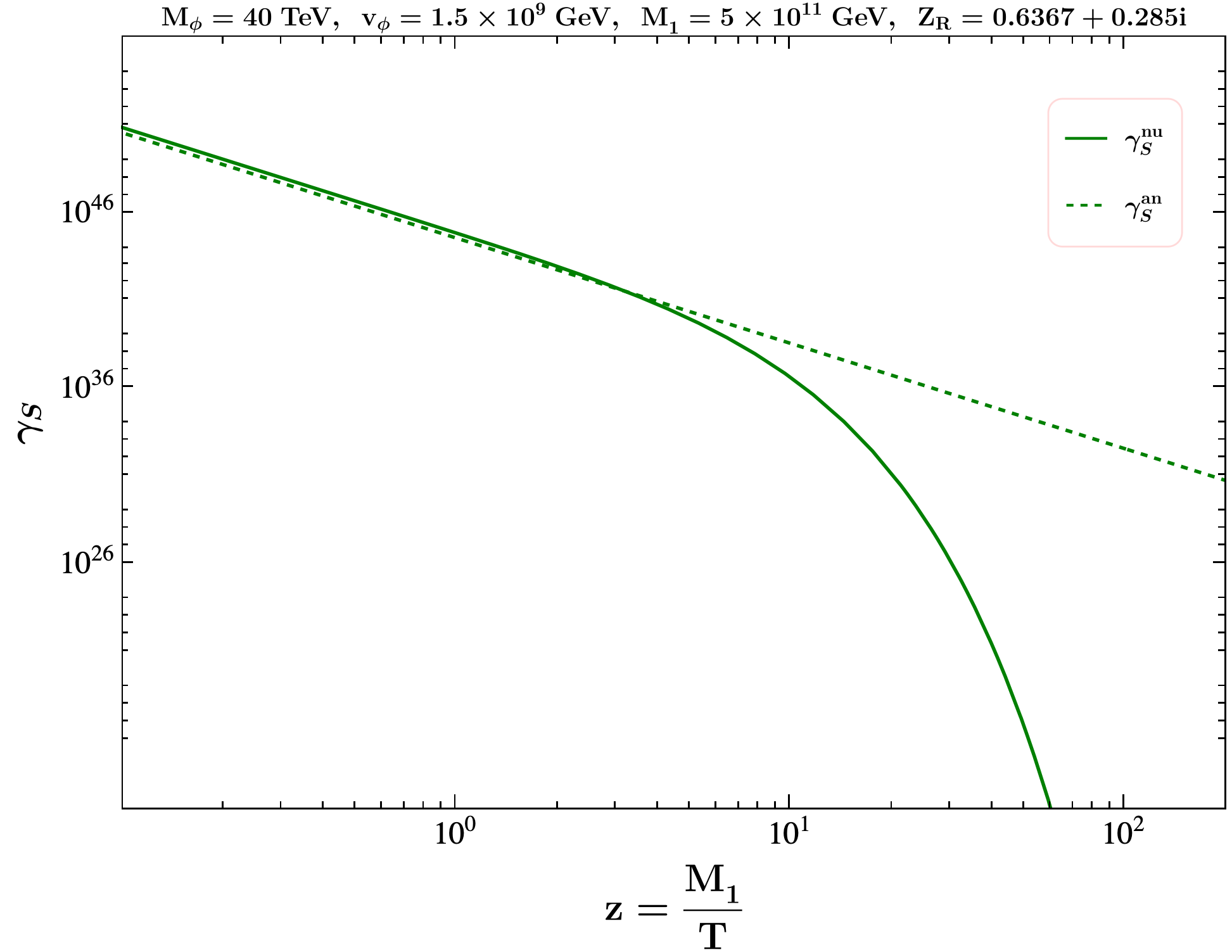}
\includegraphics[scale=0.35]{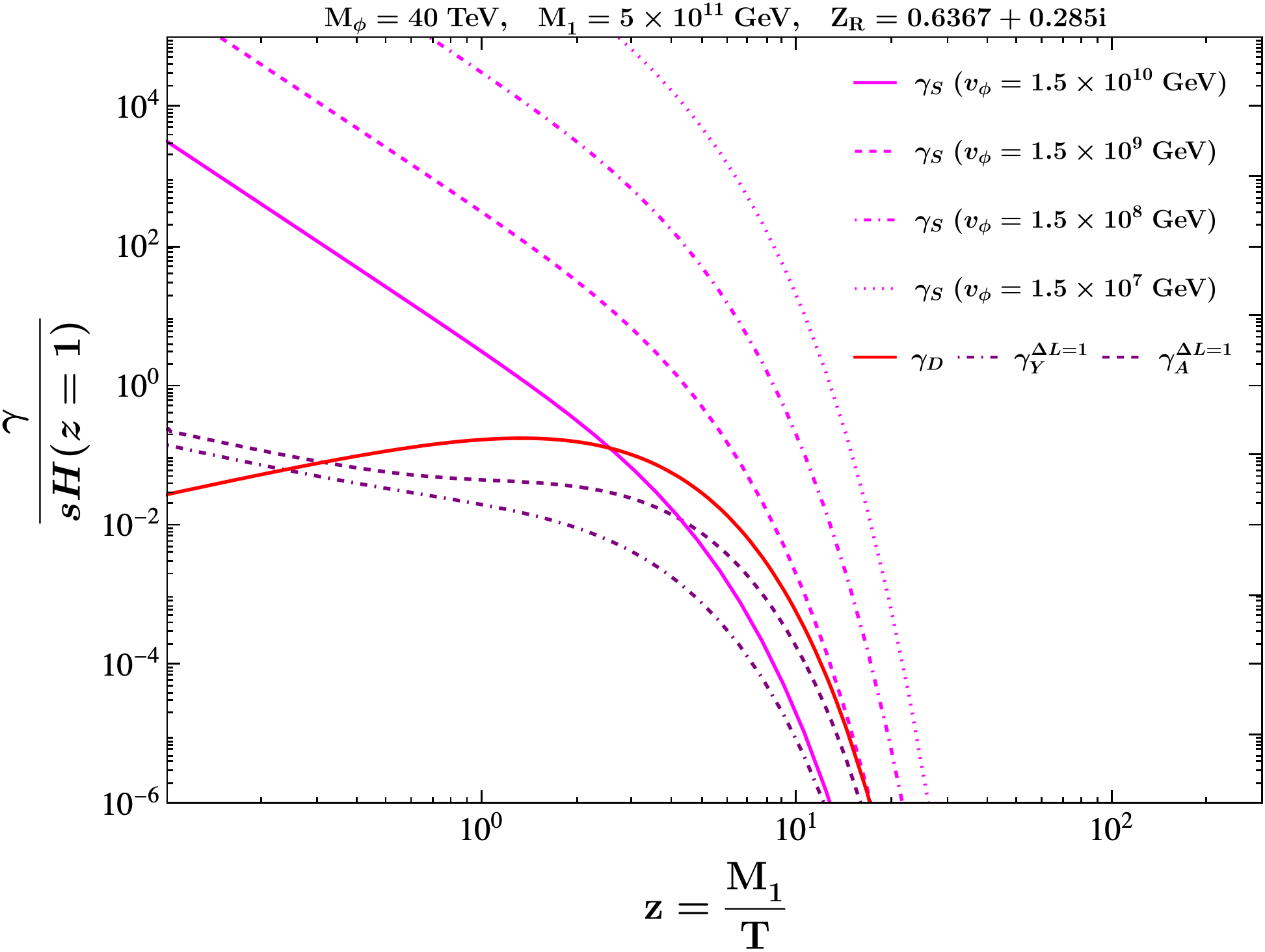}
\caption{\emph{Left:} Comparison of the analytical form of the scattering reaction density (green dashed line), valid in the high-temperature limit $z\ll1$, with the corresponding reaction density obtained from the full numerical integration (green solid line). \emph{Right:} Comparison of the reaction densities for different $\Delta L=1$ scattering processes. The processes involving quarks and $SU(2)\otimes U(1)$ gauge bosons are shown by the magenta dot-dashed and dashed lines, respectively. At the same time, the purple curves represent the $N\mathit{\Phi}\to l H$ scattering processes for different $v_{\phi}$, as mentioned in the legend.}
\label{fig:compgamma}
\end{figure}

Starting from Eq.\eqref{eq:reactscat}, the reaction density for the scattering process $N_{i}\mathit{\Phi}\to lH$ can be expressed as
\begin{equation}\label{eq:gammas}
    \gamma(N_i \phi \r l H)=\frac{T}{512\pi^6} \frac{{(y_\nu^\dagger y_\nu)}_{ii}}{v_\phi^2}\int\text{d}\tilde{E}\,\frac{{\sqrt{\lambda(\tilde{E},M^2_\phi,M^2_i)}}}{4\sqrt{\tilde{E}}}\,K_1\left(\frac{\sqrt{\tilde{E}}}{T}\right)(\tilde{E}+M^2_i-M^2_\phi)\pi\,,
\end{equation}
where $\lambda(a,b,c)\equiv(a-b-c)^2-4bc$ denotes the Källén function. In obtaining Eq.~\eqref{eq:gammas}, we have used the the center-of-mass momenta of the initial and final states, ${\rm |P|}=\sqrt{\lambda(\tilde{E},M_{1}^2,M_{\phi}^2)/4\tilde{E}}$ and ${\rm |Q|}=\sqrt{\tilde{E}/4}$, together with the angular integration of the squared matrix element, $\int{\rm |\mathcal{M}|^{2}}d\Omega=\{(y_\nu^\dagger y_\nu)_{ii}/v_{\phi}^2\}(\tilde{E}+M^2_i-M^2_\phi)\pi$~\cite{Bhattacharya:2023kws}. In the high-temperature regime $T\gg M_{i}$, corresponding to $z=M_{i}/T\ll1$, and for $M_{\phi}\ll M_{i}$, the dominant contribution to the integral arises from the region of center-of-mass energies $\sqrt{\tilde{E}}\sim T\gg M_{i}$. In this limit, the mass dependence of the kinematic factors becomes negligible, and the total reaction density, including the CP-conjugate process, reduces to the following form:
\begin{equation}\label{eq:gamsu}
    \gamma_{S}=\gamma(N\mathit{\Phi}\rightarrow LH)+\gamma(N\mathit{\Phi}\rightarrow \Bar{L}\Bar{H})\approx\frac{{(y_\nu^\dagger y_\nu)}_{ii}T}{1024\,\pi^5\,v_\phi^2}\int d\tilde{E}\,\tilde{E}^{3/2}K_1\left(\frac{\sqrt{\tilde{E}}}{T}\right)=\frac{{(y_\nu^\dagger y_\nu)}_{ii}T^6}{512\,\pi^5\,v_\phi^2}\int dx\,x^4K_1\left(x\right)\,.
\end{equation}
For $M_{\phi}\ll M_{i}$, the lower limit of the $\tilde{E}$ integral is $\tilde{E}_{min}\simeq M_{i}^2$. Upon introducing $x=\sqrt{\tilde{E}}/T$ , this corresponds to $x_{min}=M_{i}/T=z$, which approaches zero in the high-temperature regime $T\gg M_{i}$. Thus, the lower limit of the $x$ integration can be approximated by $x_{min}\simeq0$, with this the integral reduces to a standard Bessel-function integral, $\int_{0}^\infty dx\,x^4K_1\left(x\right)=16$. The scattering reaction density consequently takes the simplified analytical form
\begin{equation}
    \gamma_S\approx\frac{{(y_\nu^\dagger y_\nu)}_{ii}T^6}{32\,\pi^5\,v_\phi^2}=\frac{{(y_\nu^\dagger y_\nu)}_{ii}M_{i}^6}{32\,\pi^5\,v_\phi^2\,z^6}=\frac{(y_\nu^\dagger~y_\nu)}{32\pi^2}\frac{M_1^4}{z}f_c(M_1,v_\phi,z)\,,
\end{equation}
where $f_c(M_1,v_\phi,z)={M_1^2}/({v_\phi^2}{\pi^3 z^5})$. $\gamma_S$ has been expressed in this form to facilitate a direct comparison with the corresponding decay reaction density, $\gamma_D$. The left panel of Fig:~\ref{fig:compgamma} compares the analytical high-temperature approximation, $\gamma_S^{\rm an}$, with the full numerical result, $\gamma_S^{\rm nu}$. The analytical result agrees well with the numerical evaluation in the $z\lesssim1$ regime, confirming the validity of the high-temperature approximation. Nevertheless, in solving the Boltzmann equations for leptogenesis, we retain the full numerical reaction density throughout, thereby avoiding the high-temperature approximation outside its regime of validity. 

One may wonder why, in deriving the BEQ for $\Delta L$ (see Eq.~\eqref{eq:beql}), we retain only the $N\mathit{\Phi}\to l H$ scattering contribution while neglecting the other $\Delta L=1$ processes involving SM quarks and SM gauge bosons. These include, for instance, $N\bar{Q}\to t\bar{l},~N l\to tQ$, and the gauge-boson-involved processes $NA\to lH,~Nl\to A\bar{H},$ and $N\bar{H}\to lA,$ with $A=B_{\mu},W_{\mu}^{a}$. To verify that this approximation does not affect the resulting lepton asymmetry, we compare the corresponding reaction densities in the right panel of Fig.~\ref{fig:compgamma}, normalized in units of $sH$. The quark- and gauge-boson-involved $\Delta L=1$ processes are shown by the magenta dot-dashed ($\gamma_{Y}^{\Delta L=1}$) and dashed ($\gamma_{A}^{\Delta L=1}$) curves, respectively, while the $N\mathit{\Phi}\to lH$ contribution, $\gamma_S$, is shown by the purple curves for different values of $v_{\phi}$ indicated in the figure legend. The comparison clearly shows that $\gamma_{S}$ dominates over the other $\Delta L=1$ scattering contributions (across most of our scattering-dominated leptogenesis parameter space). Also note that, within the range $z\in[1,10]$, where the washout of the generated lepton asymmetry is most effective, the contributions from the quark- and gauge-boson-involved processes remain subdominant to the RHN decay contribution, $\gamma_D$ (depicted as the red line in Fig.~\ref{fig:compgamma}). Their impact on the evolution of $Y_{\Delta L}$ is therefore negligible, justifying their omission from the BEQ. The reaction densities of the various $\Delta L=1$ scattering processes are evaluated using the corresponding cross sections given in Refs.~\cite{Plumacher:1996kc, Pilaftsis:1997jf}.
\twocolumngrid
\bibliography{sample}
\end{document}